\documentclass[]{JHEP3}
\usepackage{graphicx}
\usepackage{amsmath,amssymb,amscd,amsfonts,bm}

\newcommand{\as}{\alpha_{\mathrm{s}}}
\newcommand{\LA}{\mathrm{A}}
\newcommand{\LB}{\mathrm{B}}
\newcommand{\LF}{\mathrm{F}}
\newcommand{\LI}{\mathrm{I}}
\newcommand{\LR}{\mathrm{R}}

\newcommand{\LT}{\mathrm{T}}
\newcommand{\La}{\mathrm{a}}
\newcommand{\Lb}{\mathrm{b}}
\newcommand{\Lc}{\mathrm{c}}
\newcommand{\Lf}{\mathrm{f}}
\newcommand{\Lg}{\mathrm{g}}

\newcommand{\Lp}{\mathrm{p}}

\def\mi{{\mathrm i}}

\def\ket#1{\big|{#1}\big\rangle}
\def\bra#1{\big\langle{#1}\big|}
\def\brax#1{\big\langle{#1}}   
\def\<>#1{\big\langle{#1}\big\rangle}
\def\[]#1{\big[{#1}\big]}

\def\sket#1{\big|{#1}\big)}
\def\sbra#1{\big({#1}\big|}
\def\sbrax#1{\big({#1}}        

\title{On the transverse momentum in $Z$-boson production in a virtuality ordered parton shower}

\author{Zolt\'an Nagy \\
DESY\\
Notkestrasse 85\\
22607 Hamburg, Germany\\
E-mail: \email{Zoltan.Nagy@desy.de}
}

\author{Davison E. Soper\\
Institute of Theoretical Science\\
University of Oregon\\
Eugene, OR  97403-5203, USA\\
E-mail: \email{soper@uoregon.edu}
}

\abstract{
Cross sections for physical processes that involve very different momentum scales in the same process will involve large logarithms of the ratio of the momentum scales when calculated in perturbation theory. One goal of calculations using parton showers is to sum these large logarithms. We ask whether this goal is achieved for the transverse momentum distribution of a $Z$-boson produced in hadron-hadron collisions when the shower is organized with higher virtuality parton splittings coming first, followed successively by lower virtuality parton splittings. We find that the virtuality ordered shower works well in reproducing the known QCD result.
}

\keywords{perturbative QCD, parton shower}
\preprint{DESY 09-201}
\begin{document}

\section{Introduction}

Parton shower algorithms with hadronization models provide a way of generating simulated events according to approximations based on QCD. Since complete final states are generated, one can generate completely exclusive cross sections in this way. By summing over variables that one chooses not to examine, one can also make predictions for inclusive observables. Of special interest are predictions that, in a perturbative expansion, involve large logarithms of ratios of different momentum scales in the physical problem. 

An important example is the distribution of the transverse momentum $\bm p_\perp$ of $Z$-bosons produced in hadron-hadron collisions at some fixed rapidity $Y$, ${d\sigma}/({d\bm p_\perp\,dY})$. When $\bm p_\perp^2 \ll M_Z^2$, the perturbative expansion of this cross section contains two powers of the large logarithm $\log(\bm p_\perp^2/M_Z^2)$ per power of $\as$. The large logarithms spoil the usefulness of fixed order perturbation theory for this observable and for observables that contain similar large logarithms. A parton shower algorithm sums contributions to the desired cross section that contain arbitrarily high powers of $\as$; thus it sums the accompanying logarithms. For this reason, one can hope that a parton shower calculation will do better than a fixed order perturbative calculation in the region $\bm p_\perp^2 \ll M_Z^2$ and analogous regions for other processes.

Indeed, the basic approximation in a parton shower is that one parton splits into two daughter partons with a probability that matches the singularities of the QCD matrix element when the two daughter partons are collinear or one of them is soft. It is just these soft/collinear configurations that give rise to the large logarithms. Thus one can hope that the cross section generated by a parton shower will be a good approximation to the true QCD result.

In many cases, including the $Z$-boson transverse momentum distribution, there are predictions based on the full field theory. That is, we know that the cross section ``exponentiates'' in a sense that one can state precisely and we know some of the leading coefficients that appear in the exponent of the formula that expresses the QCD result. 

When, for a particular process, one knows the summation of large logarithms in full QCD, then it is of significant interest to investigate whether a given shower algorithm produces matching results. To do this, one needs to derive the corresponding summation in the shower model, deriving the appropriate evolution equation for the observable in question from the general evolution equation for the shower algorithm.

We have argued above that, because parton shower algorithms are generally based on parton splitting probabilities that have the proper soft/collinear singularities, these algorithms may provide good approximations to the full QCD result in particular cases involving summing large logarithms. However, a parton shower algorithm contains several ingredients beyond the parton splitting probabilities. Among theses are the momentum mapping, the color and spin treatment, and the choice of evolution variable. Depending on the choice of these ingredients, one may obtain agreement with full QCD for a particular observable or one may fail to obtain agreement. Certainly, it is widely understood that a virtuality ordered parton shower without a proper inclusion of the effects of quantum interference can get results for some observables that do not match QCD. In contrast, there is an extensive body of literature \cite{ErmolaevQCDSummation, MuellerQCDSummation, angleorderMC, CataniEmilio32, CataniQCDSummation, angleorder, CataniMCsummation, Catani32etalQCDsummation, CacciariQCDSummation} that suggests that parton showers based on ordering in parton emission angles does better for many observables. In this paper, we will indeed see that, for the $Z$-boson transverse momentum distribution, the choice of ingredients matters.

Because the choice of ingredients matters, we think it important to validate shower evolution schemes against known results for summations of large logarithms. We believe that such a validation program could be useful for understanding the range of validity of current parton shower event generators and could provide important guidelines for the current and future development of such programs. 

In ref.~\cite{NSDGLAP}, in response to ref.~\cite{Dokshitzer}, we investigated the parton energy distribution in electron-positron annihilation as predicted by the virtuality ordered, color-dipole based parton shower algorithm of refs.~\cite{NSI, NSII, NSIII} and by $k_{\rm T}$-ordered variants of this. We concluded that the predictions of these parton shower algorithms are consistent with the field theory result represented by the well know DGLAP evolution equation \cite{DGLAP}. This conclusion was confirmed by ref.~\cite{SkandsWeinzierl}, which included numerical studies. The parton energy distribution is an example of the general program of summing large logarithms, namely the logarithm of the resolution scale for finding the partons ($\sim$ jets) divided by the electron-positron energy. However, this is a rather simple example in that there is only one logarithm per power of $\as$. Observables for which there are two logarithms per power of $\as$ present a much more subtle problem.

In this paper we investigate the transverse momentum distribution of a $Z$-boson produced in hadron-hadron collisions, the Drell-Yan process. Here, there are two logarithms per power of $\as$. We use a slightly modified version\footnote{We will find that we need to change the momentum mapping for initial state radiation from that of ref.~\cite{NSI}. In addition, one of the choices given in ref.~\cite{NSIII} for a certain function $A_{lk}'$ that was left unspecified in ref.~\cite{NSI} gives a satisfactory result, but another of the choices does not work.} of the shower evolution of ref.~\cite{NSI}. This evolution equation describes the evolution of the partonic states in a  fully exclusive way. We manipulate the shower equation to produce an evolution equation for the transverse momentum distribution of the $Z$-boson and we compare the result to the well known field theory prediction that is given in ref.~\cite{CSS}. We find that the virtuality ordered shower works well in reproducing the full QCD result.

The shower evolution equation of ref.~\cite{NSI} includes quantum interference among colors and spins. We note that this shower evolution equation is not directly practical for generating events. A simple approximation to this evolution equation is to average over spins and take the leading color approximation, which yields an evolution equation \cite{NSII} that can be written as a Markov process and is thus directly practical for generating events. We treat the full evolution equation, but we will see that the same results with some small adjustments hold for the $Z$-boson transverse momentum distribution produced by the spin-averaged, leading-color shower evolution. 

The investigation that we carry out in this paper, and that we believe would be useful for other observables and other shower algorithms, is analytical. That is, we want to test whether the transverse momentum distribution produced by the shower algorithm has the proper exponential form and, if so, whether the coefficients in the exponent are correct. A followup study, not addressed in this paper, would be numerical: how well does an actual implementation of the parton shower algorithm produce results that match numerical results given by the QCD formula. In this case, the summed QCD results, including nonperturbative parameters that are fit to experiment, could be obtained from the \textsc{Resbos} code \cite{Resbos}. We believe that one should start with an analytical study rather than a numerical study for the following reason. The parton shower, for the case of the production a $Z$-boson with small $\bm p_\perp$, represents the physics on ``hard'' scales from $M_Z$ to a few GeV. This is perturbative physics that is adjustable only to a limited extent. In particular, one can modify the parton splitting probabilities in a fashion that does not change them in the soft and collinear limits.\footnote{The \textsc{Vincia} parton shower algorithm \cite{Vincia} takes advantage of this flexibility.} A full parton shower event generator also contains elements, such as a hadronization model and a model for the underlying event, that represent soft scale physics. One can adjust the models for the soft physics by tuning various parameters. For very large $M_Z^2/\Lambda_{\rm QCD}^2$ and $M_Z^2/\bm p_\perp^2$, only the part of the shower algorithm that cannot be tuned should matter. However, for realistic values of these parameters, the tunable parts of the parton shower can make a numerical difference. One would not want to tune the parameters in order to repair a numerical disagreement that was actually caused by a mismatch between the parton shower and full QCD with respect to the hard scale physics that is fixed. The way to check the match of hard scale physics is to compare analytically. Once this is checked, the numerical comparison is appropriate and needed.
 
We organize this paper as follows. In sections~\ref{sec:states}, \ref{sec:kinematics}, and \ref{sec:framework}, we review briefly the needed features of the shower evolution of ref.~\cite{NSI} and set up the notation for our analysis. Then in section~\ref{sec:outline} we outline the derivation to come and state the nature of the approximations that we will need. The derivation is given in sections~\ref{sec:partonicevolution}, \ref{sec:realsplitting}, \ref{sec:virtualsplitting}, \ref{sec:evolutionI}, and \ref{sec:evolutionII}. The solution of the evolution equations is given in section~\ref{sec:solution} and the result is compared to the full QCD result in section~\ref{sec:result}. We discuss what one would get with other sorts of shower evolution in section~\ref{sec:other} and summarize the results in section~\ref{sec:conclusions}. In appendix \ref{sec:inclusive}, we summarize results from refs.~\cite{NSI,NSII,NSIII} that are used in this paper. In appendix \ref{sec:Bessel} we prove a certain property of integrals of the $J_0$ Bessel function.

\section{Shower states and shower evolution}
\label{sec:states}

We analyze the transverse momentum distribution of a $Z$-boson as generated by the parton shower evolution equations of ref.~\cite{NSI}. The organizing principle of the shower is that one starts at the hard interaction $q + \bar q \to Z$ and moves to softer interactions, always factoring the softer interaction from previous harder interactions.

In the notation of ref.~\cite{NSI}, states in the sense of statistical mechanics are represented by ket vectors $\sket{\rho}$, while possible measurements are represented by bra vectors $\sbra{F}$. Thus $\sbrax{F}\sket{\rho}$ is the cross section that one obtains a particular result $F$ from a measurement on an ensemble of systems represented by $\sket{\rho}$. We use basis states labelled by lists $\{p,f,s',c',s,c\}_{m}$ of parton quantum numbers for two initial state partons and $m$ final state partons. As the shower progresses, $m$ increases. The momenta of the final state partons are $\{p_1,\dots,p_m\}$. The first final state parton is the $Z$-boson, with momentum $p_1 \equiv p_Z$. The momenta of the initial state partons are specified by giving their momentum fractions, $\eta_\La$ and $\eta_\Lb$. Then their momenta are
\begin{equation}
\begin{split}
\label{eq:etadefs}
p_\La ={}& \eta_\La p_\LA
\;\;, 
\\
p_\Lb ={}& \eta_\Lb p_\LB
\;\;, 
\end{split}
\end{equation}
where $p_\LA$ and $p_\LB$ are the momenta of the incoming hadrons, treated as massless. The flavors $\Lg, u, \bar u, d, \bar d, \dots$ of the final state partons are $\{f_1,\dots,f_m\}$, while the flavors of the initial state partons are denoted by $a$ and $b$. The spins are specified by
\begin{equation}
\{s',s\}_{m} = \{s_\La, s_\Lb, s_1, \dots, s_m, 
                s'_\La, s'_\Lb, s'_1, \dots, s'_m\}
\;\;,
\end{equation}
with two spin labels for each parton because we use the quantum density matrix in spin space in order to represent possible interference among spin states. The colors are similarly specified by $\{c',c\}_{m}$. Quantum spin and color states are represented by bra and ket vectors with angle brackets, as in the quantum spin inner product $\brax{\{s'\}_{m}}\ket{\{s\}_{m}}$. The lists $\{p,f,s',c',s,c\}_{m}$ include all of these parton quantum numbers. The unit operator in the space of statistical states can be written as
\begin{equation}
\label{eq:completeness}
1 = 
\sum_m \frac{1}{m!}\int \big[d\{p,f,s',c',s,c\}_{m}\big]\
\sket{\{p,f,s',c',s,c\}_{m}}
\sbra{\{p,f,s',c',s,c\}_{m}}
\;\;,
\end{equation}
where $\big[d\{p,f,s',c',s,c\}_{m}\big]$ indicates integrating and summing over all of the quantum numbers. A much more detailed specification of our notation is provided in ref.~\cite{NSI}. In ref.~\cite{NSI}, the quarks can have non-zero masses, but in this paper we take all of the partons except for the $Z$-boson to be massless.

Let $\sket{\rho(t)} = {\cal U}(t,0)\sket{\rho(0)}$ be the statistical state at shower time $t$. Here $t=0$ at the beginning of the shower, which starts with the hard process $q + \bar q \to Z$, and $t = t_\Lf$ gives the statistical state at the end of the shower, before hadronization. (We do not discuss a hadronization model in this paper.) The total cross section for producing a $Z$-boson is expressed using the vector $\sbra{1}$, which represents the totally inclusive measurement,
\begin{equation}
\begin{split}
\sigma_Z = 
\sbrax{1}\sket{\rho(t)}
={}&
\sum_m \frac{1}{m!}\int \big[d\{p,f,s',c',s,c\}_{m}\big]
\\&\quad\times
  \sbrax{1}\sket{\{p,f,s',c',s,c\}_{m}}
  \sbrax{\{p,f,s',c',s,c\}_{m}}\sket{\rho(t)}
\\={}&
\sum_m \frac{1}{m!}\int \big[d\{p,f,s',c',s,c\}_{m}\big]\,
  \brax{\{s'\}_{m}}\ket{\{s\}_{m}}\,
  \brax{\{c'\}_{m}}\ket{\{c\}_{m}}
\\ &\times
  \sbrax{\{p,f,s',c',s,c\}_{m}}\sket{\rho(t)}
\;\;.
\end{split}
\end{equation}
Here we use the definition
\begin{equation}
\label{eq:1def}
\sbrax{1}\sket{\{p,f,s',c',s,c\}_{m}}
=
\brax{\{s'\}_{m}}\ket{\{s\}_{m}}\,
\brax{\{c'\}_{m}}\ket{\{c\}_{m}}
\;\;.
\end{equation}
The total $Z$-production cross section is independent of $t$: the shower evolution maintains $\sbra{1}{\cal U}(t,0) = \sbra{1}$.

We are interested in the differential cross section ${d\sigma}/({d\bm p_\perp\,dY})$ as obtained in the shower at the shower final time $t_\Lf$,
\begin{equation}
\frac{d\sigma}{d\bm p_\perp\,dY}
= 
\sbrax{\bm p_\perp, Y}\sket{\rho(t_\Lf)}
\;\;.
\end{equation}
Here, the measurement function $\sbra{\bm p_{\perp}, Y}$ measures the cross section for the $Z$-boson to have transverse momentum $\bm p_{\perp}$ and rapidity $Y$.\footnote{In general, we denote vectors in two transverse dimensions by boldface letters like $\bm p_\perp$. The transverse part of a four-vector $p$ is $p_\perp$. Then $p_\perp^2 = - \bm p_\perp^2 < 0$; however, to avoid confusion, we avoid writing $p_\perp^2$.} The definition is
\begin{equation}
\begin{split}
\label{eq:pThodef}
\sbrax{\bm p_{\perp}, Y}\sket{\rho(t)}
={}&
\sum_m \frac{1}{m!}\int \big[d\{p,f,s',c',s,c\}_{m}\big]
\\ &\quad\times
  \sbrax{\bm p_{\perp}, Y}\sket{\{p,f,s',c',s,c\}_{m}}
  \sbrax{\{p,f,s',c',s,c\}_{m}}\sket{\rho}
\\={}&
\sum_m \frac{1}{m!}\int \big[d\{p,f,s',c',s,c\}_{m}\big]
  \brax{\{s'\}_{m}}\ket{\{s\}_{m}}\,
  \brax{\{c'\}_{m}}\ket{\{c\}_{m}}
\\ &\quad\times
  \delta(\bm p_{Z,\perp} - \bm p_\perp)\
  \delta\!\left(\frac{1}{2}\log\left(\frac{p_Z\!\cdot\! p_B}{p_Z\!\cdot\! p_A}\right)
  -Y\right)
\\ &\quad\times
 \sbrax{\{p,f,s',c',s,c\}_{m}}\sket{\rho(t)}
\;\;.
\end{split}
\end{equation}
The starting point of shower evolution is the Born cross section,
\begin{equation}
\label{eq:born}
\left[\frac{d\sigma}{d\bm p_\perp\,dY}\right]_{\rm Born}
= 
\sbrax{\bm p_\perp, Y}\sket{\rho(0)}
\;\;,
\end{equation}
which is proportional to $\delta(\bm p_\perp)$. (See eq.~(\ref{eq:initialcondition}) in section~\ref{sec:ToStudy} for details.)

The shower evolution is specified in ref.~\cite{NSI} in the form
\begin{equation}
\label{eq:evolution0}
\frac{d}{dt}\,\sket{\rho(t)}
= [{\cal H}_\LI(t) - {\cal V}(t)]\sket{\rho(t)}
\;\;.
\end{equation}
Here ${\cal H}_\LI(t)$ is the splitting operator, which takes a basis state with $m$ final state partons and changes it to a state with $m+1$ final state partons. Next, ${\cal V}(t)$ is a ``virtual splitting'' operator that leaves number of partons and their momenta, flavors, and spins unchanged.\footnote{In general, the operator ${\cal V}(t)$ is a non-trivial operator on the partonic color space. In the leading color approximation, valid for $N_\Lc \to \infty$, it is diagonal in the color space, as described in ref.~\cite{NSII}. The derivation in this paper is given for the exact treatment of color, but works also in the leading color approximation.} The operator ${\cal V}(t)$ is determined from ${\cal H}_\LI(t)$ in such a way that
\begin{equation}
\sbra{1}{\cal V}(t) = \sbra{1}{\cal H}_\LI(t)
\;\;.
\end{equation}
With this condition,
\begin{equation}
\frac{d}{dt}\,\sbrax{1}\sket{\rho(t)} = 0
\;\;,
\end{equation}
so that the total cross section to produce a $Z$-boson remains the Born cross section, even though the $Z$-boson momentum changes as a result of recoils against parton splittings in the shower. In particular, the $Z$-boson transverse momentum distribution broadens as the shower develops from $t=0$ to $t = t_\Lf$. We need to follow the shower evolution to find how the transverse momentum distribution broadens.

\section{Initial state splitting kinematics}
\label{sec:kinematics}

We will first need some kinematics for the description of an initial state splitting. For notational convenience, we suppose that it is parton ``a'' that splits.

\FIGURE{
\centerline{\includegraphics[width = 8 cm]{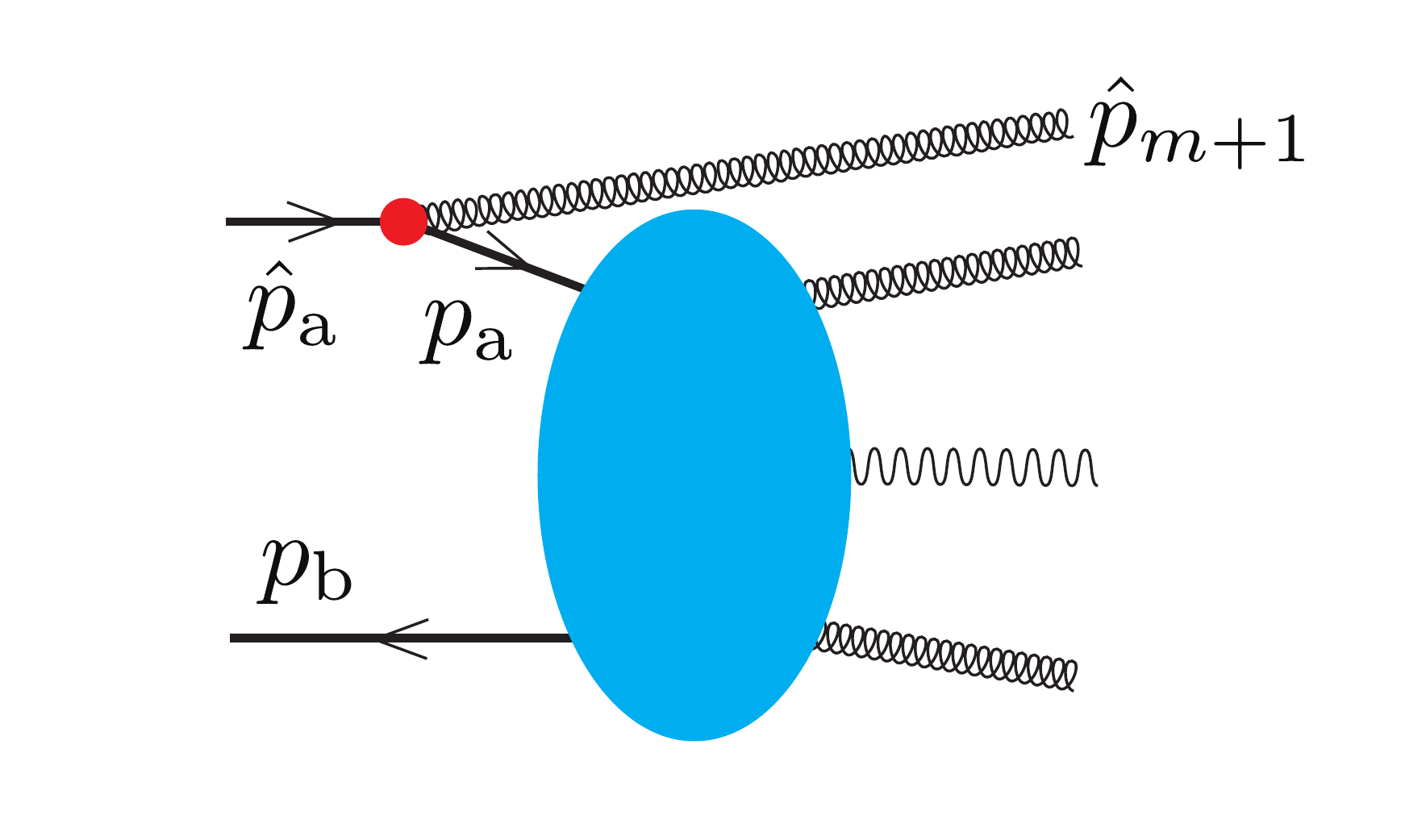}}
\caption{Illustration of the kinematics for an initial state splitting.}
\label{fig:kinematics}
} 

The initial state parton with momentum $p_\La$ splits, in backward evolution, to a new initial state parton with momentum $\hat p_\La$ and a final state parton with momentum $\hat p_{m+1}$, as illustrated in figure~\ref{fig:kinematics}. The other initial state parton has momentum $p_\Lb$ before the splitting and momentum $\hat p_\Lb$, equal to $p_\Lb$, after the splitting. In this paper,\footnote{In ref.~\cite{NSI}, the splitting operator is expressed in terms of momenta rather than splitting variables $y,z,\phi$ and we did not specify a choice of splitting variables. In ref.~\cite{NSII}, we did make a choice, but that choice is not the same as the choice used here.} we describe the splitting using splitting variables $(y,z,\phi)$ or, alternatively, $(y,\bm k_\perp)$, defined by
\begin{equation}
\begin{split}
\label{eq:splittingstart}
\hat p_{m+1} ={}& \frac{1-z}{z} (1+y)\, p_\La 
+ z\,\frac{y}{1+y}\,p_\Lb + k_\perp
\;\;,
\\
\hat p_\La ={}& \frac{1+y}{z}\,p_\La
\;\;,
\\
\hat p_\Lb ={}& p_\Lb
\;\;.
\end{split}
\end{equation}
Here $k_\perp$ is the part of $\hat p_{m+1}$ that is orthogonal to both $p_\La$ and $p_\Lb$. 

\paragraph{Virtuality.} The variable $y$ is
\begin{equation}
y = \frac{2\, \hat p_{m+1}\!\cdot\! \hat p_\La}{2\, p_\La\!\cdot\! p_\Lb}
\;\;.
\end{equation}
We will use a shower time $t$ based on virtuality,\footnote{For a final state splitting of parton $l$, the shower time is
$t = \log(M^2/(2\, \hat p_{m+1}\!\cdot\! \hat p_l))$.}
\begin{equation}
\label{eq:tdef}
t = \log\left(\frac{M^2}{2\, \hat p_{m+1}\!\cdot\! \hat p_\La}\right)
\;\;,
\end{equation}
where $M$ is the $Z$-boson mass. We will typically use $t$ as our virtuality variable instead of $y$, so that $y$ is
\begin{equation}
\label{eq:ydef}
y 
= \frac{M^2}{2\, p_\La\!\cdot\! p_\Lb}\,e^{-t}
\;\;.
\end{equation}
(See also eq.~(\ref{eq:ydef2}) below.)

\paragraph{Momentum fraction.} The fraction of the momentum $\hat p_\La$ in the direction of $p_\La$ that is carried away by the emitted final state parton $m+1$ is
\begin{equation}
\frac{\hat p_{m+1} \!\cdot\! p_\Lb}{\hat p_\La \!\cdot\! p_\Lb} = 1-z
\;\;.
\end{equation}
The variable $z$ must be in the range $0 < z < 1$. The momentum fraction $\eta_\La$ of parton ``a'' has a new value after the splitting. From eqs.~(\ref{eq:etadefs}) and (\ref{eq:splittingstart}), we have, using $2\,p_\La \cdot p_\Lb = \eta_\La \eta_\Lb s$ with $s = 2\,p_\LA \!\cdot\! p_\LB$,
\begin{equation}
\begin{split}
\label{eq:etatransformation}
\hat\eta_\La ={}& 
\frac{1+y}{z}\,\eta_\La
=
\frac{1}{z}
\left[
\eta_\La
+ \frac{M^2}{\eta_\Lb s}\,e^{-t}
\right]
\;\;,
\\
\hat\eta_\Lb ={}&
\eta_\Lb
\;\;.
\end{split}
\end{equation}
These are the exact relations. In our applications, we will generally neglect $M^2 e^{-t}$ compared to $\eta_\La \eta_\Lb s$. That is, we consider the virtuality $M^2 e^{-t}$ of a splitting to be small compared to the momentum scale of the hard process, $M^2$, which, in turn, is always smaller than $\eta_\La \eta_\Lb s$.

\paragraph{Transverse momentum.} The angle $\phi$ is the azimuthal angle of $\bm k_\perp$. The magnitude of $\bm k_\perp$ is related to $z$ and $y$:
\begin{equation}
\begin{split}
0 ={}& \hat p_{m+1}^2  
\\
={}& (1-z) y\, 2\, p_\La\!\cdot\! p_\Lb - \bm k_\perp^2
\\
={}& (1-z)M^2 e^{-t} - \bm k_\perp^2
\;\;.
\end{split}
\end{equation}
Thus
\begin{equation}
\label{eq:kperpsq}
\bm k_\perp^2 = (1-z)M^2 e^{-t}
\;\;.
\end{equation}
If we use $z$ and $\phi$ (along with $y$ or $t$) as our splitting variables, then $\bm k_\perp^2$ is a derived variable. Alternatively, we can use $\bm k_\perp$ (and thus $\bm k_\perp^2$ and $\phi$) as splitting variables. Then $z$ is a derived variable.

\paragraph{Lorentz transformation.} If $y \ne 0$, momentum difference $\hat p_\La - \hat p_{m+1}$ is not exactly equal to $p_\La$. In order to maintain momentum conservation at each step in the shower, we must, therefore, take some momentum from the partons in the final state at the time of the splitting. Each parton, with momentum $p_j$, $j \in \{ 1,2,\dots,m\}$, then gets a new momentum $\hat p_j$ after the splitting. This includes the momentum $p_Z \equiv p_1$ of the $Z$-boson. Following the shower algorithm of ref.~\cite{NSI}, the momenta $\hat p_j$ are determined by a Lorentz transformation, $\hat p_j = \Lambda p_j$ with the property
\begin{equation}
\label{eq:lorentzproperty}
\hat p_\La + \hat p_\Lb - \hat p_{m+1} = \Lambda (p_\La + p_\Lb)
\;\;.
\end{equation}
However, we use a {\em different} Lorentz transformation from that chosen in ref.~\cite{NSI}.\footnote{The choice of Lorentz transformation in ref.~\cite{NSI} takes the needed total momentum from the final state partons, but it does not properly absorb the transverse momentum recoil onto the $Z$-boson. The transformation defined here satisfies eq.~(\ref{eq:lorentzproperty}), leaves invariant any vector that is orthogonal to $p_\La$, $p_\Lb$, and $\hat p_{m+1}$ and, in addition, transforms $p_\Lb$ into a multiple of itself.} If
\begin{equation}
p = \alpha\, p_\La + \beta\, p_\Lb + p_\perp
\;\;,
\end{equation}
then $\hat p = \Lambda p$ is given by
\begin{equation}
\begin{split}
\hat p ={}& (1+y)\alpha\, p_\La 
\\ & + 
\frac{1}{1+y}
\left[
\beta
- \frac{2\, \bm p_\perp \!\cdot\! \bm k_\perp }{2\, p_\La \!\cdot\! p_\Lb}
+ \alpha\ \frac{\bm k_\perp^2}{2\, p_\La \!\cdot\! p_\Lb}
\right]
p_\Lb 
\\ & + p_\perp - \alpha k_\perp
\;\;.
\end{split}
\end{equation}
An equivalent form that is useful if $\alpha \ne 0$ is
\begin{equation}
\begin{split}
\hat p ={}& (1+y)\alpha\, p_\La 
\\ & + 
\frac{p^2 + (\bm p_\perp - \alpha \bm k_\perp)^2}
{(1+y)\alpha\, 2\, p_\La \!\cdot\! p_\Lb}\
p_\Lb 
\\ & + p_\perp - \alpha k_\perp
\;\;.
\end{split}
\end{equation}

The momenta $\hat p_\La$ and $\hat p_\Lb$ are {\em not} the Lorentz transformed versions of $p_\La$ and $p_\Lb$. It is, however, of interest to know what the Lorentz transformation does to $p_\La$ and $p_\Lb$. We have
\begin{equation}
\begin{split}
\Lambda p_\La ={}& (1+y)\, p_\La 
+ \frac{1}{1+y}\,\frac{\bm k_\perp^2}{2\, p_\La \!\cdot\! p_\Lb}\ p_\Lb
- k_\perp
\;\;,
\\
\Lambda p_\Lb ={}& \frac{1}{1+y}\ p_\Lb
\;\;.
\end{split}
\end{equation}
The important feature of this is that the emitted parton $m+1$ has transverse momentum $k_\perp$ and the momentum $\Lambda p_\La$ of parton ``a'' caries the recoil transverse momentum $- k_\perp$. We can understand what happens to this recoil transverse momentum by thinking of the shower as proceeding forward in time (oppositely to the way it is generated). Initial state parton ``a'' with momentum $\Lambda p_a$ can emit more daughter partons. But a share of the transverse momentum $- k_\perp$ is finally transmitted to the $Z$-boson.

Let us now look directly at the transformation of the momentum $p_Z$ of the $Z$-boson. We start with
\begin{equation}
\begin{split}
p_Z ={}& \sqrt{\frac{M^2 + \bm p_{Z,\perp}^2}{s}}\,
e^Y\, \frac{1}{\eta_\La}\, p_\La 
+ \sqrt{\frac{M^2 + \bm p_{Z,\perp}^2}{s}}\,
e^{-Y}\, \frac{1}{\eta_\Lb}\, p_\Lb 
+ p_{Z,\perp}
\\
={}& x_\La p_\LA + x_\Lb p_\LB + p_{Z,\perp}
\;\;,
\end{split}
\end{equation}
with
\begin{equation}
\begin{split}
x_\La ={}& \sqrt{\frac{M^2 + \bm p_{Z,\perp}^2}{s}}\,e^Y
\;\;,
\\
x_\Lb ={}& \sqrt{\frac{M^2 + \bm p_{Z,\perp}^2}{s}}\,e^{-Y}
\;\;.
\end{split}
\end{equation}
Then
\begin{equation}
\hat p_Z = \sqrt{\frac{M^2 + \hat {\bm p}_{Z,\perp}^2}{s}}\, 
e^{\hat Y}\, \frac{1}{\eta_\La}\, p_\La 
+ \sqrt{\frac{M^2 + \hat {\bm p}_{Z,\perp}^2}{s}}\, 
e^{-\hat Y}\, \frac{1}{\eta_\Lb}\, p_\Lb 
+ \hat p_{Z,\perp}
\;\;.
\end{equation}
Here
\begin{equation}
\label{eq:pZperpchange}
\hat p_{Z,\perp} = p_{Z,\perp} - \frac{x_\La}{\eta_\La}\,k_\perp
\;\;.
\end{equation}
Thus the $Z$-boson gets a share $x_\La/\eta_\La$ of the recoil transverse momentum. The new rapidity is
\begin{equation}
\label{eq:Ytransformation}
\hat Y = Y + \log(1+y)
- \frac{1}{2} \log\left(
\frac{M^2 + \hat{\bm p}_{Z,\perp}^2}{M^2 + \bm p_{Z,\perp}^2}
\right)
\;\;.
\end{equation}

This transformation law is complicated. However it simplifies in the limit that we need for this paper. Let us denote by $P$ the $Z$-boson momentum without its transverse part,
\begin{equation}
P = x_\La p_\LA + x_\Lb p_\LB
\;\;.
\end{equation}
We are interested in the development of the $Z$-boson transverse momentum in the region $\bm p_{Z,\perp}^2 \ll M^2$. Therefore we take
\begin{equation}
P^2 = M^2
\;\;.
\end{equation}
Furthermore, the development of the $Z$-boson transverse momentum distribution is controlled by splittings with $y \ll 1$. Therefore, we neglect $\bm p_{Z,\perp}^2/M^2$, $\hat{\bm p}_{Z,\perp}^2/M^2$, and $y$ compared to 1 in Eq.~(\ref{eq:Ytransformation}), giving
\begin{equation}
\hat Y = Y
\end{equation}
in each splitting. With these approximations, $x_\La$ and $x_\Lb$ are fixed:
\begin{equation}
\begin{split}
\label{eq:xapproximations}
x_\La ={}& \sqrt{\frac{M^2}{s}}\,e^Y
\;\;,
\\
x_\Lb ={}& \sqrt{\frac{M^2}{s}}\,e^{-Y}
\;\;.
\end{split}
\end{equation}
With the approximations eq.~(\ref{eq:xapproximations}), we have
\begin{equation}
\label{eq:ydef2}
y 
= \frac{x_\La x_\Lb}{\eta_\La \eta_\Lb}\,e^{-t}
\;\;.
\end{equation}
Although $x_\La$ and $x_\Lb$ are fixed, the momentum fractions $\eta_\La$ and $\eta_\Lb$ can change if there is a collinear splitting of an initial state parton.\footnote{We will see, however, that the real emissions that we need to analyze to follow the development of the $Z$-boson transverse momentum distribution have $(1-z) \ll 1$. For these emissions, the change in $\eta_\La$ and $\eta_\Lb$ is negligible.}

Although we neglect the transverse momentum of the $Z$-boson in computing its mass, we track changes in the transverse momentum as the $Z$-boson recoils against emissions from the initial state partons.  For an emission from initial parton ``a'', the new $Z$-boson transverse momentum is
\begin{equation}
\label{eq:pZperpchangeII}
\hat {\bm p}_{Z,\perp} = \bm p_{Z,\perp} - \frac{x_\La}{\eta_\La}\,\bm k_\perp
\;\;,
\end{equation}
as stated in eq.~(\ref{eq:pZperpchange}).

\section{Analysis framework}
\label{sec:framework}

The equations of ref.~\cite{NSI} specify quite precisely the evolution of a certain kind of parton shower. In order to analyze what the parton shower thus defined produces for the transverse momentum distribution of a $Z$-boson, we  develop in this section some theoretical structures beyond those presented in ref.~\cite{NSI}.

\subsection{Measurement operators ${\cal Q}$}

We are interested in the differential cross section as obtained in the shower at the shower final time $t_\Lf$,
\begin{equation}
\frac{d\sigma}{d\bm p_\perp\,dY}
= 
\sbrax{\bm p_\perp, Y}\sket{\rho(t_\Lf)}
\;\;.
\end{equation}

In this paper, we will find it useful to represent the desired measurement with the aid of an operator ${\cal Q}$ on the space of statistical states,
\begin{equation}
\sbrax{\bm p_\perp, Y}\sket{\rho(t)}
=
\sbra{1}{\cal Q}(\bm p_\perp, Y)\sket{\rho(t)}
\;\;.
\end{equation}
Here ${\cal Q}(\bm p_\perp, Y)$ is defined by
\begin{equation}
\begin{split}
{\cal Q}(\bm p_\perp, Y)&\sket{\{p,f,s',c',s,c\}_{m}}
\\
={}& \delta(\bm p_{Z,\perp} - \bm p_\perp)\
\delta\!\left(\frac{1}{2}\log\left(\frac{p_Z\!\cdot\! p_B}{p_Z\!\cdot\! p_A}\right)
-Y\right)
\sket{\{p,f,s',c',s,c\}_{m}}
\;\;.
\end{split}
\end{equation}
In the subsequent subsections, we will extend this notation in which an operator ${\cal Q}$ determines a measurement on the statistical state.

\subsection{Fourier transformation}

As was noted by Parisi and Petronzio \cite{ParisiPetronzio}, it is useful to analyze the evolution of the Fourier transform of the transverse momentum distribution. Thus we study
\begin{equation}
\label{eq:FourierTransform}
\int \frac{d\bm p_\perp}{(2\pi)^2}\ e^{-\mi \bm b \cdot \bm p_\perp}
\sbrax{\bm p_\perp, Y}\sket{\rho(t)}
=
\sbra{1}{\cal Q}(\bm b, Y)\sket{\rho(t)}
\;\;,
\end{equation}
where\footnote{We use the same letter, ${\cal Q}$, for three different operators, ${\cal Q}(\bm p_\perp, Y)$, ${\cal Q}(\bm b, Y)$ defined here, and, in the following subsection, ${\cal Q}(\bm b, Y;\eta_\La, \eta_\Lb, a, b)$. It should be clear from the context which operator is intended.}
\begin{equation}
\begin{split}
\label{eq:Qbydef}
{\cal Q}(\bm b, Y)&\sket{\{p,f,s',c',s,c\}_{m}}
\\
={}& (2\pi)^{-2}\,e^{-\mi \bm b \cdot \bm p_{Z,\perp}}\
\delta\!\left(\frac{1}{2}\log\left(\frac{p_Z\!\cdot\! p_B}{p_Z\!\cdot\! p_A}\right)
-Y\right)
\sket{\{p,f,s',c',s,c\}_{m}}
\;\;.
\end{split}
\end{equation}
We will refer to $\sbra{1}{\cal Q}(\bm b, Y)\sket{\rho(t)}$ as the $b$-space hadronic cross section.

\subsection{Tracking the momentum fractions and parton flavors}
\label{sec:distributionplain}

For our analysis, we will want to keep track of the parton momentum fractions, $\eta_\La$ and $\eta_\Lb$, and the flavors, $a$ and $b$, of the incoming partons. Thus we consider the function 
\begin{equation*}
\sbra{1}{\cal Q}(\bm b, Y;\eta_\La, \eta_\Lb, a, b)\sket{\rho(t)}
\;\;,
\end{equation*}
where
\begin{equation}
\begin{split}
\label{eq:Qbyetcdef}
{\cal Q}(\bm b, Y;\tilde\eta_\La, \tilde\eta_\Lb, \tilde a, \tilde b)&
\sket{\{p,f,s',c',s,c\}_{m}}
\\
={}& (2\pi)^{-2}\,e^{-\mi \bm p_{Z,\perp} \cdot \bm b}\
\delta\!\left(\frac{1}{2}\log\left(\frac{p_Z\!\cdot\! p_B}{p_Z\!\cdot\! p_A}\right)
-Y\right)
\\ & \times
\delta_{a, \tilde a}\,\delta_{b, \tilde b}\,
\delta\!\left(\eta_\La - \tilde\eta_\La\right)
\delta\!\left(\eta_\Lb - \tilde\eta_\Lb\right)
\sket{\{p,f,s',c',s,c\}_{m}}
\;\;.
\end{split}
\end{equation}

\subsection{Evolution of the perturbative statistical state}

The statistical state vector $\sket{\rho(t)}$ according to our definition contains a factor
\begin{equation}
\frac{
f_{a/A}(\eta_{\La},M^2 e^{-t})
f_{b/B}(\eta_{\Lb},M^2 e^{-t})}
{4n_\Lc(a) n_\Lc(b)\,2\eta_{\La}\eta_{\Lb}\,p_\LA\!\cdot\!p_\LB}
\;\;.
\end{equation}
This factor gives the parton-parton luminosity. Here $f_{a/A}(\eta_{\La},\mu^2)$ and $f_{b/B}(\eta_{\Lb},\mu^2)$ are parton distribution functions and $n_\Lc(a)$ and $n_\Lc(b)$ are the number of colors carried by partons of flavors $a$ and $b$, namely 3 for quarks and 8 for gluons. We define an alternative state vector $\sket{\rho_{\rm pert}(t)}$ in which this non-perturbative factor is removed:
\begin{equation}
\begin{split}
\label{eq:rhopertdef}
\sbrax{\{p,f,s',c',s,c\}_{m}}& 
\sket{\rho(t)}
\\
={}& 
\frac{f_{a/A}(\eta_{\La},M^2 e^{-t})
f_{b/B}(\eta_{\Lb},M^2 e^{-t})}
{4n_\Lc(a) n_\Lc(b)\,2\eta_{\La}\eta_{\Lb}p_\LA\!\cdot\!p_\LB}\
\sbrax{\{p,f,s',c',s,c\}_{m}} 
\sket{\rho_{\rm pert}(t)}
\;\;.
\end{split}
\end{equation}
A convenient notation for this is
\begin{equation}
\sket{\rho(t)} = {\cal F}(t) \sket{\rho_{\rm pert}(t)}
\;\;,
\end{equation}
where ${\cal F}(t)$ multiplies by the parton distribution factor,
\begin{equation}
\begin{split}
{\cal F}(t)\sket{\{p,f,s',c',s,c\}_{m}} ={}& 
\frac{f_{a/A}(\eta_{\La},M^2 e^{-t})
f_{b/B}(\eta_{\Lb},M^2 e^{-t})}
{4n_\Lc(a) n_\Lc(b)\,2\eta_{\La}\eta_{\Lb}p_\LA\!\cdot\!p_\LB}\
\sket{\{p,f,s',c',s,c\}_{m}} 
\;\;.
\end{split}
\end{equation}

The evolution equation for $\sket{\rho_{\rm pert}(t)}$ can be determined from the evolution equation (\ref{eq:evolution0}) for $\sket{\rho(t)}$.  We have
\begin{equation}
\begin{split}
\left[\frac{d}{dt}\,{\cal F}(t)\right]
\sket{\rho_{\rm pert}(t)}
+ {\cal F}(t) \frac{d}{dt} \sket{\rho_{\rm pert}(t)}
 ={}& 
[{\cal H}_I(t) - {\cal V}(t)]
{\cal F}(t)
\sket{\rho_{\rm pert}(t)}
\;\;,
\end{split}
\end{equation}
so
\begin{equation}
\begin{split}
 \frac{d}{dt} \sket{\rho_{\rm pert}(t)}
 ={}& 
 {\cal F}(t)^{-1}
[{\cal H}_I(t) - {\cal V}(t)]
{\cal F}(t)
\sket{\rho_{\rm pert}(t)}
\\ & 
- {\cal F}(t)^{-1}\left[\frac{d}{dt}\,{\cal F}(t)\right]
\sket{\rho_{\rm pert}(t)}
\;\;.
\end{split}
\end{equation}
We can write this as
\begin{equation}
\label{eq:rhopertevolution}
\frac{d}{dt}\sket{\rho_{\rm pert}(t)} =
[{\cal H}^{\rm pert}_I(t) - {\cal V}^{\rm pert}(t)]
\sket{\rho_{\rm pert}(t)}
\;\;.
\end{equation}
Here the revised real and virtual splitting operators are
\begin{equation}
\begin{split}
\label{eq:HpertVpert}
{\cal H}^{\rm pert}_I(t)
 ={}& 
{\cal F}(t)^{-1}
{\cal H}_I(t)
{\cal F}(t)
\;\;,
\\ 
{\cal V}^{\rm pert}(t)
 ={}& 
{\cal F}(t)^{-1}
{\cal V}(t)
{\cal F}(t)
+ {\cal F}(t)^{-1}\left[\frac{d}{dt}\,{\cal F}(t)\right]
\\ ={}&
{\cal V}(t)
+ {\cal F}(t)^{-1}\left[\frac{d}{dt}\,{\cal F}(t)\right]
\;\;.
\end{split}
\end{equation}
In the last line here, we have noted that ${\cal F}(t)$ commutes with ${\cal V}(t)$ since ${\cal V}(t)$ does not change momenta or flavors.

For the perturbative real splitting operator, we have
\begin{equation}
\begin{split}
\label{eq:HtoHpert}
\sbra{\{\hat p,\hat f,\hat s',\hat c',\hat s,\hat c\}_{m+1}}&
   {\cal H}_I(t)
  \sket{\{p,f,s',c',s,c\}_{m}} 
\\&=
\frac{n_\Lc(a) n_\Lc(b)\,\eta_{\La}\eta_{\Lb}}
{n_\Lc(\hat a) n_\Lc(\hat b)\,\hat \eta_{\La}\hat \eta_{\Lb}}\
\frac{{f_{\hat a/A}(\hat \eta_{\La},M^2 e^{-t})
f_{\hat b/B}(\hat \eta_{\Lb},M^2 e^{-t})}}
{f_{a/A}(\eta_{\La},M^2 e^{-t})
f_{b/B}(\eta_{\Lb},M^2 e^{-t})}
\
\\&\times
\sbra{\{\hat p,\hat f,\hat s',\hat c',\hat s,\hat c\}_{m+1}}
{\cal H}_I^{\rm pert}(t)
\sket{\{p,f,s',c',s,c\}_{m}}
\;\;.
\end{split}
\end{equation}
The factors in the second line appear in the definition of the matrix element of ${\cal H}_I$ in ref.~\cite{NSI}. Thus to obtain the corresponding matrix element of ${\cal H}_I^{\rm pert}(t)$, we simply omit these factors. We will present detailed formulas for ${\cal H}_I^{\rm pert}(t)$ and ${\cal V}^{\rm pert}(t)$ at the point that we need them.

\subsection{The function to study}
\label{sec:ToStudy}

The physics that we want to study is contained in the function defined in section \ref{sec:distributionplain}, in which $\bm b$ and $Y$ are measured and, in addition, we measure $\eta_\La$, $\eta_\Lb$, $a$ and $b$. As noted in the previous subsection, this function contains a nonperturbative factor that specifies the parton luminosity. We remove this factor and study
\begin{equation*}
%
\sbra{1}{\cal Q}(\bm b, Y; \eta_\La,  \eta_\Lb, a, b)
\sket{\rho_{\rm pert}(t)}
\;\;.
\end{equation*}
We will refer to this function as the $b$-space partonic cross section. Once we have the $b$-space partonic cross section, we can obtain the $b$-space hadronic cross section $\sbra{1}{\cal Q}(\bm b, Y)\sket{\rho(t)}$ by convolving it with parton distribution functions according to
\begin{equation}
\begin{split}
\label{eq:partonicvsfull}
\sbra{1}{\cal Q}(\bm b, Y)
\sket{\rho(t)} ={}& 
\sum_{a, b}\int_0^1\!d\eta_\La\int_0^1\!d\eta_\Lb\
\frac{
f_{a/A}(\eta_{\La},M^2 e^{-t})
f_{b/B}(\eta_{\Lb},M^2 e^{-t})}
{4n_\Lc(a) n_\Lc(b)\,2\eta_{\La}\eta_{\Lb}p_\LA\!\cdot\!p_\LB}
\\&\times
\sbra{1}{\cal Q}(\bm b, Y; \eta_\La,  \eta_\Lb, a, b)
\sket{\rho_{\rm pert}(t)}
\;\;.
\end{split}
\end{equation}
This relation is obtained using eqs.~(\ref{eq:completeness}), (\ref{eq:Qbydef}), (\ref{eq:Qbyetcdef}), and (\ref{eq:rhopertdef}). 

Our aim is to study how the $b$-space partonic cross section develops as the shower time $t$ increases.  At shower time 0, it is determined from the Born cross section, as in eq.~(\ref{eq:born}),
\begin{equation}
\begin{split}
\label{eq:initialcondition}
\sbra{1}{\cal Q}(\bm b, Y; \eta_\La, \eta_\Lb, 
a, b)\sket{\rho_{\rm pert}(0)}
={}& 12\, \alpha\, Q_{ab}\, x_\La x_\Lb\, 
\delta(\eta_\La - x_\La)\,
\delta(\eta_\Lb - x_\Lb)
\;\;,
\end{split}
\end{equation}
with
\begin{equation}
x_\La = \sqrt{\frac{M^2}{s}}\, e^{Y}
\;\;,
\hskip 1 cm
x_\Lb = \sqrt{\frac{M^2}{s}}\, e^{-Y}
\;\;,
\end{equation}
and
\begin{equation}
\begin{split}
Q_{ab} ={}& 
0 \qquad a = \Lg \ {\rm or}\ b = \Lg 
\;\;,
\\
Q_{ab} ={}&
\delta_{a,\bar b}\,\frac{[1 - 4\, |e_a| \sin^2(\theta_{\rm W})]^2 + 1}
{16 \sin^2(\theta_{\rm W})\cos^2(\theta_{\rm W})}
\qquad a \ne \Lg \ {\rm and}\ b \ne \Lg 
\;\;.
\end{split}
\end{equation}
Note that the partonic cross section at $t = 0$ vanishes unless $a$ is a quark or antiquark flavor and $b$ is the corresponding antiflavor. There is no dependence on $\bm b$ because the corresponding transverse momentum dependent cross section is proportional to a delta function of the transverse momentum. As the shower evolves, we expect $\sbra{1}{\cal Q}(\bm b, Y;\eta_\La, \eta_\Lb, a, b)\sket{\rho_{\rm pert}(t)}$  to develop some dependence on $\bm b$. 

We begin the study of the evolution of this function in the next subsection by outlining some key ideas that will go into the derivation.

\section{Outline of the derivation}
\label{sec:outline}

\FIGURE{
\centerline{\includegraphics[width = 8 cm]{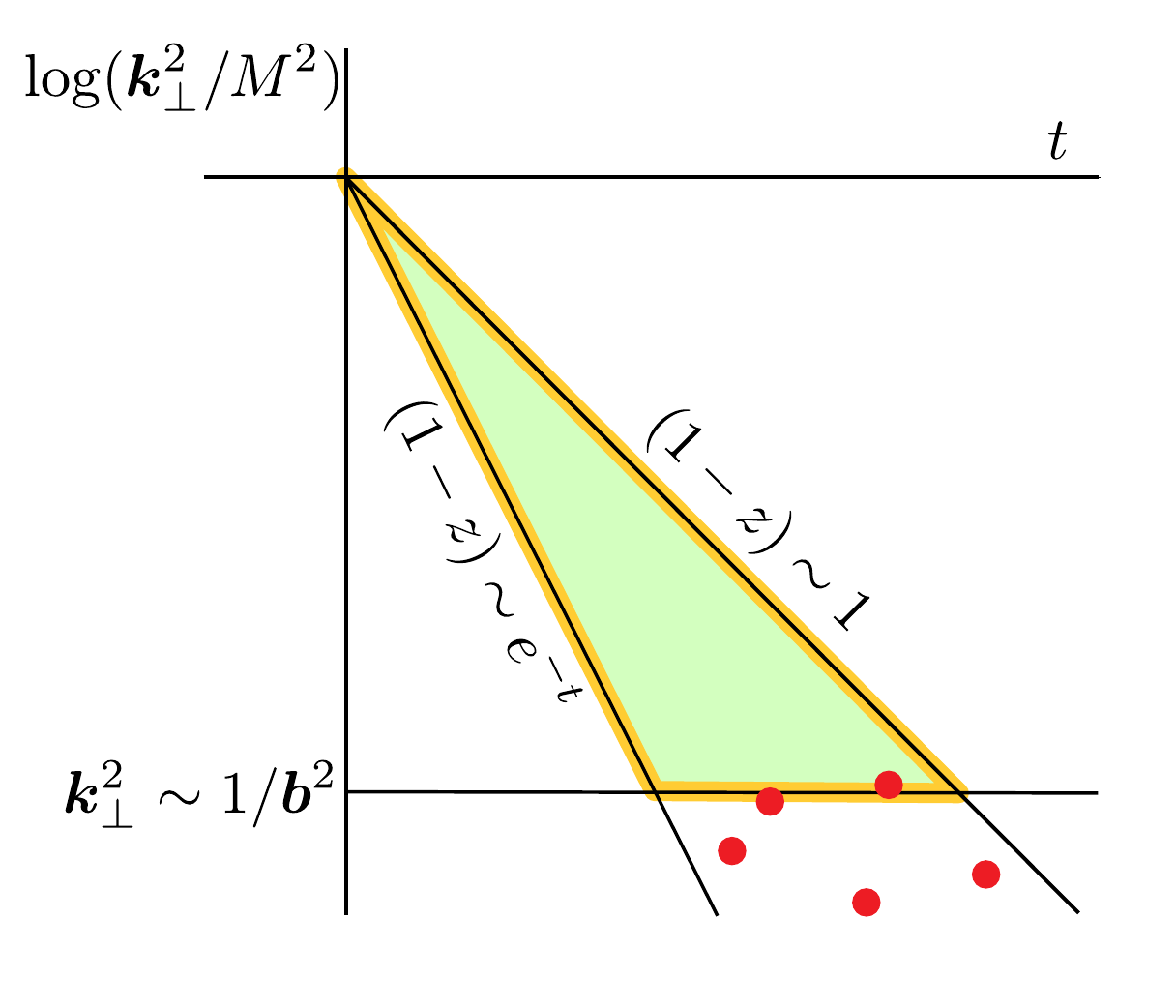}}
\caption{Integration region for initial state gluon emissions. The horizontal axis is the shower time; thus emissions are generated from left to right. The vertical axis is $\log(\bm k_\perp^2/M^2)$. The leading region for emissions lies between the lines labelled $(1-z) \sim 1$ and $(1-z) \sim e^{-t}$. The partonic $b$-space cross section for emissions with $\bm k_\perp^2 \gg 1/\bm b^2$ approximately vanishes. Thus real gluon emissions, indicated by small circles, occur near or below the horizontal line $\bm k_\perp^2 \sim 1/\bm b^2$.}
\label{fig:TrianglePlain}
} 

We are now in a position to outline the derivation that follows, leaving out most of the details and certain subtle points. One of the subtle points is the running of $\as$. For the purposes of this section, we consider $\as$ to be constant.

\subsection{Emission kinematics}
\label{sec:emissionkinematics}

We will find that the transverse momentum of the $Z$-boson is the primarily the result of recoils against emission of soft gluons from an initial state quark or antiquark. As we saw in section~\ref{sec:kinematics}, an initial state splitting can be described by splitting variables $\{t,z,\phi\}$. The shower time $t$ gives the virtuality of the splitting. The momentum fraction of the emitted gluon is $(1-z)$, so that for soft gluon emission we have $(1-z) \ll 1$. The transverse momentum $\bm k_\perp$ of the emitted gluon has azimuthal angle $\phi$ and magnitude given by eq.~(\ref{eq:kperpsq}), $\bm k_\perp^2 = (1-z)M^2 e^{-t}$.

It is useful to represent splittings by points in the plane of $t$ and $\log(\bm k_\perp^2/M^2)$, as in figure~\ref{fig:TrianglePlain}. An allowed splitting has $(1-z) \le 1$, so $\log(\bm k_\perp^2/M^2) \le - t$. That is, allowed splittings are represented by points below the line labelled ``$(1-z)\sim 1$'' in figure~\ref{fig:TrianglePlain}. One should view this line as having a thickness of order 1. As we will discuss in some detail, for a given $t$, the splitting probability $dP$ from ref.~\cite{NSI} has a term with 
\begin{equation}
\label{eq:density}
dP \sim 4 C_\LF\,\frac{\as}{2\pi}\ dt\,\frac{dz}{1-z + e^{-t}}
\;\;,
\end{equation}
including a factor 2 due to having two incoming partons that can radiate. This means that splittings are not simply concentrated along the line $(1-z) \sim 1$, but are spread over the region below this line. In fact, a splitting probability proportional to $dt\,dz/(1-z)$ would give splittings uniformly distributed in $\log(\bm k_\perp^2/M^2)$ and $t$. However, there is an effective cutoff at the line $(1-z) \sim e^{-t}$. Thus the splitting probability is approximately constant in the region $e^{-t} < (1-z) < 1$ indicated in figure~\ref{fig:TrianglePlain}.

We are interested in the partonic $b$-space cross section. For this quantity, a real emission produces a factor $\exp(\mi \bm k_\perp \cdot  \bm b)$, which simply comes from taking the Fourier transform to get to $b$-space. For this reason, the line $\bm k_\perp^2 \sim 1/\bm b^2$ in figure~\ref{fig:TrianglePlain} is significant. If we integrate the real emission probability over an interval of $\bm k_\perp$ in the region $\bm k_\perp^2 \gg 1/\bm b^2$, the factor  $\exp(\mi \bm k_\perp\!\cdot\! \bm b)$ averages to zero. That is to say, maintaining a non-zero $\sbra{1}{\cal Q}(\bm b, Y;\eta_\La, \eta_\Lb, a, b)\sket{\rho_{\rm pert}(t)}$ requires {\em not} emitting gluons with $\bm k_\perp^2 \gg 1/\bm b^2$. In the region $\bm k_\perp^2 \lesssim 1/\bm b^2$, gluons can be freely emitted into the final state. We represent gluons that might be emitted in a typical event contributing to $\sbra{1}{\cal Q}(\bm b, Y;\eta_\La, \eta_\Lb, a, b)\sket{\rho_{\rm pert}(t)}$ by filled circles in figure~\ref{fig:TrianglePlain}.

Of course, if there is an approximately uniform probability of emitting gluons in any differential unit of area $d\log(\bm k_\perp^2/M^2)\, dt$, then the probability that no gluons are emitted for $\bm k_\perp^2 \gg 1/\bm b^2$ is small when $\bm b^2$ is large. It is approximately 
\begin{equation}
\label{eq:simpleSudakov}
e^{-S} \approx
\exp\!\left(- 4 C_\LF\,\frac{\as}{2\pi}\,A\right)
\;\;,
\end{equation}
where $A$ is the area of the triangle in figure~\ref{fig:TrianglePlain}. This Sudakov factor gives, approximately, the $b$-dependence of the partonic $b$-space cross section that we seek. In the following sections, we fill in the details of this argument and make it more precise. We will find that the more precise analysis leads to a Sudakov factor similar to that in eq.~(\ref{eq:simpleSudakov}), but with running coupling effects included and an extra ``subleading'' term.

\subsection{Strategy}
\label{sec:strategy}

We study the evolution of the $b$-space partonic cross section defined in section~\ref{sec:ToStudy}. Using eq.~(\ref{eq:rhopertevolution}), the $b$-space partonic cross section evolves according to
\begin{equation}
\begin{split}
\label{eq:evolution1}
\frac{d}{dt}
\sbra{1}
{\cal Q}(\bm b, Y;\eta_\La, \eta_\Lb, a, b)&
\sket{\rho_{\rm pert}(t)}
\\
={}&
\sbra{1}
{\cal Q}(\bm b, Y; \eta_\La,  \eta_\Lb, a, b)
[ {\cal H}^{\rm pert}_\LI(t) - {\cal V}^{\rm pert}(t)
]
\sket{\rho_{\rm pert}(t)}
\;\;.
\end{split}
\end{equation}
Our aim is to use suitable approximations to turn this equation into a differential equation of the form
\begin{equation}
\begin{split}
\label{eq:evolutionform}
\frac{d}{dt}
\sbra{1}
{\cal Q}(\bm b, Y;\eta_\La, \eta_\Lb, a, b)&
\sket{\rho_{\rm pert}(t)}
\approx
- K(t,b)
\sbra{1}
{\cal Q}(\bm b, Y; \eta_\La,  \eta_\Lb, a, b)
\sket{\rho_{\rm pert}(t)}
\;\;.
\end{split}
\end{equation}
This differential equation has the solution
\begin{equation}
\begin{split}
\label{eq:resultform}
\sbra{1}
{\cal Q}(\bm b, Y;\eta_\La, \eta_\Lb, a, b)&
\sket{\rho_{\rm pert}(t)}
\\
\approx{}&
\exp\left(- \int_0^t \!d\tau\ K(\tau,b)\right)
\sbra{1}
{\cal Q}(\bm b, Y; \eta_\La,  \eta_\Lb, a, b)
\sket{\rho_{\rm pert}(0)}
\;\;.
\end{split}
\end{equation}
Here the initial $b$-space partonic cross section has the simple $b$-independent form given in eq.~(\ref{eq:initialcondition}). We will see that the evolution of $\sbra{1} {\cal Q}(\bm b, Y)\sket{\rho(t)}$ stops for $t > t_\Lc \equiv \log\left(\bm b^2 M^2\,e^{2\gamma_{\rm E}}/4\right)$, where $\gamma_{\rm E}$ is the Euler $\gamma$. Here $t_\Lc$ is approximately the shower time at which the lines $(1-z) \sim 1$ and $\bm k_\perp^2 \sim 1/\bm b^2$ in figure~\ref{fig:TrianglePlain} meet; we will see later the reason for the adjustment factor $e^{2\gamma_{\rm E}}/4$. Then 
\begin{equation}
e^{-S} = 
\exp\left(- \int_0^{t_\Lc} \!d\tau\ K(\tau,b)\right)
\end{equation}
is the Sudakov factor, for which eq.~(\ref{eq:simpleSudakov}) is a simple approximation.

\subsection{Approximations}
\label{sec:approximations}

We will need certain approximations to turn eq.~(\ref{eq:evolution1}) into the differential equation (\ref{eq:evolutionform}). We describe these approximations in general terms here.

First, we note that the behavior of the $Z$-boson $\bm p_\perp$ distribution for $\bm p_\perp^2 \ll M^2$ is controlled by the $b$-space partonic cross section for large $\bm b^2$. Thus we are interested in the $b$-space partonic cross section in the region $1/(\bm b^2 M^2) \ll 1$. Therefore, we simply neglect $1/(\bm b^2 M^2)$ compared to 1 everywhere. 

Second, we neglect $e^{-t}$ compared to 1. To justify this, imagine letting the system evolve from time 0 to a time $t_1$ and then from $t_1$ to $t_\Lc$. Let $t_1$ be large enough so that $e^{-t_1} \ll 1$, but small enough that we can treat $t_1$ as not being a large logarithm. Then evolution from 0 to $t_1$ is an approximate version of perturbation theory and gives order $\as^n$ corrections to the Born cross section with no large logarithms. We ignore these corrections. For the evolution from $t_1$ to $t_\Lc$, the approximation $e^{-t} \approx 0$ is justified. Furthermore, we can add back the evolution from 0 to $t_1$ using the approximation $e^{-t} \approx 0$, adding more order $\as^n$ corrections with no large logarithms. Then we have evolution from $0$ to $t_\Lc$ with the approximation $e^{-t} \approx 0$ at the cost of changing the result by $\as^n$ terms with no large logarithms.

For the same reason, we neglect $y$, eq.~(\ref{eq:ydef}), compared to 1 and $\bm k_\perp^2/M^2$ compared to 1.

Finally, in section~\ref{sec:angularordering}, we will analyze the structure of the splitting function near the line $\bm k_\perp^2 = 1/\bm b^2$. For this analysis, we will make what might be called a low density approximation. For initial state emissions, according to eq.~(\ref{eq:density}), the density of emission points per unit $dt$ and $d\log(\bm k_\perp^2/M^2)$ is proportional to $\as$. We treat $\as$ as small. Consider, then, two emissions, one with parameters $\{t_1,\log(\bm k_{\perp,1}^2/M^2)\}$ and the other with parameters $\{t_2,\log(\bm k_{\perp,2}^2/M^2)\}$. Suppose that $t_2 > t_1$ and that each of these emissions is not far from the line $\bm k_\perp^2 = 1/\bm b^2$. Since the density of points is small, the distance in the $\{t,\log(\bm k_{\perp}^2/M^2)\}$ plane between any two points is typically large. This suggests that we can neglect $e^{-(t_2 - t_1)}$ compared to 1. To see whether this is justified, suppose that we modify the shower algorithm so that it is not allowed to have two splittings be close to the line $\bm k_\perp^2 = 1/\bm b^2$ and close to each other. More precisely, choose a distance parameter $d_0$ and require that no two splittings have $|\log(\bm k_\perp^2\bm b^2)| < d_0$ and simultaneously $t_2 - t_1 < d_0$. Choose $d_0$ such that $e^{-d_{0}}$ is small enough to be neglected but $d_0$ is not so large that it constitutes a large logarithm. Then it is valid to replace $e^{-(t_2 - t_1)}$ by zero for any splittings that are not excluded. 

The exclusion prescription can be constructed in a different way. Generate $1 \to 2$ splittings without restriction, but add a new splitting function that describes a $1 \to 3$ splitting: incoming parton ``a'' emits partons 1 and 2 in the region $|\log(\bm k_\perp^2\bm b^2)| < d_0$ and $t_2 - t_1 < d_0$. The probability for this new splitting should be negative and just big enough to cancel the probability for parton ``a'' to first emit parton 1 and then emit parton 2 in this region. One then needs to add similar $1 \to n + 1$ splittings for $n > 2$ to cancel the probabilities to have more than two splittings that are to close the line $\bm k_\perp^2 = 1/\bm b^2$ and to one another, but we limit this discussion to the simple $1 \to 3$ case. The new $1 \to 3$ splitting probability is proportional to $\as^2$ and the excluded area, $d_0^2$. With our choice of $d_0$, this area is not large.

The new splitting function will modify the Sudakov exponent by adding a term proportional to $\as^2$ and to the length of the line $\bm k_\perp^2 = 1/\bm b^2$ between the two limits in figure~\ref{fig:TrianglePlain}, namely $\Delta t \approx \log(\bm b^2 M^2)/2$.

We conclude that replacing iterated splittings by zero in the region in which the approximation $e^{-(t_2 - t_1)} \approx 0$ is not good results in modifying the Sudakov exponent by terms of order $\as^2 \log(\bm b^2 M^2)$. What we actually do is to replace these splittings by what the inaccurate approximation $e^{-(t_2 - t_1)} \approx 0$ gives. As long as this approximation results in a fractional change of order 1 in the iterated splitting probability, we also modify the Sudakov exponent by terms of order $\as^2 \log(\bm b^2 M^2)$.

The true QCD Sudakov exponent, as discussed in section~\ref{sec:result}, has an expansion in powers of $\as$. In the coefficient of $\as^2$, the term with the most powers of $\log(\bm b^2 M^2)$ is a constant times $\as^2 \log^3(\bm b^2 M^2)$. The next-to-leading term is a constant times $\as^2 \log^2(\bm b^2 M^2)$. Thus terms of order $\as^2 \log(\bm b^2 M^2)$ are third-to-leading. The low density approximation discussed here changes these terms in an uncontrolled way.

We have not analyzed here the effect of $1 \to n + 1$ splittings for $n > 2$ that are induced by this approximation. However, it should be clear that these induce $\as^{n} \log(\bm b^2 M^2)$ changes in the Sudakov exponent.

\section{Evolution of the partonic cross section}
\label{sec:partonicevolution}

We can now begin to simplify eq.~(\ref{eq:evolution1}), which gives the evolution of the $b$-space partonic cross section.

The operator ${\cal V}^{\rm pert}(t)$ acting on a partonic basis state $\sket{\{p,f,s',c',s,c\}_{m}}$ does not add a new parton or change the parton momenta or flavors. For this reason, ${\cal Q}(\bm b, Y; \eta_\La, \eta_\Lb, a, b)$ commutes with ${\cal V}^{\rm pert}(t)$. Thus eq.~(\ref{eq:evolution1}) becomes
\begin{equation}
\begin{split}
\frac{d}{dt}
\sbra{1}
{\cal Q}(\bm b, Y; \eta_\La, \eta_\Lb, a, b)
\sket{\rho_{\rm pert}(t)}
={}&
\sbra{1}
{\cal Q}(\bm b, Y;\eta_\La, \eta_\Lb, a, b)
{\cal H}^{\rm pert}_I(t) 
\\&\qquad
- {\cal V}^{\rm pert}(t)
{\cal Q}(\bm b, Y;\eta_\La, \eta_\Lb, a, b)
\sket{\rho_{\rm pert}(t)}
\;\;.
\end{split}
\end{equation}

The operator ${\cal H}_\LI^{\rm pert}(t)$ does add a new parton and does change the parton momenta. Thus it does not commute with ${\cal Q}$. To analyze what happens, we break ${\cal H}_\LI^{\rm pert}(t)$ into three parts:
\begin{equation}
\begin{split}
\label{eq:Hexpansion}
{\cal H}_\LI^{\rm pert}(t) ={}& 
{\cal H}_{\rm FS}(t)
+ \int_0^1 \!dz\int_{-\pi}^\pi \frac{d\phi}{2\pi}\sum_{f'}\ 
{\cal H}_{\La}^{\rm pert}(t;z,\phi,f')
\\&
+ \int_0^1\! dz \int_{-\pi}^\pi \frac{d\phi}{2\pi}\sum_{f'}\ 
{\cal H}_{\Lb}^{\rm pert}(t;z,\phi,f')
\;\;.
\end{split}
\end{equation}
Here ${\cal H}_{\rm FS}(t)$ is the part of ${\cal H}_\LI^{\rm pert}(t)$ that generates the splittings of final state particles. We have dropped the ``pert'' notation here because, in the definition of ref.~\cite{NSI}, a final state splitting does not involve a factor of ratios of parton distributions, so that
the part of ${\cal H}_\LI^{\rm pert}(t)$ that creates a final state splitting is the same as the part of ${\cal H}_\LI(t)$ that creates the same final state splitting. In the second term in eq.~(\ref{eq:Hexpansion}), we let ${\cal H}_{\La}^{\rm pert}(t)$ be the part of ${\cal H}_\LI^{\rm pert}(t)$ that generates the splittings of the incoming parton from hadron A. We have decomposed this operator further as an integral and sum of ${\cal H}_{\La}^{\rm pert}(t;z,\phi, f')$, which generates the splittings of the incoming parton from hadron A in which the momentum fraction and azimuthal angle of the splitting are $z$ and $\phi$, respectively and the flavor of the emitted parton is $f'$. (For our analysis, $f' = {\rm g}$ is the most important choice.) Similarly,  ${\cal H}_{\Lb}^{\rm pert}(t;z,\phi,f')$ generates the splittings of the incoming parton from hadron B. In a similar way, we can divide the virtual splitting operator into three parts,
\begin{equation}
\begin{split}
\label{eq:Vexpansion}
{\cal V}^{\rm pert}(t) ={}& 
{\cal V}_{\rm FS}(t)
+ \int_0^1 \! dz \int_{-\pi}^\pi \frac{d\phi}{2\pi}\sum_{f'}\ 
{\cal V}_{\La}^{\rm pert}(t;z,\phi,f')
\\&
+ \int_0^1\! dz \int_{-\pi}^\pi \frac{d\phi}{2\pi}\sum_{f'}\ 
{\cal V}_{\Lb}^{\rm pert}(t;z,\phi,f')
\;\;.
\end{split}
\end{equation}

\subsection{Final state splittings}

Consider first the effect of a final state splitting. Using the definitions of ref.~\cite{NSI}, we find that a final state splitting replaces one final state parton by two daughter partons, but its effect on the rapidity and transverse momentum of the $Z$-boson is negligible. When parton $l$ with momentum $p_l$ splits to form daughter partons $l$ and $m+1$ with momenta $\hat p_l$ and $\hat p_{m+1}$, an amount of momentum $\Delta p = \hat p_l + \hat p_{m+1} - p_l$ must be taken from the other final state partons. The splitting can be characterized by the virtuality variable $y = \hat p_l \!\cdot\! \hat p_{m+1}/ p_l\!\cdot\! Q$, where $Q = p_\La + p_\Lb$. With the splitting kinematics of ref.~\cite{NSI}, the needed momentum is
\begin{equation}
\Delta p = -(1-\lambda)\, p_l + 
(1-\lambda + y)\,\frac{p_l \!\cdot\! Q}{Q^2}\ Q
\;\;,
\end{equation}
where
\begin{equation}
\lambda = \sqrt{(1+y)^2 - y\,\frac{2 Q^2}{p_l \!\cdot\! Q}}
\sim 1 - y \left(\frac{Q^2}{p_l\!\cdot\! Q} - 1\right) + \cdots
\;\;.
\end{equation}
Note that $1-\lambda \propto y$ for $y \ll 1$. The momentum $\Delta p$ is supplied by applying a Lorentz transformation to each final state parton $i$ with $i \notin \{l,m+1\}$: $\hat p_i^\mu = \Lambda^\mu_{\,\nu}\, p_i^\nu$. In particular, the momentum $p_Z$ of the $Z$-boson is transformed with $\hat p_Z^\mu = \Lambda^\mu_{\,\nu}\, p_Z^\nu$. 

As discussed in section~\ref{sec:approximations}, it suffices to consider only splittings with $y \ll 1$. For $y \ll 1$, $\Delta p$ is proportional to $y$ and hence $\Lambda^\mu_{\,\nu} - \delta^\mu_{\,\nu}$ is also proportional to $y$. Thus the change in the $Z$-boson momentum is small. In particular, the rapidity of the $Z$-boson changes very little, by an amount proportional to $y$. We can neglect this small change.

Evidently, the change in the $Z$-boson transverse momentum must also be small, but this statement is not helpful because we are trying to track small changes in the $Z$-boson transverse momentum. To see what happens, we note that the needed transverse momentum is
\begin{equation}
\Delta \bm p_\perp = - (1 - \lambda)\, \bm p_{l,\perp}
\;\;.
\end{equation}
A fraction of this transverse momentum will come from the $Z$-boson. Now, we are studying the evolution of the probability that the $Z$-boson transverse momentum $\bm p_{Z,\perp}$ is small and remains small. For this to happen, the transverse part of $p_l$ must be small, of order $\bm p_{Z,\perp}$ or smaller.\footnote{We discuss this in section \ref{sec:forbidden}.} Thus, recalling that $(1-\lambda)$ is of order $y$, we have
\begin{equation}
\hat{\bm p}_{Z,\perp} - \bm p_{Z,\perp}
\sim C\, y\, \bm p_{Z,\perp}
\;\;,
\end{equation}
where $C$ is of order 1 or smaller. That is, the fractional change in the $Z$-boson transverse momentum due to a final state splitting is negligible.

This discussion can be summarized by saying that, to a sufficient approximation,
\begin{equation}
{\cal Q}(\bm b, Y; \eta_\La, \eta_\Lb, a, b)
{\cal H}_{\rm FS}(t)
\approx
{\cal H}_{\rm FS}(t)
{\cal Q}(\bm b, Y; \eta_\La, \eta_\Lb, a, b)
\;\;.
\end{equation}
This gives
\begin{equation}
\begin{split}
\label{eq:evolution2}
\frac{d}{dt}
\sbra{1}
{\cal Q}(\bm b, Y;& \eta_\La, \eta_\Lb, a, b)
\sket{\rho_{\rm pert}(t)}
\\ 
\approx{}&
\sbra{1}
[{\cal H}_{\rm FS}(t) 
- {\cal V}_{\rm FS}(t)]
{\cal Q}(\bm b, Y; \eta_\La, \eta_\Lb, a, b)
\sket{\rho_{\rm pert}(t)}
\\ &+
\int_0^1\! dz \int_{-\pi}^\pi \frac{d\phi}{2\pi}\sum_{f'}\ 
\sbra{1}
{\cal Q}(\bm b, Y; \eta_\La, \eta_\Lb, a, b)
{\cal H}_{\La}^{\rm pert}(t;z,\phi,f')
\\&\qquad
- {\cal V}_\La^{\rm pert}(t,z,\phi,f')
{\cal Q}(\bm b, Y; \eta_\La, \eta_\Lb, a, b)
\sket{\rho_{\rm pert}(t)}
\\ &
+
\int_0^1\! dz \int_{-\pi}^\pi \frac{d\phi}{2\pi}\sum_{f'}\ 
\sbra{1}
{\cal Q}(\bm b, Y; \eta_\La, \eta_\Lb, a, b)
{\cal H}_{\Lb}^{\rm pert}(t;z,\phi,f')
\\&\qquad
- {\cal V}_\Lb^{\rm pert}(t;z,\phi,f')
{\cal Q}(\bm b, Y; \eta_\La, \eta_\Lb, a, b)
\sket{\rho_{\rm pert}(t)}
\;\;.
\end{split}
\end{equation}
This is useful because the definition of ${\cal V}_{\rm FS}(t)$ is based on the requirement, designed to insure that the shower conserves probabilities, that
\begin{equation}
\sbra{1}
[{\cal H}_{\rm FS}(t) 
- {\cal V}_{\rm FS}(t)]
= 0
\;\;.
\end{equation}
Thus the first term in eq.~(\ref{eq:evolution2}) vanishes. We must analyze initial state splittings, but we can ignore final state splittings entirely.

\subsection{Initial state splittings}

We now turn to initial state splittings. We relate ${\cal Q}$ applied after the splitting to ${\cal Q}$ applied before the splitting. The relation is
\begin{equation}
\begin{split}
\label{eq:splittingeffect}
{\cal Q}(\bm b, Y;\hat\eta_\La, \eta_\Lb, \hat a, b)
{\cal H}_{\La}^{\rm pert}(t;z,\phi, f')
\approx{}&
{\cal H}_{\La}^{\rm pert}(t;z,\phi, f')
\exp\!\left(\mi\, \frac{x_\La}{z \hat\eta_\La}\,\bm b\!\cdot\! \bm k_\perp\right)
\\&\times
z{\cal Q}(\bm b, Y; z\hat\eta_\La, \eta_\Lb, \hat a - f', b)
\;\;.
\end{split}
\end{equation}
To understand this, first look at the $\bm b$ dependence, using the definition (\ref{eq:Qbyetcdef}) of ${\cal Q}$. When we apply the operator ${\cal Q}$ after the splitting, it produces a factor
$
\exp({- \mi \bm b \!\cdot\! \hat{\bm p}_{Z,\perp}})
$
where $\hat{\bm p}_{Z,\perp}$ is the $Z$-boson transverse momentum after the splitting, which is related to the $Z$-boson transverse momentum after the splitting, ${\bm p}_{Z,\perp}$, and the transverse momentum in the splitting, $\bm k_\perp$, by eq.~(\ref{eq:pZperpchangeII}),
\begin{equation}
\hat {\bm p}_{Z,\perp} = \bm p_{Z,\perp} - \frac{x_\La}{\eta_\La}\,\bm k_\perp
\;\;.
\end{equation}
Recall that $\bm k_\perp$ is specified by $z$ and $\phi$: it has azimuthal angle $\phi$ and square $\bm k_\perp^2 = (1-z)M^2 e^{-t}$. Thus
\begin{equation}
\label{eq:bfactor}
\exp({- \mi \bm b \!\cdot\! \hat{\bm p}_{Z,\perp}}) = 
\exp({- \mi \bm b \!\cdot\! {\bm p}_{Z,\perp}})\,
\exp\left(\mi\, \frac{x_\La}{\eta_\La}\,\bm b\!\cdot\! \bm k_\perp\right)
\;\;.
\end{equation}
The factor $\exp({- \mi \bm b \!\cdot\! {\bm p}_{Z,\perp}})$ is generated by the ${\cal Q}$ operator before the splitting but the second factor in eq.~(\ref{eq:bfactor}) must be supplied. The dependence on the momentum fractions is simple. According to eq.~(\ref{eq:etatransformation}), the momentum fraction $\eta_\Lb$ is unchanged by the splitting, while he momentum fraction $\hat \eta_\La$ after the splitting is related to the momentum fraction $\eta_\La$ before the splitting by $\eta_\La \approx z \hat \eta_\La$. Thus the $\eta_\La$ argument of ${\cal Q}$ before the splitting is $z \hat \eta_\La$ and there is a jacobian factor $z$ because ${\cal Q}$ is defined with a delta function. Finally, the flavor $\hat a$ after the splitting is related to the flavor $a$ before the splitting by $a = \hat a - f'$, where we use the notation $u - \Lg = u$, $\Lg  - \bar u = u$, {\it etc}.

With these observations, our equation for the variation $b$-space partonic cross section with shower time is
\begin{equation}
\begin{split}
\label{eq:evolutionfromK}
\frac{d}{dt}
\sbra{1}
{\cal Q}(\bm b, Y;&\tilde\eta_\La, \tilde\eta_\Lb,\tilde a, \tilde b)
\sket{\rho_{\rm pert}(t)}
\\ 
\approx{}&
\int_0^1\! dz \int_{-\pi}^\pi \frac{d\phi}{2\pi}\sum_{f'}
\sbra{1} 
{\cal K}_\La(t;z,\phi,f';
\bm b, Y;\tilde\eta_\La, \tilde\eta_\Lb,\tilde a, \tilde b)
\sket{\rho_{\rm pert}(t)}
\\ &
+
\int_0^1\! dz \int_{-\pi}^\pi \frac{d\phi}{2\pi}\sum_{f'}
\sbra{1} 
{\cal K}_\Lb(t;z,\phi,f';
\bm b, Y;\tilde\eta_\La, \tilde\eta_\Lb,\tilde a, \tilde b)
\sket{\rho_{\rm pert}(t)}
\;\;,
\end{split}
\end{equation}
where ${\cal K}_\La$ describes a splitting of an initial state parton ``a'' from hadron $A$ and is given by
\begin{equation}
\begin{split}
\label{eq:evolution3}
\sbra{1} 
{\cal K}_\La(t;z,\phi,f';&
\bm b, Y;\tilde\eta_\La, \tilde\eta_\Lb,\tilde a, \tilde b)
\sket{\rho_{\rm pert}(t)}
\\ 
={}&
\sbra{1} 
{\cal H}_{\La}^{\rm pert}(t;z,\phi,f')
\exp\!\left(\mi\, \frac{x_\La}{z \tilde\eta_\La}\,\bm b\!\cdot\! \bm k_\perp\right)
z{\cal Q}(\bm b, Y; z\tilde\eta_\La, \tilde\eta_\Lb, \tilde a - f', \tilde b)
\\&\qquad
- {\cal V}_\La^{\rm pert}(t;z,\phi,f')
{\cal Q}(\bm b, Y;\tilde\eta_\La, \tilde\eta_\Lb, \tilde a, \tilde b)
\sket{\rho_{\rm pert}(t)}
\;\;.
\end{split}
\end{equation}
The operator ${\cal K}_\Lb$ for a splitting of the initial state parton from hadron $B$ is the same with the roles of ``a'' and ``b'' are interchanged.

\section{The real splitting function}
\label{sec:realsplitting}

At this point, we need to know some details about the operation of ${\cal H}_{\La}^{\rm pert}(t;z,\phi)$ on a general partonic state $\sket{\{p,f,s',c',s,c\}_{m}}$. Fortunately, we need only the inclusive splitting probability
\begin{equation}
\sbra{1}{\cal H}_{\La}^{\rm pert}(t;z,\phi, f')\sket{\{p,f,s',c',s,c\}_{m}}
\;\;.
\end{equation}
Formulas from refs.~\cite{NSI, NSII, NSIII} for this quantity are reviewed in appendix \ref{sec:inclusive}. The most important case to consider is that of a $q \to q + \Lg$ splitting or a $\bar q \to \bar q + \Lg$ splitting. However, we include all flavor choices. We treat separately two kinematic regimes: $(1-z) \ll 1$ and $(1-z) \sim 1$ since the results in these two regimes have rather different structures. Our subsequent analysis of evolution will make use of the results in these two regions.

The operator ${\cal H}_{\La}^{\rm pert}$ contains a factor $\as$. In ref.~\cite{NSI}, the argument of $\as$ was denoted by $\mu_\LR^2$ and left unspecified. In general, $\mu_\LR^2$ can be a function $\mu_\LR^2(z,t)$ of the kinematic variables that describe the splitting. Our default choice in this paper is
\begin{equation}
\label{eq:muR}
\mu_\LR^2 = \lambda_\LR (1 - z + y)  M^2 e^{-t}
\;\;,
\end{equation}
where
\begin{equation}
\label{eq:lambdaR}
\lambda_\LR
=
\exp\left(
- \frac{ C_\LA\big[67 - 3\pi^2 \big]- 10\, n_\Lf}{3\, (33 - 2\,n_\Lf)}
\right)
\;\;.
\end{equation}
Except when $(1-z)$ is very small, this is approximately the constant $\lambda_\LR$ times $\bm k_\perp^2 = (1-z) M^2 e^{-t}$. In section~\ref{sec:result}, we will see why choosing $\mu_\LR^2$ approximately proportional to $\bm k_\perp^2$ is useful and we will see why the choice given in eq.~(\ref{eq:lambdaR}) for the proportionality constant is also useful. 

Leaving these points for later, we can immediately understand why a factor $(1 - z + y)$ in eq.~(\ref{eq:muR}) is preferable to the simpler choice $(1-z)$. As we will see, having a running scale $\mu_\LR^2$ as in eq.~(\ref{eq:muR}) with either a factor $(1 - z + y)$ or a factor a factor $(1 - z)$ affects the Sudakov exponent that we obtain in section~\ref{sec:result}. When $\as(\mu_\LR^2)$ is expanded in powers of $\as(M^2)$ one obtains terms proportional to logarithms of $(1 - z + y)$ or $(1 - z)$ times extra powers of $\as$. These terms improve the matching between the Sudakov exponent obtained with the shower and the true QCD Sudakov exponent. We will find that, with the accuracy of matching that we can obtain, logarithms of $(1 - z + y)$ or of $(1 - z)$ are equivalent. However, we can still ask which is more desirable in general. The running $\as(\mu_\LR^2)$ incorporates some features of the singularity structure of higher order graphs into the leading order splitting. We note that the leading order splitting kernel from ref.~\cite{NSI} has a singularity $1/(1 - z + y)$.  This suggests that the higher order contributions might naturally contain a logarithm of this same variable, $(1 - z + y)$. In contrast, the leading order splitting kernel does {\em not} have a singularity when $(1-z)\to 0$ at fixed $y$, so it would not be natural to introduce a logarithm of $(1-z)$ into the expansion of $\as(\mu_\LR^2)$. Indeed, soft gluon emissions correspond to $y \to 0$ and $(1-z)\to 0$ together, while $(1-z)\to 0$ at fixed $y$ corresponds to an anticollinear emission in which incoming parton ``a'' emits a gluon in the direction of incoming parton ``b''. There are such singularities, but they are associated with emissions from incoming parton ``b'' rather than from incoming parton ``a''. For this reason, our default choice (\ref{eq:muR}) for $\mu_\LR^2$ has a factor $(1 - z + y)$ rather than $(1 - z)$.

\subsection{$(1-z) \ll 1$}
\label{sec:1mzll1}

We use the splitting operator for an initial state splitting from ref.~\cite{NSI}, using the splitting variables $t,z,\phi$ defined in section~\ref{sec:kinematics} of this paper. When the emitted parton is a gluon, the splitting probability has a ``soft gluon emission'' singularity that corresponds to a factor $1/(1-z)$ when $y \ll (1-z) \ll 1$. In this section, we extract the terms that have this soft gluon factor; other terms will be included in the following subsection, where we study the regime $(1-z) \sim 1$. We note, in particular, that contributions in which the emitted parton is not a gluon do not give a $1/(1-z)$ contribution.

We use the results of ref.~\cite{NSI}, particularly eqs.~(12.20), (12.21), and (12.22), as reviewed in appendix \ref{sec:inclusive}.\footnote{We note that the first factor on the right hand side of eq.~(12.22) of ref.~\cite{NSI} needs to be complex conjugated; the * is missing.} Part of eq.~(12.21) is a function $A_{lk}$. There is some freedom in choosing this function. We use the definition in eqs. (7.2) and (7.12) of ref.~\cite{NSIII}, which are equivalent to eqs.~(\ref{eq:AlktoAlkprime}) and (\ref{eq:Alkangle}); we discuss the reason for this choice in section~\ref{sec:CSmod}. We neglect $y$ compared to 1 and $(1-z)$ compared to 1. However, we do not neglect $y$ compared to $(1-z)$. In the shower formalism of ref.~\cite{NSI}, when parton ``a'' splits by emitting a gluon, we include interference graphs. The soft gluon is emitted from parton ``a'' in the amplitude and by some other parton $k$ in the complex conjugate amplitude, or by parton ``a'' in the complex conjugate amplitude and by parton $k$ in the amplitude. Here parton $k$ could be parton ``b'' or could be a final state parton. For this reason, there is a sum over the indices $k$ of helper partons, with $k \ne \La$. 

After some calculation, we find in appendix \ref{sec:inclusive} that
\begin{equation}
\begin{split}
\label{eq:realsmallzsplitting}
\sbra{1}z{\cal H}_{\La}^{\rm pert}(t;z,\phi;f'))&
\sket{\{p,f,s',c',s,c\}_{m}}
\\
\approx{}& \delta_{f',\Lg}\,\sum_{k \ne \La} 
\brax{\{s'\}_{m}}\ket{\{s\}_{m}}
(-1)\bra{\{c'\}_{m}}\bm T_k\!\cdot\! \bm T_\La\ket{\{c\}_{m}}
\\& \times
\frac{\as\!\left(\mu_\LR^2\right)}{2\pi}\,
\frac{2}{1 - z + y}\,f(z,y,\phi;r_k)
\;\;.
\end{split}
\end{equation}
Here $r_k$ is the rapidity of the helper parton $k$ relative to the rest frame of $p_\La + p_\Lb$ and $\phi_k$ is its azimuthal angle; then $f$ is
\begin{equation}
f(z,y,\phi;r_k)
\approx
\left[
1
- e^{r_k}\,\frac{ 2\sqrt{(1-z) y}}{1 - z + y}\, \cos(\phi - \phi_k)
+e^{2r_k}\,\frac{2 y}{1 - z + y}
\right]^{-1}
\;\;.
\end{equation}
Because we are calculating an inclusive splitting probability, indicated by the measurement function $\sbra{1}$, there is a quantum inner product $\brax{\{s'\}_{m}}\ket{\{s\}_{m}}$ between the spin states in the amplitude and the conjugate amplitude.\footnote{In a spin averaged shower, $\{s'\}_{m} = \{s\}_{m}$ at every step and this factor is 1.} There is also a color inner product, which is non-trivial because it contains a color matrix $T_\La^c$ for emitting a gluon from line ``a'' and a color matrix $T_k^c$ for emitting the gluon from line $k$, summed over the eight colors $c$ of the emitted gluon. This same factor appears in the Catani-Seymour dipole subtraction formalism for next-to-leading order perturbative calculations \cite{CataniSeymour}.

This splitting function is complicated because of the function $f$. To understand $f$, denote the rapidity of the emitted gluon relative to the rest frame of $p_\La + p_\Lb$ by $r$. For a soft gluon, $(1-z) \ll 1$ and $y \ll 1$, we have 
\begin{equation}
(1-z) \approx e^{2r} y
\;\;.
\end{equation}
For a given splitting time $t$, $y$ is fixed and we integrate over $z$. There is a near singularity in this integral for $(1-z) \to 0$, but this near singularity is cut off when $(1-z)$ becomes comparable to $y$. That is, $r \gtrsim 1$ is favored in the integration. Writing $e^{2r} y$ for $(1-z)$ in $f(z,y,\phi;r_k)$ gives
\begin{equation}
f(z,y,\phi;r_k)
=
\left[
1
- e^{r_k - r}\,\frac{2}{1 + e^{-2r}}\, \cos(\phi-\phi_k)
+e^{2(r_k - r)}\,\frac{2}{1 + e^{-2r}}
\right]^{-1}
\;\;.
\end{equation}
We see that when $r \sim r_k$, all three terms in $f(z,y,\phi;r_k)$ are comparable, so that we have quite a complicated function. However, 
\begin{equation}
f(z,y,\phi;r_k)
\sim 1
\hskip 1 cm {\rm when}\ r \gg r_k
\;\;.
\end{equation}

We thus see that when the rapidity of the emitted gluon is large compared to the rapidities of previously emitted gluons, the splitting function simplifies to 
\begin{equation}
\begin{split}
\label{eq:realsplitting2}
\sbra{1}z{\cal H}_{\La}^{\rm pert}&(t;z,\phi;f')
\sket{\{p,f,s',c',s,c\}_{m}}
\\
\approx{}&
\delta_{f',\Lg}
\sum_{k \ne \La} \brax{\{s'\}_{m}}\ket{\{s\}_{m}}
(-1)\bra{\{c'\}_{m}}\bm T_k\!\cdot\! \bm T_\La\ket{\{c\}_{m}}\,
\frac{\as\!\left(\mu_\LR^2\right)}{2\pi}\,
\frac{2}{1 - z + y}
\;\;.
\end{split}
\end{equation}
Now the only dependence on the helper parton index $k$ is through the color factor. This enables us to perform the color sum as described in ref.~\cite{NSI},
\begin{equation}
\begin{split}
\label{eq:colorsum}
-\sum_{k \ne \La} \bra{\{c'\}_{m}}\bm T_k\!\cdot\! \bm T_\La\ket{\{c\}_{m}}={}& 
\bra{\{c'\}_{m}}\bm T_\La\!\cdot\! \bm T_\La\ket{\{c\}_{m}}
= C_a \brax{\{c'\}_{m}}\ket{\{c\}_{m}}
\;\;,
\end{split}
\end{equation}
where
\begin{equation}
\label{eq:Cadef}
C_a = 
\begin{cases}
C_\LF  & a\ne \Lg \\
C_\LA  & a = \Lg 
\end{cases}
\;\;.
\end{equation}
Thus
\begin{equation}
\begin{split}
\label{eq:realsimplesplittingA}
\sbra{1}z{\cal H}_{\La}^{\rm pert}&(t;z,\phi;f')
\sket{\{p,f,s',c',s,c\}_{m}} 
\\
\approx{}& \delta_{f',\Lg}\,\brax{\{s'\}_{m}}\ket{\{s\}_{m}}
\brax{\{c'\}_{m}}\ket{\{c\}_{m}}\
\frac{\as\!\left(\mu_\LR^2\right)}{2\pi}\,C_a\,
\frac{2}{1 - z + y}
\;\;.
\end{split}
\end{equation}
That is, using eq.~(\ref{eq:1def}),
\begin{equation}
\begin{split}
\label{eq:realsimplesplitting}
\sbra{1}z{\cal H}_{\La}^{\rm pert}(t;z,\phi;f')&
\sket{\{p,f,s',c',s,c\}_{m}} 
\\
\approx{}& \delta_{f',\Lg}\,
\frac{\as\!\left(\mu_\LR^2\right)}{2\pi}\,C_a\,
\frac{2}{1 - z + y}\,
\sbrax{1}\sket{\{p,f,s',c',s,c\}_{m}}
\;\;.
\end{split}
\end{equation}

\subsection{$(1-z) \sim 1$}

Now consider the collinear limit $(1-z)\sim 1$. We use the general result from ref.~\cite{NSI}, reviewed in appendix \ref{sec:inclusive}. We neglect $e^{-t}$ compared to 1 as discussed in section \ref{sec:approximations}. As described in  appendix \ref{sec:inclusive}, the result is a simple color structure that multiplies the standard (unregulated) DGLAP splitting kernels $P_{a,a'}(z)$ and the ratio of the number of colors for parton flavor $a' = a + f$ to the number of colors for parton flavor $a$,
\begin{equation}
\begin{split}
\label{eq:realcollinearsplitting}
\sbra{1}z{\cal H}_{\La}^{\rm pert}(t;z,\phi, f')&\sket{\{p,f,s',c',s,c\}_{m}}
\\
\approx{}&
\frac{n_\Lc(a+f')}{n_\Lc(a)}\,
\frac{\as\!\left(\mu_\LR^2\right)}{2\pi}\,
\frac{1}{z}\,P_{a,a+f'}(z)\,
\sbrax{1}\sket{\{p,f,s',c',s,c\}_{m}}
\;\;.
\end{split}
\end{equation}

\section{The virtual splitting function}
\label{sec:virtualsplitting}

We now look at the virtual splitting function. We need 
\begin{equation}
\sbra{1}{\cal V}_{\La}^{\rm pert}(t;z,\phi, f')\sket{\{p,f,s',c',s,c\}_{m}}
\;\;.
\end{equation}
Eq.~(\ref{eq:HpertVpert}) gives
\begin{equation}
\begin{split}
\label{eq:Vpert}
\sbra{1}{\cal V}^{\rm pert}(t)
={}&
\sbra{1}
\left[{\cal V}(t)
+ {\cal F}(t)^{-1}\,\frac{d}{dt}\,{\cal F}(t)
\right]
\;\;.
\end{split}
\end{equation}
Following ref.~\cite{NSI}, we define the virtual splitting operator ${\cal V}(t)$ from the requirement
\begin{equation}
\sbra{1}{\cal V}(t) = \sbra{1}{\cal H}_\LI(t)
\;\;,
\end{equation}
which guarantees that shower evolution does not change the cross section when we integrate over all final states that start from the hard scattering. We break ${\cal H}_\LI$ into pieces as in eq.~(\ref{eq:Hexpansion}),
\begin{equation}
\begin{split}
\label{eq:Hexpansionnonpert}
{\cal H}_\LI(t) ={}& 
{\cal H}_{\rm FS}(t)
+ \int_0^1 \! dz \int_{-\pi}^\pi \frac{d\phi}{2\pi}\sum_{f'}\ 
\Big[
{\cal H}_{\La}(t;z,\phi,f')
+
{\cal H}_{\Lb}(t;z,\phi,f')
\Big]
\;\;.
\end{split}
\end{equation}
This differs from eq.~(\ref{eq:Hexpansion}) in only one respect: the parton distribution factors that were omitted in the pieces of ${\cal H}^{\rm pert}_\LI(t)$ are included in the pieces of ${\cal H}_\LI(t)$, as indicated in eq.~(\ref{eq:HtoHpert}); for a splitting of parton ``a'', using $\hat \eta_\La \approx \eta_\La/z$ and $\hat a = a + f'$, the relation is 
\begin{equation}
\begin{split}
\label{eq:HtoHpert2}
\sbra{1}
{\cal H}_{\La}&(t;z,\phi,f')
\sket{\{p,f,s',c',s,c\}_{m}} 
\\&=
\frac{n_\Lc(a)\,f_{(a+f')/A}(\eta_{\La}/z,M^2 e^{-t})}
{n_\Lc(a+f')\,f_{a/A}(\eta_{\La},M^2 e^{-t})}
\,
\sbra{1}
z{\cal H}_{\La}^{\rm pert}(t;z,\phi,f')
\sket{\{p,f,s',c',s,c\}_{m}}
\;\;.
\end{split}
\end{equation}
We can similarly write ${\cal V}(t)$ as a sum and integral in the form
\begin{equation}
\begin{split}
\label{eq:Vexpansionnonpert}
{\cal V}(t) ={}& 
{\cal V}_{\rm FS}(t)
+ \int_0^1 \! dz \int_{-\pi}^\pi \frac{d\phi}{2\pi}\sum_{f'}\ 
\Big[
{\cal V}_{\La}(t;z,\phi,f')
+
{\cal V}_{\Lb}(t;z,\phi,f')
\Big]
\;\;.
\end{split}
\end{equation}
With this notation, the definition in ref.~\cite{NSI} is
\begin{equation}
\sbra{1}{\cal V}_{\La}(t;z,\phi,f') =
\sbra{1}{\cal H}_{\La}(t;z,\phi,f')
\;\;.
\end{equation}

The needed matrix element involving the derivative of ${\cal F}(t)$ is 
\begin{equation}
\begin{split}
\sbra{1}{\cal F}(t)^{-1}\,\frac{d}{dt}\,{\cal F}(t)
\sket{\{p,f,s',c',s,c\}_{m}}
={}& 
\left[
\frac{\frac{d}{dt}\,f_{a/A}(\eta_\La,M^2e^{-t})}
{f_{a/A}(\eta_\La,M^2e^{-t})} 
+
\frac{\frac{d}{dt}\,f_{b/B}(\eta_\La,M^2e^{-t})}
{f_{b/B}(\eta_\La,M^2e^{-t})}
\right]
\\&\times
\sbrax{1}\sket{\{p,f,s',c',s,c\}_{m}}
\;\;.
\end{split}
\end{equation}
The parton distribution functions obey the DGLAP equation\footnote{If the parton distributions used in shower generation obey the second or higher order DGLAP equation, then there are more terms with extra powers of $\as$. We do not display possible higher order terms in order to keep the notation from becoming complicated, but in the end we will find that their inclusion would change the $Z$-boson transverse momentum distribution at a level that is beyond the accuracy that we aim for. }
\begin{equation}
\begin{split}
\label{eq:DGLAP}
\frac{d}{dt}\,f_{a/A}(\eta_\La,M^2e^{-t}) ={}&
- \int_0^1\!dz\sum_{f'}\ \frac{\as\!\left(\mu_\LR^2\right)}{2\pi}\
\bigg\{ \frac{1}{z}\, P_{a,a+f'}(z)\,
f_{(a+f')/A}(\eta_\La/z,M^2e^{-t})
\\&\quad
-
\delta_{f',\Lg}
\left[\frac{2C_a}{1-z} - \gamma_a
\right]f_{a/A}(\eta_\La,M^2e^{-t})
\bigg\}
+ {\cal O}(\alpha_s^2)
\;\;.
\end{split}
\end{equation}
Here $P_{a,\hat a}(z)$ are the standard (unregularized) DGLAP kernels, $C_a$ is either $C_\LF$ or $C_\LA$ as in eq.~(\ref{eq:Cadef}), and 
\begin{equation}
\label{eq:gammaa}
\gamma_a = 
\begin{cases}
\frac{3}{2}\,C_\LF  & a\ne \Lg \\
\frac{1}{6}\,[11 C_\LA - 2 n_\Lf]   & a = \Lg 
\end{cases}
\;\;.
\end{equation}
Following our default choice (\ref{eq:muR}) for the argument of $\as$, we have used 
\begin{equation*}
\as\!\left(\mu_\LR^2\right) = \as\left(\lambda_\LR (1 - z + y) M^2 e^{-t}\right)
\end{equation*}
in the evolution equation for the parton distribution functions. This is not the standard choice, but it can be accommodated without changing the parton distribution functions by modifying the evolution kernel at next-to-leading order and beyond. That is, the terms indicated by ${\cal O}(\alpha_s^2)$ are modified from what they would have been had we used $\as\!\left(M^2 e^{-t}\right)$. These terms do not affect our analysis.

With these results, we can write the complete ${\cal V}^{\rm pert}(t)$ as a sum and integral in the form used in eq.~(\ref{eq:Vexpansion}),
\begin{equation}
\begin{split}
\label{eq:Vpertexpansion}
{\cal V}^{\rm pert}(t) ={}& 
{\cal V}_{\rm FS}^{\rm pert}(t)
+ \int_0^1 \! dz \int_{-\pi}^\pi \frac{d\phi}{2\pi}\sum_{f'}\ 
\Big[
{\cal V}_{\La}^{\rm pert}(t;z,\phi,f')
+
{\cal V}_{\Lb}^{\rm pert}(t;z,\phi,f')
\Big]
\;\;.
\end{split}
\end{equation}
For ${\cal V}_{\La}^{\rm pert}(t;z,\phi,f')$, we have
\begin{equation}
\begin{split}
\label{eq:VfromH}
\sbra{1}{\cal V}_{\La}^{\rm pert}&(t;z,\phi, f')\sket{\{p,f,s',c',s,c\}_{m}} 
\\\approx{}& 
\frac{n_\Lc(a)\,f_{(a+f')/A}(\eta_{\La}/z,M^2 e^{-t})}
{n_\Lc(a+f')\,f_{a/A}(\eta_{\La},M^2 e^{-t})}
\sbra{1}z{\cal H}_{\La}^{\rm pert}(t;z,\phi, f')\sket{\{p,f,s',c',s,c\}_{m}} 
\\&
- \frac{\as\!\left(\mu_\LR^2\right)}{2\pi}\
\bigg\{
P_{a,a+f'}(z)\,
\frac{f_{(a+f')/A}(\eta_\La/z,M^2e^{-t})}
{z\, f_{a/A}(\eta_\La,M^2e^{-t})}
-
\delta_{f',\Lg}
\left[\frac{2C_a}{1-z} - \gamma_a
\right]
\bigg\}
\\&\times
\sbrax{1}\sket{\{p,f,s',c',s,c\}_{m}}
\;\;.
\end{split}
\end{equation}
This simplifies in two limits, which we now discuss.

\subsection{$(1-z) \ll 1$}

Consider the limit $(1-z) \ll 1$. Then the leading terms in ${\cal V}_{\La}^{\rm pert}$ contain a factor that equals $1/(1-z)$ as long as $1-z \gg y$; contributions without this factor can be neglected. There is no such factor unless $f' = \Lg$. Thus only $f' = \Lg$ is important. In the factors with ratios of parton distributions, $a + f' = a$ and $\eta_\La/z \approx \eta_\La$; thus these factors are well approximated by 1. This gives
\begin{equation}
\begin{split}
\sbra{1}{\cal V}_{\La}^{\rm pert}&(t;z,\phi, f')\sket{\{p,f,s',c',s,c\}_{m}} 
\\\approx{}& 
\delta_{f',\Lg}
\sbra{1}z{\cal H}_{\La}^{\rm pert}(t;z,\phi, f')\sket{\{p,f,s',c',s,c\}_{m}} 
\\&
- \delta_{f',\Lg}\,\frac{\as\!\left(\mu_\LR^2\right)}{2\pi}\
\bigg\{
P_{a,a}(z)
-
\frac{2C_a}{1-z} + \gamma_a
\bigg\}
\sbrax{1}\sket{\{p,f,s',c',s,c\}_{m}}
\;\;.
\end{split}
\end{equation}
In the second line on the right hand side of this equation, $P_{a,a}(z)$ contains a term proportional to $1/(1-z)$. Other terms, which do not contain this factor, can be ignored. Similarly, we can ignore the term $-\gamma_a$ since it has no $1/(1-z)$. In fact, the term in $P_{a,a}(z)$ proportional to $1/(1-z)$ is $2 C_a/(1-z)$, which cancels the term $-2 C_a/(1-z)$. This leaves the very simple result,
\begin{equation}
\begin{split}
\label{eq:Vsmallz}
\sbra{1}{\cal V}_{\La}^{\rm pert}(t;z,\phi, f')&\sket{\{p,f,s',c',s,c\}_{m}} 
\\ \approx{}& 
\delta_{f',\Lg}
\sbra{1}z{\cal H}_{\La}^{\rm pert}(t;z,\phi, f')\sket{\{p,f,s',c',s,c\}_{m}} 
\;\;.
\end{split}
\end{equation}

\subsection{$(1-z)\sim 1$}

Now consider the collinear limit $(1-z)\sim 1$. For the matrix element of ${\cal H}_{\La}^{\rm pert}$ in this limit, we use eq.~(\ref{eq:realcollinearsplitting}). This gives
\begin{equation}
\begin{split}
\sbra{1}{\cal V}_{\La}^{\rm pert}&(t;z,\phi, f')\sket{\{p,f,s',c',s,c\}_{m}} 
\\\approx{}& 
\frac{f_{(a+f')/A}(\eta_{\La}/z,M^2 e^{-t})}
{z f_{a/A}(\eta_{\La},M^2 e^{-t})}
\frac{\as\!\left(\mu_\LR^2\right)}{2\pi}\,P_{a,a+f'}(z)\,
\sbrax{1}\sket{\{p,f,s',c',s,c\}_{m}}
\\&
- \frac{\as\!\left(\mu_\LR^2\right)}{2\pi}\
\bigg\{
P_{a,a+f'}(z)\,
\frac{f_{(a+f')/A}(\eta_\La/z,M^2e^{-t})}
{z\, f_{a/A}(\eta_\La,M^2e^{-t})}
-
\delta_{f',\Lg}
\left[\frac{2C_a}{1-z} - \gamma_a
\right]
\bigg\}
\\&\times
\sbrax{1}\sket{\{p,f,s',c',s,c\}_{m}}
\;\;.
\end{split}
\end{equation}
The two terms involving ratios of parton distributions cancel, leaving the very simple result,
\begin{equation}
\begin{split}
\label{eq:Vcollinear}
\sbra{1}{\cal V}_{\La}^{\rm pert}(t;z,\phi, f')&\sket{\{p,f,s',c',s,c\}_{m}} 
\\ \approx{}& 
\delta_{f',\Lg}\,\frac{\as\!\left(\mu_\LR^2\right)}{2\pi}\
\left[\frac{2C_a}{1-z} - \gamma_a
\right]
\sbrax{1}\sket{\{p,f,s',c',s,c\}_{m}}
\;\;.
\end{split}
\end{equation}

\section{Evolution for $t < t_\Lc$}
\label{sec:evolutionI}

The lines $\bm k_\perp^2 \bm b^2 \sim 1$ and $(1-z) \sim 1$ in figure~\ref{fig:TrianglePlain} cross at a shower evolution time given approximately by $t = \log(\bm b^2 M^2)$. At this time, the nature of the evolution of the $b$-space partonic cross section changes. We define a critical shower evolution time $t_\Lc$ by
\begin{equation}
\label{eq:tcdef}
t_\Lc = \log\left(\frac{1}{4}\,\bm b^2 M^2\,e^{2\gamma_{\rm E}}\right)
\;\;.
\end{equation}
Here $\gamma_{\rm E}$ is the Euler constant. We will see later the reason for the factor $\exp(2 \gamma_{\rm E})/4$. In this section, we make use of the results from the previous section to analyze the evolution before $t_\Lc$, $e^{-t} \gg e^{- t_\Lc}$.

Recall from eq.~(\ref{eq:evolutionfromK}) that that the evolution is specified by operators that we called ${\cal K}_\La$ and ${\cal K}_\Lb$,
\begin{equation}
\begin{split}
\label{eq:evolutionfromKbis}
\frac{d}{dt}
\sbra{1}
{\cal Q}(\bm b, Y;&\eta_\La, \eta_\Lb, a, b)
\sket{\rho_{\rm pert}(t)}
\\ 
\approx{}&
\int_0^1\! dz \int_{-\pi}^\pi \frac{d\phi}{2\pi}\sum_{f'}
\sbra{1} 
{\cal K}_\La(t;z,\phi,f';\bm b, Y;\eta_\La, \eta_\Lb, a, b)
\sket{\rho_{\rm pert}(t)}
\\ &
+
({\rm b\ splitting\ term})
\;\;.
\end{split}
\end{equation}
The operator ${\cal K}_\La$ is defined in eq.~(\ref{eq:evolution3}). We will analyze ${\cal K}_\La$; the analysis of ${\cal K}_\Lb$ is the same.

In the end, ${\cal K}_\La$ is very simple. However, we will have to analyze it in stages, pruning away complications at each stage.

\subsection{Distribution of final state partons}
\label{sec:forbidden}

We are concerned with the evolution of the $b$-space partonic cross section, 
\begin{equation*}
\sbra{1} {\cal Q}(\bm b, Y; \eta_\La, \eta_\Lb, a, b) \sket{\rho_{\rm pert}(t)}
\;\;.
\end{equation*}
This is an inclusive quantity, but we first note something about the structure of the partonic final states that contribute to it. Recall from its definition (\ref{eq:Qbyetcdef}) that ${\cal Q}$ contains a factor
\begin{equation}
\begin{split}
\label{eq:Qfactor}
e^{-\mi \bm p_{Z,\perp} \cdot \bm b}
= \prod_{j>1} e^{\mi \bm p_{j,\perp} \cdot \bm b}\
\;\;,
\end{split}
\end{equation}
where $\bm p_{j,\perp}$ is the transverse momentum of the $j$th final state parton (other than the $Z$-boson). When we form the $b$-space partonic cross section, we integrate over all of these transverse momenta. The integration region in which one or more partons have $\bm p_{j,\perp}^2 \gg 1/ \bm b^2$ gives a negligible contribution because the exponential factor $\exp({\mi \bm p_{j,\perp} \cdot \bm b})$ averages to zero. It is as if ${\cal Q}$ contained a factor $\prod_j \theta(\bm p_{j,\perp}^2 < C/\bm b^2)$, where $C$ is a constant that is large compared to 1 but with $\log C$ not large. That is, ${\cal Q}$ effectively projects onto partonic final states in which no partons have been emitted in the region above the line $\bm k_{\perp}^2 \sim 1/\bm b^2$ in figure~\ref{fig:TrianglePlain}.

One simple consequence of this concerns any real parton splitting that has the potential to change the flavor $a$ or momentum fraction $\eta_\La$ of parton ``a''. Such a splitting must be close to the line $(1-z) \sim 1$ in figure~\ref{fig:TrianglePlain}. To see this, note first that in order to significantly change $\eta_\La$ it is necessary that $(1-z)$ not be tiny compared to 1. Second, in order to change the flavor $a$, the parton emitted into the final state must not be a gluon, but the splitting functions for these splittings do not have $(1-z) \to 0$ singularities and hence have negligible probabilities of occurring with $(1-z) \ll 1$. 

We now note from eqs.~(\ref{eq:kperpsq}) and (\ref{eq:tcdef}) that 
\begin{equation}
\label{eq:kperpsqmod}
\bm k_\perp^2 = 4 e^{2\gamma_{\rm E}}\,\frac{1-z}{\bm b^2} e^{t_\Lc - t}
\;\;.
\end{equation}
Thus any emission with $(1-z) \sim 1$ and $e^{-t} \gg e^{-t_\Lc}$ has $\bm k_\perp^2 \gg 1/\bm b^2$. We conclude that all emissions that contribute to the $b$-space partonic cross section for $t < t_\Lc$ leave the flavor $a$ of parton ``a'' unchanged and leave its momentum fraction $\eta_\La$ approximately unchanged. Thus the initial conditions
\begin{equation}
\label{eq:etaconservation}
\eta_\La = x_\La
\;\;,
\hskip 1 cm
\eta_\Lb = x_\Lb
\end{equation}
remain (approximately) true and the flavors $a$ and $b$ of the incoming partons do not change. With eq.~(\ref{eq:etaconservation}), eq.~(\ref{eq:ydef2}) for the splitting variable $y$ reads
\begin{equation}
\label{eq:approximatey}
y = e^{-t}
\;\;.
\end{equation}

In the following section, we note another consequence of the fact that ${\cal Q}$ effectively projects onto partonic final states in which no partons have been emitted with $\bm k_{\perp}^2 \gg 1/\bm b^2$.

\subsection{Angular ordering}
\label{sec:angularordering}

In the range $e^{-t} \gg e^{-t_\Lc}$ and $\bm k_\perp^2 \sim {1}/{\bm b^2}$, both real and virtual emissions contribute to the evolution. For $e^{-t} \gg e^{-t_\Lc}$ and $\bm k_\perp^2 \gg {1}/{\bm b^2}$, only virtual emissions contribute. In either case, it appears that the analysis of the evolution will be very complicated. The virtual emission operator ${\cal V}_{\La}^{\rm pert}(t;z,\phi,\Lg)$ is given in eq.~(\ref{eq:VfromH}) in terms of the real emission operator ${\cal H}_{\La}^{\rm pert}(t;z,\phi,\Lg)$. The real emission operator is given in eqs.~(\ref{eq:realsmallzsplitting}), (\ref{eq:realsimplesplitting}), and (\ref{eq:realcollinearsplitting}). Here is where the complications are. There is a certain algebraic complexity and there is a non-trivial color structure. Worse, there is a sum over helper partons with labels $k$, with a different splitting function for each $k$. Thus, in order follow the evolution of the inclusive quantity $\sbrax{\bm b,Y}\sket{\rho^{\rm pert}(t)}$, it appears that we need to track the structure of the complete partonic state.\footnote{This would not be quite so bad in the leading color approximation generally used for parton showers. Then, we would need only the momentum of the parton that is color connected to the incoming quark or antiquark, which would generally be the gluon previously emitted from that incoming parton line.}

\FIGURE{
\centerline{\includegraphics[width = 8 cm]{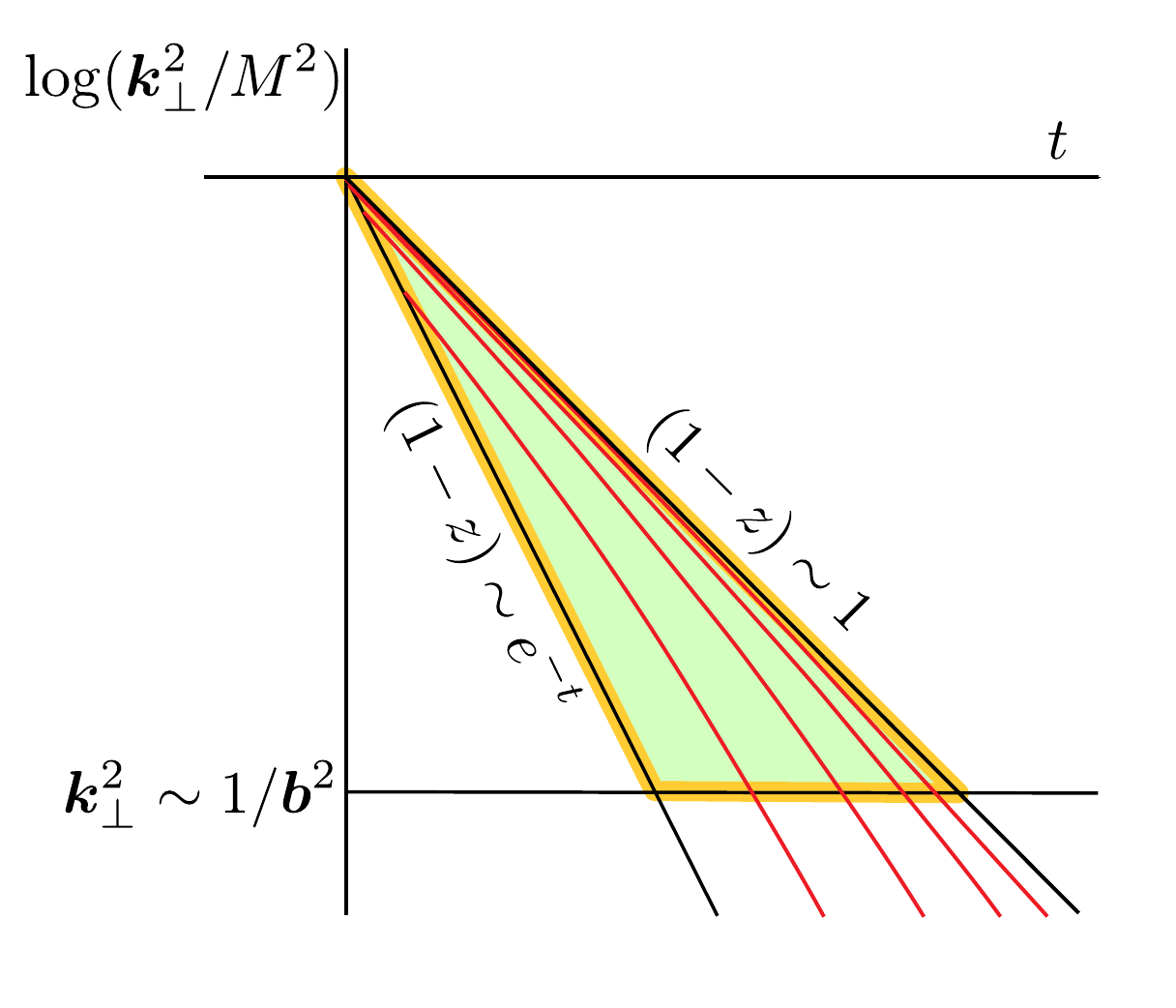}}
\caption{Integration region for initial state gluon emissions, as in Figure \ref{fig:TrianglePlain}. The slightly curved lines indicate lines of constant rapidity for the emitted gluon, with rapidity near 0 at the left and large rapidity on the right.}
\label{fig:TriangleAngle}
} 

Fortunately, there is a simplification available. We consider the real or virtual emission of a gluon with $\bm k_\perp^2 \gtrsim {1}/{\bm b^2}$ at shower time $t$ with $e^{-t} \gg e^{-t_\Lc}$. The rapidity $r$ in the $Z$-boson rest frame of the gluon is approximately given by
\begin{equation}
\label{eq:rapidityA0}
r = t + \frac{1}{2}\log\left(\frac{\bm k_\perp^2}{M^2}\right) 
+ \log\left(\frac{\eta_\La }{z x_\La }\right)
\end{equation}
as long as $e^{-t} \ll 1$. Using eq.~(\ref{eq:etaconservation}), this becomes
\begin{equation}
\label{eq:rapidityA}
r = t + \frac{1}{2}\log\left(\frac{\bm k_\perp^2}{M^2}\right) 
+ \log\left(\frac{1}{z}\right)
\;\;.
\end{equation}
Curves of constant $r$ are shown in Figure~\ref{fig:TriangleAngle}.

The rapidity of a previously emitted gluon (that is, a real gluon in the final state) is
\begin{equation}
\label{eq:rapidityAk}
r_k = t_k + \frac{1}{2}\log\left(\frac{\bm k_{\perp,k}^2}{M^2}\right) 
\;\;.
\end{equation}
Here we have omitted the $\log\left({1}/{z_k}\right)$ because any previously emitted gluon would have had $(1-z_k)\ll 1$, so $\log\left({1}/{z_k}\right) \approx 0$. Since the gluon was previously emitted, we have $t_k < t$. Since it was a real emission, the transverse momentum satisfied $\bm k_{\perp,k}^2 \lesssim 1/\bm b^2$. On the other hand,  $\bm k_{\perp}^2 \gtrsim 1/\bm b^2$. Additionally, $\log(1/z) \ge 0$. From these inequalities, we conclude that
\begin{equation}
r_k \lesssim r
\;\;.
\end{equation}
If either ${\bm k_{\perp}^2} \gg \bm b^2$ or ${\bm k_{\perp,k}^2} \ll \bm b^2$, we have 
\begin{equation}
\label{eq:rapidityapprox}
e^{2r} \gg e^{2r_k}
\;\;.
\end{equation}
In the event that $\bm k_\perp^2$ and $\bm k_{\perp,k}^2$ are both of order $1/\bm b^2$, this strong inequality may not hold. However, the emission probability per unit $t$ and per unit $\log(\bm k_\perp^2/M^2)$ is small, of order $\as$. Thus we can apply the low density approximation developed in section~\ref{sec:approximations} to conclude that in this event we can approximate
\begin{equation}
\label{eq:tapprox}
e^{2t} \gg e^{2t_k}
\end{equation}
at the cost of affecting terms in the Sudakov exponent that we will obtain at the level of third-to-leading terms and beyond. Eq.~(\ref{eq:tapprox}) then implies eq.~(\ref{eq:rapidityapprox}).

For this reason, we can make the approximation eq.~(\ref{eq:rapidityapprox}) in the expression (\ref{eq:realsmallzsplitting}) for ${\cal H}_{\La}^{\rm pert}(t;z,\phi,\Lg)$. This gives the much simpler formula eq.~(\ref{eq:realsimplesplitting}), for which the details of the state of the shower at time $t$ are not needed.

We are now prepared to write an evolution equation for $\sbra{1}
{\cal Q}(\bm b, Y;\eta_\La, \eta_\Lb, a, b) \sket{\rho_{\rm pert}(t)}$ for $t < \log(\bm b^2 M^2)$. We consider first $z \ll 1$, then $z \sim 1$. Then we combine these two cases.

\subsection{Contribution from $(1-z) \ll 1$}
\label{sec:small1mzsimplified}

In this subsection, we consider ${\cal K}_\La$ in the region
\begin{equation}
1-z \ll 1\;\;,\hskip 1 cm
e^{-t} \gg e^{-t_\Lc}
\;\;.
\end{equation}
That is, we consider splittings that are represented by points well below the line $(1-z) \sim 1$ in figure~\ref{fig:TrianglePlain} and well to the left of $t = t_\Lc$ that passes through the lower right vertex of the triangle. In this region, gluon emission from line ``a'' is important because the splitting function has a term with a $1/(1-z)$ factor. However, splittings with $f' \ne \Lg$ lack this factor and may be omitted. With this approximation, eq.~(\ref{eq:evolution3}) becomes
\begin{equation}
\begin{split}
\label{eq:evolution4}
\sbra{1} 
{\cal K}_\La(t;& z,\phi,f';\bm b, Y;\eta_\La, \eta_\Lb, a, b)
\sket{\rho_{\rm pert}(t)}
\\ 
\approx{}&
\delta_{f',\Lg}\,
\sbra{1} 
{\cal H}_{\La}^{\rm pert}(t;z,\phi,\Lg)
\exp\!\left(\mi\, \frac{x_\La}{z\eta_\La}\,\bm b\!\cdot\! \bm k_\perp\right)
z{\cal Q}(\bm b, Y; z \eta_\La, \eta_\Lb, a, b)
\\&\qquad
- {\cal V}_\La^{\rm pert}(t;z,\phi,\Lg)
{\cal Q}(\bm b, Y; \eta_\La,  \eta_\Lb, a, b)
\sket{\rho_{\rm pert}(t)}
\;\;.
\end{split}
\end{equation}
Since $1-z \ll 1$, we can also replace $z$ by 1 in the exponential factor and in the argument of ${\cal Q}(\bm b, Y; z \eta_\La, \eta_\Lb, a, b)$. We can also use eq.~(\ref{eq:etaconservation}) to replace $x_\La/\eta_\La$ in the exponent by 1. Finally, we use eq.~(\ref{eq:Vsmallz}) for ${\cal V}_\La^{\rm pert}$. With these changes, we have
\begin{equation}
\begin{split}
\label{eq:evolution6}
\sbra{1}& 
{\cal K}_\La(t; z,\phi,f';\bm b, Y;\eta_\La, \eta_\Lb, a, b)
\sket{\rho_{\rm pert}(t)}_{a,b \ne \Lg}
\\ 
\approx{}&
\delta_{f',\Lg}\,
\sbra{1} z{\cal H}_\La^{\rm pert}(t;z,\phi,\Lg)
{\cal Q}(\bm b, Y; \eta_\La,  \eta_\Lb, a, b)
\sket{\rho_{\rm pert}(t)}
\left[
\exp\!\left(\mi\,\bm b\!\cdot\! \bm k_\perp\right)
- 1 \right]
\;\;.
\end{split}
\end{equation}

We can make one more simplification. The right hand side of eq.~(\ref{eq:evolution6}) vanishes when $\bm k_\perp^2 \ll {1}/{\bm b^2}$. In the alternative case that $\bm k_\perp^2 \gtrsim {1}/{\bm b^2}$, we can apply the lesson of section~\ref{sec:angularordering} so as to use the simple form of $\sbra{1}z{\cal H}_\La^{\rm pert}$ given in eq.~(\ref{eq:realsimplesplitting}), 
\begin{equation}
\begin{split}
\label{eq:realsimplesplitting2}
\sbra{1}z{\cal H}_{\La}^{\rm pert}(t;z,\phi;\Lg)&
\sket{\{p,f,s',c',s,c\}_{m}} 
\\
\approx{}& 
\frac{\as\!\left(\mu_\LR^2\right)}{2\pi}\,C_a\,
\frac{1}{1 - z + e^{-t}}\,
\sbrax{1}\sket{\{p,f,s',c',s,c\}_{m}}
\;\;.
\end{split}
\end{equation}
Here we have replaced $y$ by $e^{-t}$ according to eq.~(\ref{eq:approximatey}). Thus
\begin{equation}
\begin{split}
\label{eq:evolution7}
\sbra{1} 
{\cal K}_\La(t; z,\phi,f';\bm b, Y;&\eta_\La, \eta_\Lb, a, b)
\sket{\rho_{\rm pert}(t)}
\\
\approx{}&
\delta_{f',\Lg}\,
\frac{\as\!\left(\mu_\LR^2\right)}{2\pi}\,C_a\,
\frac{1}{1 - z + e^{-t}}\,
\left[
\exp\!\left(\mi\, \bm b\!\cdot\! \bm k_\perp\right)
- 1 \right]
\\ & \quad \times
\sbra{1}{\cal Q}(\bm b, Y; \eta_\La,  \eta_\Lb, a, b)
\sket{\rho_{\rm pert}(t)}
\;\;.
\end{split}
\end{equation}
Of course, the expression that we have used for $\sbra{1}z{\cal H}_{\La}^{\rm pert}$ is not accurate in the region $\bm k_\perp^2 \ll {1}/{\bm b^2}$; nevertheless, both the exact form and the new form vanish in this region, so this inaccuracy is not a problem. We can use eq.~(\ref{eq:evolution7}) for both $\bm k_\perp^2 \ll {1}/{\bm b^2}$  and $\bm k_\perp^2 \gtrsim {1}/{\bm b^2}$ so long as $(1-z) \ll 1$ and $e^{-t} \gg e^{-t_\Lc}$.

Notice the key feature of eq.~(\ref{eq:evolution7}) that the object for which we seek an evolution equation,
\begin{equation*}
\sbra{1}{\cal Q}(\bm b, Y; \eta_\La,  \eta_\Lb, a, b)
\sket{\rho_{\rm pert}(t)}
\;\;,
\end{equation*}
appears on the right hand side of this result.

We can make one more simplification in this. Recall that we need the average over the emission angle $\phi$ of $\sbra{1} {\cal K}_\La\sket{\rho_{\rm pert}(t)}$. The only $\phi$ dependence is in the factor $\exp\!\left(\mi\, \bm b\!\cdot\! \bm k_\perp\right)$. When we take the average over $\phi$, we get a Bessel function:
\begin{equation}
\begin{split}
\label{eq:evolution8}
\int_{-\pi}^{\pi} \frac{d\phi}{2\pi}\,
\sbra{1} 
{\cal K}_\La(t; z,\phi,f';\bm b, Y;&\eta_\La, \eta_\Lb, a, b)
\sket{\rho_{\rm pert}(t)}
\\
\approx{}&
\delta_{f',\Lg}\,
\frac{\as\!\left(\mu_\LR^2\right)}{2\pi}\,C_a\,
\frac{1}{1 - z + e^{-t}}\,
\left[
J_0(|\bm k_\perp| |\bm b|)
- 1 \right]
\\ & \quad \times
\sbra{1}{\cal Q}(\bm b, Y; \eta_\La,  \eta_\Lb, a, b)
\sket{\rho_{\rm pert}(t)}
\;\;.
\end{split}
\end{equation}

We will want to integrate this over $z$. Consider the integral between limits $z_1$ and $z_2$, with $e^{-t} \ll (1-z_1) < (1-z_2) \ll 1$.
The denominator $(1-z + e^{-t})$ can then be approximated by simply $(1-z)$. Similarly, the argument of $\as$ is, using eq.~(\ref{eq:approximatey}) and then $e^{-t} \ll (1-z)$, 
\begin{equation}
\mu_\LR^2 = \lambda_\LR (1-z+e^{-t})M^2 e^{-t} \approx 
\lambda_\LR (1-z)M^2 e^{-t} = \lambda_\LR \bm k_\perp^2
\;\;.
\end{equation}

Using $\bm k_\perp^2 = (1-z) M^2 e^{-t}$, we can change the integration variable to $\bm k_\perp^2$. This gives
\begin{equation}
\begin{split}
\label{eq:Bessel1}
\int_{z_1}^{z_2}\!dz
\int_{-\pi}^{\pi} \frac{d\phi}{2\pi}\,
\sbra{1} 
{\cal K}_\La(t; z,\phi,f';\bm b, Y;&\eta_\La, \eta_\Lb, a, b)
\sket{\rho_{\rm pert}(t)}
\\
\approx{}&
\delta_{f',\Lg}\,C_a\,
\int_{Q_1^2}^{Q_2^2}\!\frac{d\bm k_\perp^2}{\bm k_\perp^2}\,
\frac{\as\!\left(\lambda_\LR \bm k_\perp^2\right)}{2\pi}\,
\left[
J_0(|\bm k_\perp| |\bm b|)
- 1 \right]
\\ & \quad \times
\sbra{1}{\cal Q}(\bm b, Y; \eta_\La,  \eta_\Lb, a, b)
\sket{\rho_{\rm pert}(t)}
\;\;.
\end{split}
\end{equation}
Thus we encounter an integral of the form
\begin{equation}
\label{eq:BesselA}
F(Q_2^2,Q_1^2) =
\int_{Q_1^2}^{Q_2^2}\!\frac{d\bm k_\perp^2}{\bm k_\perp^2}
f(\bm k_\perp^2)
\left[
J_0(|\bm k_\perp| |\bm b|)
- 1 \right]
\;\;,
\end{equation}
where the function $f(\bm k_\perp^2)$ allows for the dependence of $\as$ on $\bm k_\perp^2$. As we have previously argued, if $Q_1^2$ and $Q_2^2$ are both much smaller than $1/\bm b^2$, then we can replace $[J_0(|\bm k_\perp| |\bm b|) - 1]$ by zero since $J_0(|\bm k_\perp| |\bm b|) \sim 1$ for $|\bm k_\perp| |\bm b| \to 0$. Similarly, we have argued that if $Q_1^2$ and $Q_2^2$ are both much larger than $1/\bm b^2$, then we can replace $[J_0(|\bm k_\perp| |\bm b|) - 1]$ by $-1$ because $J_0(|\bm k_\perp| |\bm b|)$ is a rapidly oscillating function in the integration range. But what happens if $Q_1^2 \ll 1/\bm b^2$ while $Q_2^2 \gg 1/\bm b^2$? There is a simple approximation that is accurate provided that $f(\bm k_\perp^2)$ is a slowly varying function. We can replace $[J_0(|\bm k_\perp| |\bm b|) - 1]$ by $-\theta(|\bm k_\perp| |\bm b| > 2e^{-\gamma_E})$, where $\gamma_E$ is the Euler constant. Thus the integral becomes
\begin{equation}
\label{eq:BesselB}
F(Q_2^2,Q_1^2)_{\rm approx} =
-\int_{4e^{-2\gamma_E}/\bm b^2}^{Q_2^2}\!\frac{d\bm k_\perp^2}{\bm k_\perp^2}\,
f(\bm k_\perp^2)
\;\;.
\end{equation}
The integrals in eq.~(\ref{eq:BesselA}) and (\ref{eq:BesselB}) differ by terms proportional to a power of $1/(Q_2^2 \bm b^2)$ or a power of $Q_1^2 \bm b^2$ and by a term proportional to the second derivative of $f(\bm k_\perp^2)$ with respect to $\log(\bm k_\perp^2)$ in the region near $\bm k_\perp^2 \sim 1/\bm b^2$. The exact statement can be found in appendix \ref{sec:Bessel}. Note that the second derivative of $\as\!\left(\lambda_\LR \bm k_\perp^2\right)$ with respect to $\log(\bm k_\perp^2)$ is of order $\as\!\left(\lambda_\LR \bm k_\perp^2\right)^3$, so that this is a good approximation for our purposes.

With this approximation,
\begin{equation}
\begin{split}
\label{eq:evolution9}
\sbra{1} 
{\cal K}_\La(t; z,\phi,f';\bm b, Y;\eta_\La, \eta_\Lb, a, b)&
\sket{\rho_{\rm pert}(t)}
\\
\sim{}&
-\delta_{f',\Lg}\,
\frac{\as\!\left(\mu_\LR^2\right)}{2\pi}\,C_a\,
\frac{1}{1 - z + e^{-t}}\,
\theta(\bm k_\perp^2 \bm b^2 > 4e^{-2\gamma_E})
\\ & \quad \times
\sbra{1}{\cal Q}(\bm b, Y; \eta_\La,  \eta_\Lb, a, b)
\sket{\rho_{\rm pert}(t)}
\;\;,
\end{split}
\end{equation}
where the $\sim$ in this case indicates that this approximation works inside the integration over $\phi$ and $z$.

Specifically, the derivation so far covers a range of $z$ between limits $e^{-t} \ll (1-z_1) < (1-z_2) \ll 1$. We have in mind that $(1 - z_2)$ is chosen small enough that the approximation $(1 - z) \ll 1$ applies within the integration range. The region $(1 - z) \sim 1$ is treated separately in the following section. 

Similarly, we choose $(1 - z_1) = C_1 e^{-t}$, where $C_1 \gg 1$ but $\log C_1$ is not large. Then the integration range $0 < (1-z) < C_1 e^{-t}$ needs a separate treatment, which we now provide. For that range, the denominator $1 - z + e^{-t}$ in eq.~(\ref{eq:evolution8}) provides a lower cutoff on $(1-z)$ at around $(1-z) \approx e^{-t}$. Thus the integration range that needs a separate treatment is really $e^{-t} \lesssim (1-z) < C_1 e^{-t}$. What happens here depends on the factor $\left[ J_0(|\bm k_\perp| |\bm b|) - 1 \right]$ in eq.~(\ref{eq:evolution8}). Since $\bm k_\perp^2 = (1-z) M^2 e^{-t}$, define the value $(1-z_b)$ of $(1-z)$ that corresponds to $|\bm k_\perp| |\bm b| = 1$ by
\begin{equation}
(1-z_b) M^2 b^2 e^{-t} = 1
\;\;.
\end{equation}
If $(1-z_b) \ll e^{-t}$, then $|\bm k_\perp| |\bm b| \gg 1$ in the integration range $e^{-t} \lesssim (1-z) < C_1 e^{-t}$ and we can approximate $\left[
J_0(|\bm k_\perp| |\bm b|) - 1 \right] \to -1$. If $(1-z_b) \gg C_1 e^{-t}$, then $|\bm k_\perp| |\bm b| \ll 1$ in the integration range $e^{-t} \lesssim (1-z) < C_1 e^{-t}$ and we can approximate $\left[J_0(|\bm k_\perp| |\bm b|) - 1 \right] \to 0$. Both cases are covered by the approximation
\begin{equation}
\left[J_0(|\bm k_\perp| |\bm b|) - 1 \right] \to \theta(\bm k_\perp^2 \bm b^2 > 4e^{-2\gamma_E})
\;\;.
\end{equation}
In the remaining case, $e^{-t} \lesssim (1-z_b) \lesssim C_1 e^{-t}$, we will use this same replacement. This is not an accurate approximation. However, this inaccuracy occurs only near the point at the intersection of the line $(1-z) = e^{-t}$ and the line $\bm k_\perp^2 = 1/\bm b^2$ in figure~\ref{fig:TrianglePlain}. Following an argument like that in section \ref{sec:approximations}, we recognize that this inaccuracy does not matter because it does not lead to contributions with a large logarithm in the Sudakov exponent. Thus eq.~(\ref{eq:evolution9}) is also a sufficient approximation when applied inside an integration over $z$ that includes $(1-z) \to 0$.

\subsection{Contribution from $(1-z) \sim 1$}

The approximations in the previous section cover the integration region $(1-z) \ll 1$ for $e^{-t} \gg e^{-t_\Lc}$. For the integration region $(1-z) \sim 1$ with $e^{-t} \gg e^{-t_\Lc}$, only virtual splittings contribute:
\begin{equation}
\begin{split}
\label{eq:zsim1}
\sbra{1} 
{\cal K}_\La(t;& z,\phi,f';\bm b, Y;\eta_\La, \eta_\Lb, a, b)
\sket{\rho_{\rm pert}(t)}
\\ 
\approx{}&
-
\sbra{1} 
{\cal V}_\La^{\rm pert}(t;z,\phi,f')
{\cal Q}(\bm b, Y; \eta_\La,  \eta_\Lb, a, b)
\sket{\rho_{\rm pert}(t)}
\;\;.
\end{split}
\end{equation}
For the matrix element of ${\cal V}_{\La}^{\rm pert}$, we can use eq.~(\ref{eq:Vcollinear}). This gives

\begin{equation}
\begin{split}
\label{eq:zsim1final}
\sbra{1} 
{\cal K}_\La(t;& z,\phi,f';\bm b, Y;\eta_\La, \eta_\Lb, a, b)
\sket{\rho_{\rm pert}(t)} 
\\ \approx{}& 
-\delta_{f',\Lg}\,\frac{\as\!\left(\mu_\LR^2\right)}{2\pi}\
\left[\frac{2C_a}{1-z} - \gamma_a
\right]
\sbra{1} 
{\cal Q}(\bm b, Y; \eta_\La,  \eta_\Lb, a, b)
\sket{\rho_{\rm pert}(t)}
\;\;.
\end{split}
\end{equation}
As in the previous section, we notice that the object for which we seek an evolution equation, $\sbra{1} {\cal Q}(\bm b, Y; \eta_\La,  \eta_\Lb, a, b)\sket{\rho_{\rm pert}(t)}$, appears on the right hand side of this result. 

\subsection{Evolution for all $z$}

Compare eq.~(\ref{eq:evolution9}), which applies for $(1-z) \ll 1$, with eq.~(\ref{eq:zsim1final}), which applies for $(1-z) \sim 1$. Notice that the theta function $\theta(\bm k_\perp^2 \bm b^2 > 4e^{-2\gamma_E})$ is always 1 when $(1-z) \sim 1$ and $e^{-t} \gg e^{-t_\Lc}$, that $(1-z+e^{-t})$ is the same as $(1-z)$ when $(1-z) \sim 1$ and $e^{-t} \ll 1$, and that the term $-\gamma_a$ with no $1/(1-z)$ factor is negligible (compared to $1/(1-z)$) for $(1-z) \ll 1$. Thus these two forms match and we can combine them in the form
\begin{equation}
\begin{split}
\label{eq:evolution10}
\sbra{1} 
{\cal K}_\La&(t; z,\phi,f';\bm b, Y;\eta_\La, \eta_\Lb, a, b)
\sket{\rho_{\rm pert}(t)}
\\
\sim{}&
-\delta_{f',\Lg}\,
\frac{\as\!\left(\mu_\LR^2\right)}{2\pi}
\left[\frac{2C_a}{1-z+ e^{-t}} - \gamma_a
\right]
\theta(\bm k_\perp^2 \bm b^2 > 4e^{-2\gamma_E})
\\ & \quad \times
\sbra{1}{\cal Q}(\bm b, Y; \eta_\La,  \eta_\Lb, a, b)
\sket{\rho_{\rm pert}(t)}
\;\;.
\end{split}
\end{equation}
This gives the evolution equation, for  $e^{-t} \gg e^{-t_\Lc}$ and assuming that parton ``b'' caries the opposite flavor from parton ``a'', $b = \bar a$,
\begin{equation}
\begin{split}
\label{eq:evolutionresult}
\frac{d}{dt}
\sbra{1}
{\cal Q}&(\bm b, Y;\eta_\La, \eta_\Lb, a, b)
\sket{\rho_{\rm pert}(t)}
\\
\approx{}&
-\int_0^1\!dz\
\frac{\as\!\left(\mu_\LR^2\right)}{2\pi}\,2
\left[\frac{2 C_a}{1-z+ e^{-t}} - \gamma_a
\right]
\theta(\bm k_\perp^2 \bm b^2 > 4e^{-2\gamma_E})
\\ & \quad \times
\sbra{1}{\cal Q}(\bm b, Y; \eta_\La,  \eta_\Lb, a, b)
\sket{\rho_{\rm pert}(t)}
\;\;.
\end{split}
\end{equation}
There is a factor $2$ here because emissions from both initial state lines contribute equally when $b = \bar a$.

This analysis could apply when $a = b = \Lg$, which would be relevant for computing the transverse momentum distribution in Higgs boson production. However, the case of interest in this paper is that $a$ is a quark flavor and $b$ is the corresponding antiquark flavor, or vice versa. In this case, $C_a = C_\LF$ and $\gamma_a = (3/2)\, C_\LF$. Then
\begin{equation}
\begin{split}
\label{eq:quarkevolution}
\frac{d}{dt}
\sbra{1}
{\cal Q}&(\bm b, Y;\eta_\La, \eta_\Lb, a, b)
\sket{\rho_{\rm pert}(t)}_{a,b \ne \Lg}
\\
\approx{}&
-\int_0^1\!dz\
\frac{\as\!\left(\mu_\LR^2\right)}{2\pi}\,2C_\LF
\left[\frac{2}{1-z+ e^{-t}} - \frac{3}{2}
\right]
\theta(\bm k_\perp^2 \bm b^2 > 4e^{-2\gamma_E})
\\ & \quad \times
\sbra{1}{\cal Q}(\bm b, Y; \eta_\La,  \eta_\Lb, a, b)
\sket{\rho_{\rm pert}(t)}
\;\;.
\end{split}
\end{equation}

\section{Evolution for $t > t_\Lc$}
\label{sec:evolutionII}

After shower time $t_\Lc$, the evolution changes character. To see what happens, it is simplest to consider $\sbra{1} {\cal Q}(\bm b, Y) \sket{\rho(t)}$ instead of $\sbra{1} {\cal Q}(\bm b, Y;\eta_\La, \eta_\Lb, a, b) \sket{\rho_{\rm pert}(t)}$. We have
\begin{equation}
\begin{split}
\frac{d}{dt} \sbra{1} {\cal Q}(\bm b, Y) \sket{\rho(t)} ={}& 
\sbra{1} {\cal Q}(\bm b, Y)[{\cal H}_\LI(t) - {\cal V}(t)] \sket{\rho(t)}
\\
\approx{}&
\sbra{1} {\cal Q}(\bm b, Y) {\cal H}_\LI(t) 
- {\cal V}(t){\cal Q}(\bm b, Y) \sket{\rho(t)}
\;\;.
\end{split}
\end{equation}
Consider the relation of ${\cal Q}(\bm b, Y) {\cal H}_\LI(t)$ with ${\cal H}_\LI(t) {\cal Q}(\bm b, Y)$. As we have noted in our previous analysis, each emission from an initial state line with transverse momentum $\bm k_\perp$ produces a phase factor $\exp(\mi\bm b \!\cdot\! \bm k_\perp)$ in the result of applying ${\cal Q}(\bm b, Y)$ after the splitting; this phase factor is not present in the result of applying ${\cal Q}(\bm b, Y)$ before the splitting. However, all emissions for $e^{-t} \ll e^{- t_\Lc}$ have $\bm k_\perp^2 \ll 1/ \bm b^2$. Thus this phase factor is simply 1. Thus
\begin{equation}
{\cal Q}(\bm b, Y) {\cal H}_\LI(t) 
\approx 
{\cal H}_\LI(t) {\cal Q}(\bm b, Y)
\;\;.
\end{equation}
This result gives
\begin{equation}
\begin{split}
\frac{d}{dt} \sbra{1} {\cal Q}(\bm b, Y) \sket{\rho(t)} \approx{}& 
\sbra{1} [{\cal H}_\LI(t) - {\cal V}(t)] {\cal Q}(\bm b, Y)\sket{\rho(t)}
\\
={}&
0
\;\;.
\end{split}
\end{equation}

Thus $\sbra{1} {\cal Q}(\bm b, Y) \sket{\rho(t)}$ stops evolving at $t = t_\Lc$.
The b-space partonic cross section, $\sbra{1}{\cal Q}(\bm b, Y; \eta_\La,  \eta_\Lb, a, b) \sket{\rho_{\rm pert}(t)}$, does evolve for $t > t_\Lc$ because initial state emissions can be collinear and thus change the momentum fractions and flavors $\eta_\La,  \eta_\Lb, a, b$. What happens is that the partonic cross section evolves and the parton distribution functions evolve with opposite evolution kernels, so that the net change of $\sbra{1} {\cal Q}(\bm b, Y) \sket{\rho(t)}$ with $t$ vanishes.

\section{Solution at $t = t_\Lc$}
\label{sec:solution}

We can solve eq.~(\ref{eq:quarkevolution}) for the $b$-space partonic cross section with initial condition (\ref{eq:initialcondition}), evolving from shower time $t = 0$ to shower time $t = t_\Lc$. This gives
\begin{equation}
\begin{split}
\label{eq:result1}
\sbra{1}{\cal Q}(\bm b, Y; \eta_\La, \eta_\Lb, 
a, b)\sket{\rho_{\rm pert}(t_\Lc)}
={}& 12\, \alpha\, Q_{ab}\, x_\La x_\Lb\,
\delta(\eta_\La - x_\La)\,
\delta(\eta_\Lb - x_\Lb)
\\ &\times 
\exp(- S_0(M^2,\bf b^2))
\;\;,
\end{split}
\end{equation}
where the Sudakov exponent $S_0$ is
\begin{equation}
\label{eq:sudakovdef}
S_0(M^2,\bm b^2) = \int_0^{t_\Lc}\!dt
\int_0^1\!dz\
\frac{\as\!\left(\mu_\LR^2\right)}{2\pi}\,2C_\LF
\left[\frac{2}{1-z+ e^{-t}} - \frac{3}{2}
\right]
\theta(\bm k_\perp^2 \bm b^2 > 4e^{-2\gamma_E})
\;\;.
\end{equation}
It is useful to change variables from $t$ to $\bm k_\perp^2 = (1-z)M^2 e^{-t}$, giving
\begin{equation}
\label{eq:sudakovdresult}
S_0(M^2,\bm b^2) = \int_{4e^{-2\gamma_E}/\bm b^2}^{M^2}
\!\frac{d\bm k_\perp^2}{\bm k_\perp^2}\
\int_0^1\!dz\
\frac{\as\!\left(\mu_\LR^2\right)}{2\pi}\,2C_\LF
\left[\frac{2 (1-z)}{(1-z)^2 + \bm k_\perp^2/M^2} - \frac{3}{2}
\right]
\;\;.
\end{equation}

We can now approximate $S_0(M^2,\bm b^2)$ in a fashion that makes it simpler. We are interested in the behavior of $S_0(M^2,\bm b^2)$ for $M^2 \bm b^2 \gg 1$. In particular, we are interested in terms with logarithms $\log(M^2 \bm b^2)$. The region of the $\bm k_\perp^2$ integration that dominates for large $\bm b^2$ is $\bm k_\perp^2 \ll M^2$. In contrast, the region $\bm k_\perp^2 \sim M^2$ cannot contribute logarithms $\log(M^2 \bm b^2)$ and is thus not of interest for us. For this reason, we can make the approximation $\bm k_\perp^2 \ll M^2$ inside $S_0(M^2,\bm b^2)$. Additionally, it is useful to expand $\as(\mu_\LR^2)$ in powers of $\as(\bm k_\perp^2)$. Then, after performing the $z$-integral, we obtain contributions to the integrand of the $\bm k_\perp^2$ integral of the form
\begin{equation*}
\as(\bm k_\perp^2)^n  
\left[\log\!\left({M^2}/{\bm k_\perp^2}\right)\right]^m
\;\;.
\end{equation*}
The leading terms in this counting are those with $m = n$. We need to keep track of those, along with the next-to-leading terms with $m = n-1$. However, we will not be interested in the coefficients of terms with $m \le n-2$. For this reason, we drop all such terms in the approximations below.

To proceed with this program, we note that the renormalization scale in $\as$ is, using the definition (\ref{eq:muR}) and then eq.~(\ref{eq:approximatey}),
\begin{equation}
\mu_\LR^2 = \lambda_\LR (1-z+e^{-t}) M^2 e^{-t}
= \lambda_\LR \bm k_\perp^2 
\left\{1 + \frac{\bm k_\perp^2}{(1-z)^2 M^2}\right\}
\;\;.
\end{equation}
Expanding $\as(\mu_\LR^2)$ in powers of $\as(\bm k_\perp^2)$ and displaying only the order $\as$ and $\as^2$ terms, we have
\begin{equation}
\begin{split}
\label{eq:asexpansion}
\frac{\as(\mu_\LR^2)}{2\pi}
={}& \frac{\as(\bm k_\perp^2)}{2\pi}
- 2\beta_1 \log(\lambda_\LR)
\left(\frac{\as(\bm k_\perp^2)}{2\pi}\right)^2
\\ & 
- 2\beta_1 \log\!\left(1 + \frac{\bm k_\perp^2}{(1-z)^2 M^2}\right)
\left(\frac{\as(\bm k_\perp^2)}{2\pi}\right)^2
+ {\cal O}(\as^3)
\;\;.
\end{split}
\end{equation}
Here
\begin{equation}
\beta_1 = \frac{33 - 2\,n_\Lf}{12}
\;\;.
\end{equation}
The $z$-integral that multiplies the order $\as$ term is
\begin{equation}
\label{eq:zintegral}
\int_0^1\!dz
\left[\frac{2 (1-z)}{(1-z)^2 + \bm k_\perp^2/M^2} - \frac{3}{2}
\right]
= \log\!\left(\frac{M^2}{\bm k_\perp^2}\right) - \frac{3}{2}
+ \log\!\left(1+\frac{\bm k_\perp^2}{M^2}\right)
\;\;.
\end{equation}
The most important term here is the logarithm $\log\left({M^2}/{\bm k_\perp^2}\right)$. The term $-3/2$ is next-to-leading, so we keep it. The third term in eq.~(\ref{eq:zintegral}) contributes to the $\bm k_\perp^2$ integral only for $\bm k_\perp^2 \sim M^2$ but is power suppressed in the dominant region $\bm k_\perp^2 \ll M^2$. Consequently, we neglect this term. This same integral multiplies the order $\as^2$ term proportional to $\log(\lambda_\LR)$. Here, we keep the large logarithm $\log\left({M^2}/{\bm k_\perp^2}\right)$ but neglect the term $-3/2$ in which the extra power of $\as$ is not multiplied by a large logarithm. For the remaining $\as^2$ term, we need the integral
\begin{equation}
\begin{split}
\int_0^1\!dz
\left[\frac{2 (1-z)}{(1-z)^2 + \bm k_\perp^2/M^2} - \frac{3}{2}
\right]
\log\!\left(1 + \frac{\bm k_\perp^2}{(1-z)^2 M^2}\right)
={}& \frac{\pi^2}{6} + {\cal O}\!\left(\frac{\bm k_\perp^2}{M^2}\right)
\;\;.
\end{split}
\end{equation}
Here there is no logarithm, $\log\left({M^2}/{\bm k_\perp^2}\right)$, multiplying the extra power of $\as$, so we neglect this contribution entirely. An analogous analysis shows that we can neglect entirely the order $\as^3$ and higher order terms in the expansion eq.~(\ref{eq:asexpansion}).

With these approximations, $S_0 \approx S$ where
\begin{equation}
\begin{split}
\label{eq:showerSudakov}
S(M^2,\bm b^2) ={}& \int_{4e^{-2\gamma_E}/\bm b^2}^{M^2}
\!\frac{d\bm k_\perp^2}{\bm k_\perp^2}
\Bigg\{
\frac{\as\!\left(\bm k_\perp^2\right)}{2\pi}\,2C_\LF
\left[\log\!\left(\frac{M^2}{\bm k_\perp^2}\right) - \frac{3}{2}
\right]
\\&\qquad
- \left(\frac{\as\!\left(\bm k_\perp^2\right)}{2\pi}\right)^2
4\beta_1 C_\LF \log(\lambda_\LR)\,
\log\!\left(\frac{M^2}{\bm k_\perp^2}\right)
\Bigg\}
\;\;. 
\end{split}
\end{equation}

\section{Result}
\label{sec:result}

We have seen that the $b$-space hadronic cross section $\sbra{1} {\cal Q}(\bm b, Y)\sket{\rho(t)}$ evolves from $t=0$ to $t = t_\Lc \equiv \log\left(\bm b^2 M^2\,e^{2\gamma_{\rm E}}/4\right)$ and then stops evolving. Therefore the Fourier transform of the physical $p_\perp$-space cross section is given by $\sbra{1} {\cal Q}(\bm b, Y)\sket{\rho(t_\Lc)}$. Furthermore, $\sbra{1} {\cal Q}(\bm b, Y)\sket{\rho(t_\Lc)}$ is given as a convolution of parton distributions and the $b$-space partonic cross section $\sbra{1}{\cal Q}(\bm b, Y; \eta_\La, \eta_\Lb, a, b)\sket{\rho_{\rm pert}(t_\Lc)}$ by eq.~(\ref{eq:partonicvsfull}),
\begin{equation}
\begin{split}
\label{eq:partonicvsfullattc}
\sbra{1}{\cal Q}(\bm b, Y)
\sket{\rho(t_\Lc)} 
={}& 
\sum_{a b}\int_0^1\!d\eta_\La\int_0^1\!d\eta_\Lb\
\frac{
f_{a/A}(\eta_{\La},4 e^{-2\gamma_{\rm E}}/ \bm b^2)
f_{b/B}(\eta_{\Lb},4 e^{-2\gamma_{\rm E}}/ \bm b^2)}
{4n_\Lc(a) n_\Lc(b)\,2\eta_{\La}\eta_{\Lb}p_\LA\!\cdot\!p_\LB}
\\&\times
\sbra{1}{\cal Q}(\bm b, Y; \eta_\La,  \eta_\Lb, a, b)
\sket{\rho_{\rm pert}(t_\Lc)}
\;\;.
\end{split}
\end{equation}
We have found that the $b$-space partonic cross section at $t = t_\Lc$ is given by
\begin{equation}
\begin{split}
\label{eq:showerresult}
\sbra{1}{\cal Q}(\bm b, Y; \eta_\La, \eta_\Lb, 
a, b)\sket{\rho_{\rm pert}(t_\Lc)}
={}& 12\, \alpha\, Q_{ab}\,
\delta(1 - x_\La/\eta_\La)\,
\delta(1 - x_\Lb/\eta_\Lb)
\\ &\times 
\exp(- S(M^2,\bf b^2))
\;\;,
\end{split}
\end{equation}
where the Sudakov exponent $S$ given in eq.~(\ref{eq:showerSudakov}). This has been obtained with the approximations discussed in section~\ref{sec:approximations} and in section \ref{sec:solution}. In particular, terms in $S$ of the form $\as^n\, \log^{n-1}(M^2 \bm b^2)$ are affected by the approximations. 

We can compare this to the result of ref.~\cite{CSS} for QCD. In ref.~\cite{CSS}, there are two arbitrary parameters, $C_1$ and $C_2$, that do not affect the result summed to all orders of perturbation theory, but do affect the perturbative expansion. Here, we take the simplest choice, $C_1 = 2 e^{-\gamma_{\rm E}}$ and $C_2 = 1$. Additionally, the result is given for an arbitrary choice of the scale $\mu^2 = M^2 e^{-t}$ that appears in the parton distribution functions. Here $t$ should be close to $t_\Lc$. We choose $t = t_\Lc$. With these choices, the result of ref.~\cite{CSS} is that the hadronic $b$-space cross section is given by eq.~(\ref{eq:partonicvsfullattc}) with
\begin{equation}
\begin{split}
\label{eq:resultcss}
\sbra{1}{\cal Q}(& \bm b, Y; \eta_\La,  \eta_\Lb, a, b)\sket{\rho(t_\Lc)}_{\rm QCD}
\\
\approx{}&
12\alpha
\sum_{a' b'}Q_{a'b'}\
 C_{a'a}\!\left(\frac{x_\La}{\eta_\La}, \as\!\left(\frac{4e^{-2\gamma_E}}{\bm b^2}\right)\right)
 C_{b'b}\!\left(\frac{x_\Lb}{\eta_\Lb}, \as\!\left(\frac{4e^{-2\gamma_E}}{\bm b^2}\right)
 \right)
 \\ &\times
\exp\left(-\int_{4e^{-2\gamma_E}/\bm b^2}^{M^2}
\!\frac{d\bm k_\perp^2}{\bm k_\perp^2}
\left[
A(\as(\bm k_\perp^2))\log\left(\frac{M^2}{\bm k_\perp^2}\right)
+ B(\as(\bm k_\perp^2))
\right]
\right)
\;\;.
\end{split}
\end{equation}
Here $A$, $B$, and $C$ have perturbative expansions in powers of $\as$ of the indicated arguments. The first terms in these expansions are \cite{CSS, CSBacktoBackJets, Kodaira32, DaviesStirling, EllisMartinelliPetronzio}
\begin{equation}
\begin{split}
A(\as) ={}& 
2\,C_\LF\,\frac{\as}{2\pi} +
2\,C_\LF\,K
\left(\frac{\as}{2\pi}\right)^2 + \cdots \;\;,
\\
B(\as) ={}& 
-4\,\frac{\as}{2\pi} +
\left[
-\frac{197}{3}+\frac{34 n_\Lf}{9}+\frac{20 \pi^2}{3}
- \frac{8 n_\Lf \pi^2}{27} + \frac{8\zeta(3)}{3}
\right]
\left(\frac{\as}{2\pi}\right)^2 + \cdots \;\;,
\\
C_{a'a}\!\left(z, \as\right) ={}& \delta_{a'a}\delta(1-z)
\\&
+ \frac{\as}{2\pi}
\bigg[ \delta_{a'a}\left\{\frac{4}{3}\,(1-z)
+ \delta(1-z)
\left(-\frac{3}{2} + \frac{2\pi^2}{3} - \frac{23}{6}\right)\right\}
\\ &\qquad 
+ \delta_{a\Lg}\, z(1-z) \bigg]
\;\;.
\end{split}
\end{equation}
Here we follow the notation of \cite{CataniMCsummation} in defining
\begin{equation}
K = C_\LA \left[\frac{67}{18}-\frac{\pi^2}{6}\right] - \frac{5\,n_\Lf}{9}
\;\;.
\end{equation}

We begin the comparison of the shower result (\ref{eq:showerresult}) with the QCD result (\ref{eq:resultcss}) by noting that they agree at the Born level, order $\as^0$. This agreement was built in when we chose the Born cross section as the starting point for shower evolution.

We next note that the QCD cross section exponentiates in $b$-space. This is a statement about $\sbra{1}{\cal Q}(\bm b, Y; \eta_\La,  \eta_\Lb, a, b)\sket{\rho(t_\Lc)}_{\rm QCD}$. We consider the logarithm\footnote{In taking the logarithm, we consider the functions $C$ to be operators on the space of parton distribution functions. Thus the first term, $\delta_{a'a}\delta(1-z)$, represents the unit operator. The coefficients $D_{nm}(Y)$ in eq.~(\ref{eq:pertexpansion}) are then, in general, operators on the space of parton distribution functions. One could diagonalize these operators by taking moments, but there is little reason to do so.} of this function and expand it in powers of $\as(M^2)$ and $\log(M^2 \bm b^2)$. To obtain this expansion, we need to expand $\as(\bm k_\perp^2)$ in powers of $\as(M^2)$ and $\log(M^2/\bm k_\perp^2)$ using the evolution equation for $\as$, then perform the integration over $\bm k_\perp^2$ in the Sudakov exponent $S$. Similarly, in the coefficients $C$, we need to expand $\as\!\left(4e^{-2\gamma_E}/{\bm b^2}\right)$ in powers of $\as(M^2)$ and $\log(M^2/{\bm b^2})$. The general form of the perturbative expansion would be
\begin{equation}
\begin{split}
\label{eq:pertexpansion}
\log\big[\sbra{1}{\cal Q}(\bm b, Y; \eta_\La, \eta_\Lb, &  a, b) \sket{\rho(t_\Lc)}_{\rm QCD}\big]
\\
={}&
- \sum_{n=1}^\infty \sum_{m=0}^{2n} D_{nm}(Y)\,
\left(\frac{\as(M^2)}{2\pi}\right)^n \Big(\log(M^2 \bm b^2)\Big)^m
\;\;.
\end{split}
\end{equation}
We know that the maximum power $m$ of $\log(M^2 \bm b^2)$ must be no larger than $2n$ because each loop can give at most two logarithms, one from a collinear singularity and one from a soft singularity. If we look at the perturbative expansion of the partonic $b$-space cross section instead of its logarithm, then terms with $m = 2n$ actually occur. By the statement that the shower cross section exponentiates in $b$-space we mean that, for all $n$,
\begin{equation}
\label{eq:exponentiates}
D_{nm} = 0 \qquad {\rm for}\ m > n+1
\;\;.
\end{equation}
This property of the QCD result (\ref{eq:resultcss}) is shared by the shower result (\ref{eq:showerresult}). For the shower, it is a simple consequence of having a differential equation in which the derivative with respect to shower time of the partonic $b$-space cross section is a kernel times this same partonic $b$-space cross section, where the kernel has one $\as$ and one logarithm. 

One sometimes finds discussions of summations of large perturbative logarithms expressed in terms of the directly observed cross section, which can be written as a convolution of parton distribution functions and a partonic $p_\perp$-space cross section defined analogously to the partonic $b$-space cross section. Thus one may write
\begin{equation}
\begin{split}
\label{eq:pertexpansionkt}
\sbra{1}{\cal Q}(\bm p_\perp, Y; \eta_\La,  \eta_\Lb, &  a, b)
\sket{\rho_{\rm pert}(t_\Lc)}
\\
={}&
\sum_{n=1}^\infty \sum_{m=1}^{2n} E_{nm}(Y)\,
\left(\frac{\as(M^2)}{2\pi}\right)^n 
\frac{1}{\bm p_\perp^2}\Big(\log(M^2 \bm p_\perp^2)\Big)^{m-1}
\;\;.
\end{split}
\end{equation}
One can then discuss the coefficients $E_{nm}$. However, this is not very useful. For instance, no condition on the $E_{nm}$ for $m = 2n$ and $m = 2n - 1$ will imply eq.~(\ref{eq:exponentiates}). It is for this reason that we compare the shower result and the QCD result with respect to the logarithm of the partonic $b$-space cross section.

Let us now compare the $n=1$ terms in eq.~(\ref{eq:pertexpansion}) between the QCD result (\ref{eq:resultcss}) and the shower result (\ref{eq:showerresult}). We first note that the terms $D_{10}$ do {\em not} match between the two results: the order $\as$ contribution to $C_{a'a}\!\left(z, \as\right)$ in the QCD result is lacking in the shower result. These terms arise in part from the one loop virtual graphs, which are not included in the shower splitting functions. For that reason, the shower does not get these terms right. In our analysis we have have, accordingly, completely neglected contributions to the evolution kernel that have a factor $\as$ with no logarithms $\log(M^2\bm b^2)$.

We now compare the $\as^1$ terms with one or two powers of $\log(M^2\bm b^2)$. For the shower result (\ref{eq:showerresult}), we have
\begin{equation}
\begin{split}
D_{12} ={}& \frac{4}{3}
\;\;,
\\
D_{11} ={}& -\frac{16}{3}\,[\log 2 - \gamma_{\rm E}] - 4
\;\;.
\end{split}
\end{equation}
This is the same as $D_{12}$ and $D_{11}$ in the QCD result. It is not surprising that the leading coefficient, $D_{12}$, matches. The contribution $D_{12}\log^2(M^2\bm b^2)$ is $4 C_\LF$ times the area of the triangle in figure~\ref{fig:TrianglePlain}. The contribution $D_{11}\log(M^2\bm b^2)$ is more subtle. This contribution is associated with the edges of the triangle. The bottom edge of the triangle is associated with the lower limit on the $\bm k_\perp^2$ integral, which is $4e^{-2\gamma_E}/\bm b^2$ instead of simply $1/\bm b^2$. In the shower evolution, this value arises as a property in integrals involving Bessel functions $J_0(|\bm k_\perp| |\bm b|)$. In the analysis of the contribution to evolution associated with $\bm k_\perp^2$ near this limit, it was important that the rapidity of a potential gluon emission was large compared to the rapidities of previous gluon emissions. In the analysis of the contribution associated with the left hand edge of the triangle, it was important that the denominator of the splitting function is $1 - z + e^{-t}$ and not, say, $1 - z + 2 e^{-t}$. This denominator is related to the treatment of the soft gluon interference diagrams. Finally, the right hand edge of the triangle is associated with the term $-3/2$ in eq.~(\ref{eq:showerSudakov}) for the Sudakov exponent. This term comes from the $-\gamma_a$ in the virtual part of the DGLAP evolution equation for the parton distribution functions, eq.~(\ref{eq:DGLAP}).

Let us now look beyond the order $\as$ contribution to the Sudakov exponent. 

We note that the QCD result and the shower result share the contribution inside the $\bm k_\perp^2$ integral,
\begin{equation}
\label{eq:leadingterm}
\frac{\as\!\left(\bm k_\perp^2\right)}{2\pi}\,2C_\LF
\log\left(\frac{M^2}{\bm k_\perp^2}\right)
\;\;.
\end{equation}
In the shower result, this form reflects our choice to use $\mu_\LR^2$ defined in eq.~(\ref{eq:muR}) as the argument of $\as$; this is a choice that is not required by the logic of the shower, in which we factor softer interactions from harder interactions. With this choice, we match the contribution in eq.~(\ref{eq:leadingterm}) between the shower result and the QCD result. Expanding in powers of $\as(M^2)$, we see that 
\begin{equation}
D_{mn}^{\rm shower} = D_{mn}^{\rm QCD} \qquad n \ge 1,\ \ m = n+1
\;\;.
\end{equation}

We note also that the QCD result and the shower result share the contribution inside the $\bm k_\perp^2$ integral,
\begin{equation}
\label{eq:Bterm}
\frac{\as\!\left(\bm k_\perp^2\right)}{2\pi}\,2C_\LF \left(- \frac{3}{2}\right)
\;\;.
\end{equation}
This term contributes to the coefficients $D_{nm}$ for $n = m$. However, the QCD result has a contribution
\begin{equation}
\label{eq:A2term}
\left(\frac{\as\!\left(\bm k_\perp^2\right)}{2\pi}\right)^2
2 C_\LF K
\log\!\left(\frac{M^2}{\bm k_\perp^2}\right)
\;\;,
\end{equation}
which is to be compared to
\begin{equation}
- \left(\frac{\as\!\left(\bm k_\perp^2\right)}{2\pi}\right)^2
4\beta_1 C_\LF \log(\lambda_\LR)\,
\log\!\left(\frac{M^2}{\bm k_\perp^2}\right)
\;\;.
\end{equation}
It is not a surprise that these do not match for $\lambda_\LR = 1$, since we have used a leading order shower and this is a second order contribution. However, this term in the Sudakov exponent contributes to the coefficients $D_{nm}$ for $n = m$ and $n \ge 2$. If we would like to match these terms, we can, by means of a trick introduced in ref.~\cite{CataniMCsummation}. We choose $\lambda_\LR$ such that
\begin{equation}
- 4\beta_1 C_\LF \log(\lambda_\LR)
=
2 C_\LF K
\;\;.
\end{equation}
This is what we have done in our default choice given in eq.~(\ref{eq:lambdaR}). For $n_\Lf = 5$, this is $\lambda_\LR \approx 0.41$. If we take this default choice, then
\begin{equation}
D_{mn}^{\rm shower} = D_{mn}^{\rm QCD} \qquad n \ge 1,\ \ m = n
\;\;.
\end{equation}

This scale choice applied to $q \to q + \Lg$ splittings produces the second order term in $A(\as)$ for $Z$-boson production. The same formalism applies to the transverse momenta of Higgs bosons produced in hadron collisions. In that case, the incoming partons are two gluons instead of a quark and an antiquark. The Sudakov exponent has the same form, but with different functions $A$, $B$, and $C$. For $A$, the first two terms are now $2C_\LA (\as/(2\pi)) + 2 C_A K (\as/(2\pi))^2$, but the ratio of the first and second coefficients in $A(\as)$ is unchanged \cite{CataniEmilio32}. Thus the same factor $\lambda_\LR$ applied to $\Lg \to \Lg + \Lg$ splittings produces the second order term in $A(\as)$ for the Higgs boson transverse momentum distribution. Of course, this is a trick that applies for the particular purpose at hand. It is not equivalent to using a second order splitting kernel. 

For the Higgs boson $p_\perp$ distribution, we also replace $B_1 = -2 \gamma_q = - 4$ by $B_1 = -2 \gamma_\Lg$, where $\gamma_\Lg$ is given in eq.~(\ref{eq:gammaa}) \cite{CataniEmilio32}.

\section{Other choices}
\label{sec:other}

In this section, we briefly investigate the $Z$-boson transverse momentum distribution that would result if we were to make other choices for the construction of the shower.

\subsection{Spin-averaged, leading-color shower}
\label{sec:SpinAverageLeadingColor}

We have organized this paper as an analysis of the shower evolution equation in ref.~\cite{NSI}, with some minor modifications. This evolution equation contains the effects of spin correlations and applies to arbitrary color states. It contains the physics that we believe should be approximately represented in a parton shower event generator. However, the nature of the evolution is such that finding a way to represent the equations in a practical computer program is challenging. To start with a base approximation, one can average over the spins and take the standard leading color approximation. We analyzed this case in ref.~\cite{NSII}. With a spin average and a leading color approximation for each splitting, the evolution equations have the right form to be implemented as a Markov process, as is commonly used for parton showers.\footnote{We are, in fact, currently engaged in writing computer code for this purpose.} We are thus led to ask whether the results of this paper still hold if one averages over spins and takes the leading color approximation at each parton splitting step. The answer to this question is that the results do still hold. 

Spin does not matter because the observable of interest, the $Z$-boson transverse momentum, is independent of parton spins. The spin dependence is represented in the spin factor in eq.~(\ref{eq:realsmallzsplitting}), $\brax{\{s'\}_{m}}\ket{\{s\}_{m}}$: the parton spins $\{s\}_{m}$ in the quantum amplitude must equal those in the quantum conjugate amplitude, $\{s'\}_{m}$, but the splitting probability is independent of what the spin values are. If we average over spins, this factor goes away.

Color is of some importance. However if we use the exact color dependence then eq.~(\ref{eq:colorsum}) allowed us to reduce the color dependence to a simple factor of $C_\LF$ or $C_\LA$. If we use the leading color approximation, then each gluon is treated as carrying color $\bm 3 \otimes \bar {\bm 3}$ instead of color $\bm 8$. The allowed parton pairs forming a dipole are then the color connected partners. Then eq.~(\ref{eq:colorsum}) still holds with $C_\LF$ replaced by $C_\LA/2$. One can get back to $C_\LF$ by simply adjusting the quark-quark-gluon couplings in the splitting functions. With this understanding, the results of the previous section calculated with full color remain unchanged in the leading-color approximation.

\subsection{Alternative choice of dipole partitioning function}
\label{sec:CSmod}

Within the framework of a virtuality ordered shower as presented in ref.~\cite{NSI}, there is a choice to make. Consider the emission of a soft gluon from a color dipole composed of the initial state parton ``a'' and one other parton, $k$. The total emission probability is proportional to
\begin{equation}
\label{eq:helperk}
H_{\La k}(\{\hat p\}_{m+1}) + H_{k\La}(\{\hat p\}_{m+1})\equiv 
\frac{\as}{\pi}\
\frac{\hat p_\La\!\cdot\!\hat p_k}
{\hat p_{m+1}\!\cdot\!\hat p_\La\ \hat p_{m+1}\!\cdot\!\hat p_k}
\;\;.
\end{equation}
Here we use the eikonal approximation, which is valid for very small $\hat p_{m+1}$. The shower algorithm divides this into pieces, $H_{\La k}$ and $H_{k \La}$, defined by
\begin{equation}
\begin{split}
\label{eq:Alkprimedef}
H_{\La k}(\{\hat p\}_{m+1}) ={}& 
\frac{\as}{\pi}\
 A_{\La k}'(\{\hat p\}_{m+1})\
\frac{\hat p_\La\!\cdot\!\hat p_k}
{\hat p_{m+1}\!\cdot\!\hat p_\La\ \hat p_{m+1}\!\cdot\!\hat p_k}
\;\;,
\\
H_{k \La}(\{\hat p\}_{m+1}) ={}& 
\frac{\as}{\pi}\
 A_{k\La}'(\{\hat p\}_{m+1})\
\frac{\hat p_\La\!\cdot\!\hat p_k}
{\hat p_{m+1}\!\cdot\!\hat p_\La\ \hat p_{m+1}\!\cdot\!\hat p_k}
\;\;.
\end{split}
\end{equation}
Here $A_{\La k}'$ and $A_{k\La}'$ are both positive and obey
\begin{equation}
A_{\La k}' + A_{k\La}' = 1
\;\;.
\end{equation}
The contribution proportional to $A_{\La k}'$ is considered to be an emission from parton $\La$ and is treated using the momentum mapping associated with emission from parton $\La$. The contribution proportional to $A_{k \La}'$ is considered to be an emission from parton $k$ and is treated using the momentum mapping associated with emission from parton $k$.\footnote{In eq.~(12.21) of ref.~\cite{NSI}, the splitting probabilities are expressed in terms of functions $A_{\La k}$ and $A_{k \La}$, which are related to $A_{\La k}'$ and $A_{k \La}'$ by eq.~(7.2) of ref.~\cite{NSIII}.} Thus the function $A_{\La k}'$ tells how to partition the soft emission from a dipole into two parton splitting contributions.

In this paper, we use the $A_{\La k}'$ defined in eq.~(7.12) of ref.~\cite{NSIII},
\begin{equation}
\label{eq:Alkangle}
A_{\La k}' = 
\frac{\hat p_{m+1}\!\cdot\!\hat p_k\  
\hat p_\La\!\cdot\!(\hat p_\La + \hat p_\Lb)}
{\hat p_{m+1}\!\cdot\!\hat p_k\  \hat p_\La\!\cdot\!(\hat p_\La + \hat p_\Lb)
+
\hat p_{m+1}\!\cdot\!\hat p_\La\  \hat p_k\!\cdot\!(\hat p_\La + \hat p_\Lb)}
\;\;,
\end{equation}
with $A_{k \La}'$ given by the same expression with $\La \leftrightarrow k$. In the $\hat p_\La + \hat p_\Lb$ rest frame, this is
\begin{equation}
\label{eq:Alkangle2}
A_{\La k}' = 
\frac{1 - \cos \theta_{m+1,k}}
{(1 - \cos \theta_{m+1,k})
+
(1 - \cos \theta_{m+1,\La})}
\;\;.
\end{equation}
We have seen that with this choice the shower will produce the proper structure for the Drell-Yan transverse momentum distribution.

One can consider using instead the $A_{\La k}'$ defined in eq.~(7.13) of ref.~\cite{NSIII},
\begin{equation}
\label{eq:AlkCS}
A_{\La k}' = 
\frac{\hat p_{m+1}\!\cdot\!\hat p_k}
{\hat p_{m+1}\!\cdot\!\hat p_k
+
\hat p_{m+1}\!\cdot\!\hat p_\La}
\;\;.
\end{equation}
This is the form used by Catani and Seymour and by others for the purpose of partitioning the dipole splitting probability in defining perturbative dipole subtractions. The most important difference between the two choices is that $A_{\La k}'$ as given by eq.~(\ref{eq:Alkangle}) is invariant under the scaling $\hat p_k \to \lambda \hat p_k$. Thus this $A_{\La k}'$ depends on the angle between the emitted gluon and the momentum of parton $k$, but not on the energy of parton $k$.

We have investigated the possibility of using $A_{\La k}'$ defined by eq.~(\ref{eq:AlkCS}) to generate the parton shower. We find that if we do so, the argument given in sections~\ref{sec:1mzll1} and \ref{sec:angularordering}  fails. Now the probability of emitting a gluon $m+1$ from the initial state parton ``a'' with virtuality smaller than that of a previously emitted gluon $k$ can depend on the momentum $p_k$. As a result, we no longer obtain a differential equation for the $b$-space partonic cross section that has only a kernel times the $b$-space partonic cross section on the right hand side. The $Z$-boson will get less transverse momentum recoil from emitted gluons, but we cannot say how much less.

We can draw a lesson from this: a parton shower built on splitting functions that reflect the collinear and soft singularities of QCD does not automatically sum large logarithms that appear in a physical cross section of interest. Seemingly minor details like the choice of $A_{\La k}'$ matter.

\subsection{Catani-Seymour dipole shower}

The shower discussed in this paper is the virtuality ordered shower of ref.~\cite{NSI} with an improved momentum mapping for initial state radiation and with the renormalization scale for $\as$ given in eq.~(\ref{eq:muR}). A close relative is the Catani-Seymour (CS) dipole shower, proposed in ref.~\cite{NSRingberg} and implemented in refs.~\cite{Weinzierl, Schumann}. Here one takes the splitting functions and momentum mappings of the Catani-Seymour algorithm \cite{CataniSeymour} for defining subtractions in NLO perturbative calculations and uses them instead to define a shower algorithm. We examine this choice here, with the evolution variable taken to be
\begin{equation}
\label{eq:tperp}
t_\perp = \log\left( \frac{M^2}{\bm k_\perp^2}\right)
\;\;,
\end{equation}
as in refs.~\cite{NSRingberg, Weinzierl, Schumann}.

The most important difference between the shower of this paper and the CS dipole shower lies in its treatment of momentum conservation. Let us consider the (backward) evolution of an initial state parton by emission of a gluon into the final state. The initial state parton is on shell with zero transverse momentum. It is replaced by a prior on-shell parton with zero transverse momentum and the final state gluon that has non-zero transverse momentum. Thus there is a transverse momentum imbalance. In the shower of this paper, the recoil from the final state gluon is taken by all of the other final state partons collectively, by means of a Lorentz transformation $\hat k_i = \Lambda k_i$. With this choice, we have seen that the recoil transverse momentum is predominately taken by the $Z$-boson. In the CS shower, for each splitting there is a partner parton, the other parton in the CS dipole. Let us suppose that we use the leading color approximation, which is commonly applied in parton showers. Then the partner of a radiating initial state quark is the parton that is color connected to the quark. For the first gluon emission, the partner of incoming parton ``a'' is the other incoming parton ``b''. In this case, the CS scheme assigns the recoil transverse momentum to the $Z$-boson. However, once one gluon has been emitted, the partner of ``a'' is the previously emitted gluon. Then the CS scheme assigns the recoil transverse momentum to the partner parton. That is, the previously emitted gluon recoils against the newly emitted gluon and the $Z$-boson does not take any recoil.

This leads to a curious situation. With a transverse momentum ordered shower, the shower first evolves by not emitting any gluons. The probability not to emit any gluons up to a certain shower time $t_\perp$ is a Sudakov form factor. At some point, one real gluon is emitted. Then evolution of the $Z$-boson transverse momentum distribution stops because the recoil from further gluons emitted is taken by one of the gluons previously emitted. This problem has been analyzed and addressed in ref.~\cite{PlatzerGieseke}.

We can formulate the issue in the style of our analysis of the previous sections if we supply one additional variable in the measurement operator ${\cal Q}$, a variable ${\tt recoil}$ that takes values $\LT$ and $\LF$. Including this variable, our operator expressing possible measurements is ${\cal Q} (\bm b, Y;\eta_\La, \eta_\Lb, a, b;{\tt recoil})$. If ${\tt recoil} = \LT$, then when a gluon is emitted from either of the initial state partons, the $Z$-boson will share in the recoil against the gluon transverse momentum. If ${\tt recoil} = \LF$, then, when a gluon is emitted from an initial state parton, the $Z$-boson does not share in the recoil. 

The evolution equation of the ${\tt recoil} = \LT$ cross section is
\begin{equation}
\begin{split}
\label{eq:evolutionCST}
\frac{d}{dt_\perp}
\sbra{1}
{\cal Q}(\bm b, Y;\eta_\La,& \eta_\Lb, a, b;\LT)
\sket{\rho_{\rm pert}(t_\perp)}
\\
={}&
-\sbra{1}
{\cal V}^{\rm pert}(t)
{\cal Q}(\bm b, Y; \eta_\La,  \eta_\Lb, a, b; \LT)
\sket{\rho_{\rm pert}(t_\perp)}
\;\;.
\end{split}
\end{equation}
This does not involve the ${\tt recoil} = \LF$ cross section. Using the approximations developed in the preceding sections, we can solve this equation. We find
\begin{equation}
\begin{split}
\label{eq:resultCST}
\sbra{1}
{\cal Q}(\bm b, Y;\eta_\La,& \eta_\Lb, a, b;\LT)
\sket{\rho_{\rm pert}(t_\perp)}
\\
={}&
12\, \alpha\, Q_{ab}\, x_\La x_\Lb\,
\delta(\eta_\La - x_\La)\,
\delta(\eta_\Lb - x_\Lb)
\\ &\times 
\exp\left(- 
\int_{\bm k_\perp^2}^{M^2}
\!\frac{d\mu^2}{\mu^2}
\frac{\as\!\left(\lambda_\LR \mu^2\right)}{2\pi}\,2C_\LF
\left[\log\left(\frac{M^2}{\mu^2}\right) - \frac{3}{2}
\right]
\right)
\;\;.
\end{split}
\end{equation}
Note that for very large $t_\perp$, the ${\tt recoil} = \LT$ cross section tends to zero.

The evolution equation of the ${\tt recoil} = \LF$ cross section has only the ${\tt recoil} = \LT$ cross section on the right-hand side:
\begin{equation}
\begin{split}
\label{eq:evolutionCSF}
\frac{d}{dt_\perp}
\sbra{1}
{\cal Q}(\bm b, Y;&\eta_\La, \eta_\Lb, a, b; \LF)
\sket{\rho_{\rm pert}(t_\perp)}
\\ 
\approx{}&
\int_0^1\! dz \int_{-\pi}^\pi \frac{d\phi}{2\pi}\sum_{f'}
\exp\!\left(\mi\, \frac{1}{z}\,\bm b\!\cdot\! \bm k_\perp\right)
\\&\times
\sbra{1} 
z{\cal H}_{\La}^{\rm pert}(t;z,\phi,f')
{\cal Q}(\bm b, Y; z \eta_\La,  \eta_\Lb,  a - f',  b; \LT)
\sket{\rho_{\rm pert}(t_\perp)}
\\ &
+
({\rm b\ splitting\ term})
\;\;.
\end{split}
\end{equation}

The total partonic $b$-space cross section is
\begin{equation}
{\cal Q}(\bm b, Y;\eta_\La, \eta_\Lb, a, b)
=
{\cal Q}(\bm b, Y;\eta_\La, \eta_\Lb, a, b;\LT) + 
{\cal Q}(\bm b, Y;\eta_\La, \eta_\Lb, a, b;\LF)
\;\;.
\end{equation}
For very large $t_\perp$, the ${\tt recoil} = \LT$ contribution vanishes and we are left with the ${\tt recoil} = \LF$ contribution, which we can obtain by solving eq.~(\ref{eq:evolutionCSF}) with the use of eq.~(\ref{eq:resultCST}). The result for large $t_\perp$ is
\begin{equation}
\begin{split}
\label{eq:evolutionCSFsoln}
\sbra{1}
{\cal Q}(\bm b, Y;&\eta_\La, \eta_\Lb, a, b; \LF)
\sket{\rho_{\rm pert}(\infty)}
\\ 
\approx{}&
\int_0^{M^2}\!\frac{d \bm k_\perp^2}{\bm k_\perp^2}
\int_0^1\!dz
\sum_{f'}
J_0\!\left(\frac{|\bm k_\perp| |\bm b|}{z}\right)
\\&\times
12\, \alpha\, Q_{a-f',b}\, x_\La x_\Lb\,
\delta(z\eta_\La - x_\La)\,
\delta(\eta_\Lb - x_\Lb)
\\ &\times 
\exp\left(- 
\int_{\bm k_\perp^2}^{M^2}
\!\frac{d\mu^2}{\mu^2}
\frac{\as\!\left(\lambda_\LR \mu^2\right)}{2\pi}\,2C_\LF
\left[\log\left(\frac{M^2}{\mu^2}\right) - \frac{3}{2}
\right]
\right)
\\ &
\frac{n_\Lc(a)}{n_\Lc(a-f')}\,
\frac{\as\!\left(\mu_\LR^2\right)}{2\pi}\,
\frac{1}{z}\,
\left[P_{a-f',a}(z)
- \frac{2 z\, C_\LF\, \delta_{f',\Lg}}{1-z}
+ \frac{2 z\, C_\LF\, \delta_{f',\Lg}}{1-z+e^{-t}}
\right]
\\ &
+
({\rm b\ splitting\ term})
\;\;.
\end{split}
\end{equation}
We see that the partonic cross section does not exponentiate in $b$-space. In fact, it is simplest in $p_\perp$-space instead: if we take a Fourier transform back to $\bm p_\perp$, the factor $J_0(|\bm k_\perp| |\bm b| /z)$ provides a delta function $2(2\pi)^2 \delta(\bm p_\perp^2 - \bm k_\perp^2/z^2)$. This leaves a relatively simple expression consisting of a Sudakov factor and a splitting function for the single allowed splitting.

Again, we see that a parton shower built on splitting functions that reflect the collinear and soft singularities of QCD does not automatically sum large logarithms that appear in a physical cross section of interest. Seemingly minor details like the momentum mapping matter.

\subsection{Transverse momentum ordered shower}

The shower examined in this paper is based on virtuality ordering. What would happen if we kept everything else the same but used transverse momentum ordering? That is, what if we replace the evolution time $t$ of eq.~(\ref{eq:tdef}) by $t_\perp$ defined in eq.~(\ref{eq:tperp})? We keep the momentum mapping the same as in the virtuality ordered shower, so that the transfer of recoil transverse momentum to the $Z$-boson is the same as in the main body of this paper. Transverse momentum ordering is used in \textsc{Pythia} \cite{SjostrandSkands, Pythia8}, but the algorithm that we consider in this section differs in some respects from that of \textsc{Pythia}.

With $k_\perp$ ordering, the first stage of shower evolution is simple. For $t_\perp \ll \log(M^2 \bm b^2)$, only virtual splittings contribute to the evolution of the $b$-space partonic cross section; the rapidly oscillating $J_0(|\bm k_\perp| |\bm b|)$ factor that multiplies ${\cal H}_\LI^{\rm pert}$ eliminates the real emission contribution.

The later stages of shower evolution are also simple. The argument of section \ref{sec:evolutionII} implies that the rate of change of the $b$-space hadronic cross section vanishes for $t_\perp \gg \log(M^2 \bm b^2)$. 

We are left with the region $t_\perp \sim \log(M^2 \bm b^2)$. This is the bottom of the triangle in figure~\ref{fig:TrianglePlain}. With a virtuality ordered shower, this region is sampled from left to right, which is in order of increasing rapidity of emitted gluons -- that is, decreasing emission angles. Then the splitting function in eq.~(\ref{eq:realsmallzsplitting}) took the simple form in which the complicated function $f$ was replaced by 1 because the rapidity of the real or virtual gluon was much larger than the rapidity of previously emitted gluons. With $t_\perp$ as the evolution variable, there is no guarantee that a previously emitted gluon had smaller rapidity than a new real or virtual daughter gluon. Thus the form of the evolution equation changes and the evolution of the partonic $b$-space cross section depends on the previous emission history.

Because the evolution is more complicated than we have dealt with, we do not know what the result is. We do note that the difficulty arises only for a limited region, $t_\perp \sim \log(M^2 \bm b^2)$.

\subsection{Angle ordered shower}
\label{sec:angleordered}

What would happen if we used angular ordering instead of virtuality ordering? That is, what if we replace the evolution time $t$ of eq.~(\ref{eq:tdef}) by $t_\angle = - \log(\tan(\theta/2))$ where $\theta$ is the splitting angle. Then ordering in $t_\angle$ is ordering in splitting angle, starting from large angles and progressing to smaller angles as $t_\angle$ increases. At the same time, we would use a revised version of the dipole partitioning function $A_{lk}'$, as specified below. This gives us two of the main features of \textsc{Herwig} \cite{HerwigDescribed, Herwig}, although if we start with the shower described in this paper and change only these features, we certainly do not have exactly \textsc{Herwig}.

To be more precise, let $P = x_\La p_\La + x_\Lb p_\Lb$ be the $Z$-boson momentum approximated as having no transverse part but as obeying $P^2 = M_Z^2$. Consider the splitting of a parton with label $l$ and momentum $p_l$ and let $n_l$ be the lightlike vector
\begin{equation}
n_l = p_l - \frac{M_Z^2}{2\, p_l\!\cdot\! P}\,P
\;\;.
\end{equation}
Let the daughter partons have momenta $\hat p_l$ and $\hat p_{m+1}$. Then define
\begin{equation}
\label{eq:tangledef}
t_\angle = \frac{1}{2}\,\log\left(
\frac{\hat p_{m+1}\!\cdot\! n_l\ P\!\cdot\! p_l}
{\hat p_{m+1}\!\cdot\! p_l\ P\!\cdot\! n_l}
\right)
\;\;.
\end{equation}
For an emission from initial state line ``a'', $t_\angle$ is the rapidity of the splitting that we used in section~\ref{sec:angularordering}. From eq.~(\ref{eq:rapidityA}) we have, for $y \ll 1$ and $\eta_\La \approx x_\La$,
\begin{equation}
\label{eq:tangleA}
t_\angle = t + \frac{1}{2}\log\left(\frac{\bm k_\perp^2}{M^2}\right) 
+ \log\left(\frac{1}{z}\right)
\;\;.
\end{equation}
Curves of constant $t_\angle$ are shown in figure~\ref{fig:TriangleAngle}. Over most of the region of the figure, $(1-z)\ll 1$ so $\log(1/z) \approx 0$. This gives straight lines in the $t$-$\log(\bm k_\perp^2/M^2)$ plane. As these lines approach the line $(1-z)\sim 1$ in the figure, the $\log(1/z)$ term becomes important and the $t_\angle$ curves bend.

The curves of constant $t_\angle$ extend down to arbitrarily small virtuality, $M^2 e^{-t}$. In a virtuality ordered shower, one would simply stop the evolution when $M^2 e^{-t}$ reaches some infrared cutoff at which application of a hadronization model is more appropriate than continued use of perturbative showering. For the angle ordered shower, we should impose this same cutoff as a upper limit on integrations over $t$ at fixed $t_\angle$. Equivalently, if we parameterize the curves of fixed $t_\angle$ using $\bm k_\perp^2$, then we impose a lower limit on the integrations over $\bm k_\perp^2$.

One should do more than simply switching the evolution variable. One can also modify the dipole partitioning function $A'_{\La k}$ as used in eq.~(\ref{eq:Alkprimedef}). For an angle ordered shower, it is standard \cite{angleorder} to begin with 
\begin{equation}
\begin{split}
&A^{(0)\prime}_{\La k} = 
\frac{1}{2}\left[
1+
\frac{\cos\theta_{m+1,\La} - \cos\theta_{m+1,k}}{1 - \cos\theta_{\La,k}}
\right]
\;\;,
\\
&A^{(0)\prime}_{k\La} = 
\frac{1}{2}\left[
1 -
\frac{\cos\theta_{m+1,\La} - \cos\theta_{m+1,k}}{1 - \cos\theta_{\La,k}}
\right]
\;\;,
\end{split}
\end{equation}
where the angles are defined, for instance, in the $Z$-boson rest frame. These functions add to one, but are not everywhere positive. One can then replace the functions $H_{\La,k}$ and $H_{k,\La}$ defined in eq.~(\ref{eq:Alkprimedef}) by their averages over the azimuthal angles defined by rotating $\hat p_{m+1}$ about, respectively, the directions of $\hat p_\La$  and $\hat p_k$. As shown in ref.~\cite{angleorder}, this gives a simple result,
\begin{equation}
\begin{split}
\label{eq:angleaveraged}
H_{\La k}(\{\hat p\}_{m+1}) ={}& 
\frac{\as}{\pi}\
A^{(0)\prime}_{\La k}(\{\hat p\}_{m+1})\
\frac{1}{E_{m+1}^2}
\frac{1 - \cos\theta_{\La,k}}
{(1 - \cos\theta_{m+1,\La})\,(1 - \cos\theta_{m+1,k})}
\\
&\to 
\frac{\as}{\pi}\
\frac{1}{E_{m+1}^2}
\frac{\theta(\theta_{\La, m+1} < \theta_{\La, k})}
{1 - \cos\theta_{m+1,\La}}
\;\;,
\\
H_{k \La}(\{\hat p\}_{m+1}) ={}& 
\frac{\as}{\pi}\
A^{(0)\prime}_{k \La}(\{\hat p\}_{m+1})\
\frac{1}{E_{m+1}^2}
\frac{1 - \cos\theta_{\La,k}}
{(1 - \cos\theta_{m+1,\La})\,(1 - \cos\theta_{m+1,k})}
\\
&\to 
\frac{\as}{\pi}\
\frac{1}{E_{m+1}^2}
\frac{\theta(\theta_{k, m+1} < \theta_{\La, k})}
{1 - \cos\theta_{m+1,k}}
\;\;.
\end{split}
\end{equation}
The procedure is equivalent to replacing $A^{(0)\prime}_{\La k}$ and $A^{(0)\prime}_{k \La}$ by
\begin{equation}
\begin{split}
\label{eq:AlkWebber}
A_{\La k}' ={}&
\theta(\theta_{\La, m+1} < \theta_{\La, k})\,
\frac{1 - \cos\theta_{m+1,k}}{1 - \cos\theta_{\La,k}}
\;\;,
\\
A_{k \La}' ={}&
\theta(\theta_{k, m+1} < \theta_{\La, k})\,
\frac{1 - \cos\theta_{m+1,\La}}{1 - \cos\theta_{\La,k}}
\;\;.
\end{split}
\end{equation}
These functions are positive. They do not sum to 1, but, as just noted, one gets the right result after suitable averaging over azimuthal angles.

It is now easy to see how the partonic $b$-space cross section ${\cal Q}(\bm b, Y;\eta_\La, \eta_\Lb, a, b)$ evolves as $t_\angle$ increases. At each step $d t_\angle$, there are contributions from all values of $\bm k_\perp^2$ down to the infrared cutoff. We divide the integration over $\bm k_\perp^2$ into three distinct regions: $\bm k_\perp^2 \gg 1/\bm b^2$, $\bm k_\perp^2 \sim 1/\bm b^2$, and $\bm k_\perp^2 \ll 1/\bm b^2$. In the region $\bm k_\perp^2 \ll 1/\bm b^2$, the real emission term with a factor $J_0(|\bm k_\perp| |\bm b|)$ cancels the virtual emission term with a factor 1 because $J_0(|\bm k_\perp| |\bm b|) - 1 \approx 0$, as in our other analyses. Thus this region does not contribute to the evolution of ${\cal Q}(\bm b, Y;\eta_\La, \eta_\Lb, a, b)$. In the region $\bm k_\perp^2 \gg 1/\bm b^2$, only virtual emissions contribute to the evolution of ${\cal Q}(\bm b, Y;\eta_\La, \eta_\Lb, a, b)$. The rapidly oscillating $J_0(|\bm k_\perp| |\bm b|)$ factor that multiplies ${\cal H}_\LI^{\rm perp}$ eliminates the real emission contribution. In the region $\bm k_\perp^2 \sim 1/\bm b^2$, both real and virtual emissions contribute. 

We have by no means carried out a detailed investigation of this angle ordered scheme. However, it appears to us that it will give equivalent results for the $Z$-boson transverse momentum distribution as does the virtuality ordered shower.

\subsection{Dipole antenna shower}

In this paper so far we have discussed dipole parton showers in which the creation of a new gluon is attributed to the splitting of one of the previously existing partons. There is an ambiguity because, for soft gluon emissions, interference diagrams are important: one has to include emission from a parton $l$ squared, emission from a parton $k$ squared, and the $l$-$k$ interference graphs. Using the eikonal approximation, the sum of these is simple, as in eq.~(\ref{eq:helperk}). 
We have partitioned the total emission probability into a fraction $A'_{lk}$ associated with splitting of parton $l$ and a fraction $A'_{kl} = 1-A'_{lk}$ associated with the splitting of parton $k$, as in eq.~(\ref{eq:Alkprimedef}). In the nomenclature of section 16 of ref.~\cite{dipoleshowernames}, this constitutes a {\em partitioned dipole shower}.\footnote{Note that from the point of view adopted in section \ref{sec:angleordered}, an angle ordered shower also can be considered as a partitioned dipole shower.} There is a separate momentum mapping ${\cal P}_{l}$ for each parton $l$ that splits. Here ${\cal P}_{l}$ is an operator, defined in ref.~\cite{NSI}, on the space of parton states. It is not necessary that this momentum mapping depends on the choice of $k$, and the momentum mapping used in this paper and ref.~\cite{NSI} does not depend on $k$. Thus, in a highly compressed notation, we can write that the emission from the $l$-$k$ dipole is partitioned into two terms,
\begin{equation}
\label{eq:partitioned}
{\cal H}^{\rm part}_{lk}(t) \propto 
{\cal P}_{l}\,
A'_{lk}\
\frac{\hat p_l\!\cdot\!\hat p_k}
{\hat p_{m+1}\!\cdot\!\hat p_l\ \hat p_{m+1}\!\cdot\!\hat p_k}
+
{\cal P}_{k}\,
A'_{kl}\
\frac{\hat p_l\!\cdot\!\hat p_k}
{\hat p_{m+1}\!\cdot\!\hat p_l\ \hat p_{m+1}\!\cdot\!\hat p_k}
\;\;.
\end{equation}

In section \ref{sec:CSmod}, we have seen that the summation of large logarithms can be sensitive to the choice of the partitioning function $A'_{lk}$; a choice based on the angles between parton $l$ and $k$ and the emitted soft gluon is preferred.

One might be tempted to get rid of this ambiguity, particularly in the leading color approximation where one can use pairs of color connected partons as the dipoles. The idea is to consider each $l$-$k$ dipole as a unit that can emit a gluon. Thus the basic building blocks are $2 \to 3$ parton splittings. A shower based on this approach may be called a {\em dipole antenna shower}. The pioneering development along these lines is the final state shower of \textsc{Ariadne} \cite{ariadne}. More recent examples include those in ref.~\cite{Vincia}. There is a corresponding subtraction scheme for next-to-leading order calculations, antenna subtraction \cite{antennasubtract}.

In a dipole antenna shower, there is no $A'_{lk}$. There is a separate momentum mapping ${\cal P}_{lk}$ for each dipole $l$-$k$. Thus eq.~(\ref{eq:partitioned}) is replaced by
\begin{equation}
\label{eq:antenna}
{\cal H}^{\rm ant}_{lk}(t) \propto 
{\cal P}_{lk}\,
\frac{\hat p_l\!\cdot\!\hat p_k}
{\hat p_{m+1}\!\cdot\!\hat p_l\ \hat p_{m+1}\!\cdot\!\hat p_k}
\;\;.
\end{equation}
The freedom to choose $A'_{lk}$ now resides in the freedom to choose ${\cal P}_{lk}$. It must be symmetric under the interchange of partons $l$ and $k$. It is usually defined in such a way that momentum is conserved locally in the $2\to 3$ splitting, without taking momentum from any of the other partons. For a final state dipole, this means
\begin{equation}
\label{eq:localmapping}
p_{l} + p_{k} = \hat p_{l} + \hat p_{k} + \hat p_{m+1}
\;\;.
\end{equation}
Such a mapping is defined in ref.~\cite{antenna}. We note that it is possible to have a shower that is simultaneously a partitioned dipole shower and an antenna dipole shower. We get that case if we define
\begin{equation}
{\cal P}_{lk} = \theta(\vartheta_{l,m+1} < \vartheta_{k,m+1}) \, {\cal P}_{l} + \theta(\vartheta_{k,m+1} < \vartheta_{l,m+1})\, {\cal P}_{k}
\;\;,
\end{equation}
where the $\vartheta_{ij}$ is the angle between momenta $i$ and $j$ and ${\cal P}_{l}$ represents the momentum mapping that is used in this paper. Then $A'_{lk} = \theta(\vartheta_{l,m+1} < \vartheta_{k,m+1})$.

As far as we can see, a dipole antenna shower can reproduce the proper summation of large logarithms for the $Z$-boson transverse momentum distribution studied in this paper. However, the momentum mapping for initial state splittings cannot be local in the sense of eq.~(\ref{eq:localmapping}). For a dipole consisting of the two initial state partons, eq.~(\ref{eq:localmapping}) becomes
\begin{equation}
\label{eq:localmappingII}
p_{\La} + p_{\Lb} = \hat p_{\La} + \hat p_{\Lb} - \hat p_{m+1}
\;\;.
\end{equation}
We want $p_{\La,\perp} = \hat p_{\La,\perp} = 0$ and also $p_{\Lb,\perp} = \hat p_{\Lb,\perp} = 0$. This leaves no place for the transverse momentum of parton $m+1$ to go. One might hope to use a non-local momentum mapping for this case, but use a local mapping in the case of a dipole consisting of one initial state parton, say ``a'', and a final state colored parton $k$. In this case, eq.~(\ref{eq:localmapping}) becomes
\begin{equation}
\label{eq:localmappingIII}
p_{\La} - p_{k} = \hat p_{\La} - \hat p_{k} - \hat p_{m+1}
\;\;.
\end{equation}
However, that would mean that $\hat p_{k,\perp} - p_{k,\perp} = -\hat p_{m+1,\perp}$: the recoil transverse momentum from the emission of parton $m+1$ is taken up by parton $k$ instead of by the $Z$-boson. Thus a non-local momentum mapping is needed for all dipoles that include an initial state parton.

\section{Conclusions}
\label{sec:conclusions}

In this paper, we have examined the differential cross section $d\sigma/(d\bm p_\perp dY)$ for producing a $Z$-boson in hadron-hadron collisions as a function of the transverse momentum $\bm p_\perp$ and rapidity $Y$ of the $Z$-boson. For $\bm p_\perp^2 \ll M_Z^2$, the perturbative expansion of $d\sigma/(d\bm p_\perp dY)$ contains large logarithms, $\log(\bm p_\perp^2/ M_Z^2)$. There are known QCD results for the summation of these logarithms, quoted in section~\ref{sec:result}. We have asked to what extent a virtuality ordered parton shower algorithm of the sort defined in ref.~\cite{NSI} (with some small modifications as discussed in this paper) correctly reproduces the known results. 

The parton shower evolves in shower time $t$, defined to be minus the logarithm of the virtuality of a parton splitting, so that hard splittings come first, soft splittings last. The state of the shower at time $t$, in the sense of the distribution of parton configurations, is represented by a vector $\sket{\rho(t)}$. In this notation, $d\sigma/(d\bm p_\perp dY)$ as it has developed by shower time $t$ is denoted by $\sbrax{\bm p_\perp, Y}\sket{\rho(t)}$. Then, at the end of the shower at time $t_\Lf$, the predicted cross section is 
\begin{equation}
\frac{d\sigma}{d\bm p_\perp\,dY}
= 
\sbrax{\bm p_\perp, Y}\sket{\rho(t_\Lf)}
\;\;.
\end{equation}
We have examined whether the cross section obtained from the shower algorithm has the structure of the QCD result by examining how $\sbrax{\bm p_\perp, Y}\sket{\rho(t)}$ evolves with $t$.

To find $d\sbrax{\bm p_\perp, Y}\sket{\rho(t)}/dt$, we used eq.~(\ref{eq:evolution0}), which expresses how the completely exclusive partonic state evolves. This expresses $d\sbrax{\bm p_\perp, Y}\sket{\rho(t)}/dt$ in terms of the complete state of all of the partons at time $t$. Fortunately, we found that, with appropriate approximations, $d\sbrax{\bm p_\perp, Y}\sket{\rho(t)}/dt$ is expressed as a convolution of a certain kernel with $\sbrax{\bm p_\perp, Y}\sket{\rho(t)}$. That is, the detailed configuration of all of the partons does not matter, only the inclusive distribution $\sbrax{\bm p_\perp, Y}\sket{\rho(t)}$ matters. Thus we obtain a closed form differential-integral equation for $\sbrax{\bm p_\perp, Y}\sket{\rho(t)}$.

The result can be simply stated in terms of the Fourier transform of transverse momentum distribution, defined in eq.~(\ref{eq:FourierTransform}),
\begin{equation}
\sbra{1}{\cal Q}(\bm b, Y)\sket{\rho(t)}
=
\int \frac{d\bm p_\perp}{(2\pi)^2}\ e^{-\mi \bm b \cdot \bm p_\perp}
\sbrax{\bm p_\perp, Y}\sket{\rho(t)}
\;\;.
\end{equation}
We found that $\sbra{1}{\cal Q}(\bm b, Y)\sket{\rho(t)}$ stops evolving at a certain evolution time $t_\Lc$, with $t_\Lc = \log\left(\bm b^2 M^2\,e^{2\gamma_{\rm E}}/4\right)$. We found in eq.~(\ref{eq:partonicvsfullattc}) that $\sbra{1}{\cal Q}(\bm b, Y)\sket{\rho(t_\Lc)}$ is a convolution of parton distribution functions evaluated at the appropriate scale times a function that does not involve parton distribution functions,
\begin{equation}
\begin{split}
\label{eq:partonicvsfullattcbis}
\sbra{1}{\cal Q}(\bm b, Y)
\sket{\rho(t_\Lc)} 
={}& 
\sum_{a b}\int_0^1\!d\eta_\La\int_0^1\!d\eta_\Lb\
\frac{
f_{a/A}(\eta_{\La},4 e^{-2\gamma_{\rm E}}/ \bm b^2)
f_{b/B}(\eta_{\Lb},4 e^{-2\gamma_{\rm E}}/ \bm b^2)}
{4n_\Lc(a) n_\Lc(b)\,2\eta_{\La}\eta_{\Lb}p_\LA\!\cdot\!p_\LB}
\\&\times
\sbra{1}{\cal Q}(\bm b, Y; \eta_\La,  \eta_\Lb, a, b)
\sket{\rho_{\rm pert}(t_\Lc)}
\;\;.
\end{split}
\end{equation}
The important structure is thus expressed in the function $\sbra{1}{\cal Q}(\bm b, Y; \eta_\La,  \eta_\Lb, a, b) \sket{\rho_{\rm pert}(t_\Lc)}$.

Consider the logarithm of $\sbra{1}{\cal Q}(\bm b, Y; \eta_\La,  \eta_\Lb, a, b)
\sket{\rho_{\rm pert}(t_\Lc)}$. On general grounds, one would expect that this function has two powers of the large logarithm $\log(M^2 \bm b^2)$ per power of $\as(M^2)$:
\begin{equation}
\begin{split}
\label{eq:pertexpansionencore}
\log[\sbra{1}{\cal Q}(\bm b, Y; \eta_\La,&  \eta_\Lb, a, b)
\sket{\rho_{\rm pert}(t_\Lc)}]
\\
={}&
- \sum_{n=1}^\infty \sum_{m=0}^{2n} D_{nm}(Y)\,
\left(\frac{\as(M^2)}{2\pi}\right)^n \Big(\log(M^2 \bm b^2)\Big)^m
\;\;.
\end{split}
\end{equation}
However, we found that the partonic cross section produced by the shower algorithm exponentiates in $b$-space in the sense that
\begin{equation}
\label{eq:exponentiatesagain}
D_{nm} = 0 \qquad {\rm for}\ m > n+1
\;\;.
\end{equation}
This property is shared with the full QCD result.

This exponentiation is not guaranteed by simply having a parton splitting probability that matches QCD in the soft and collinear limits, including the proper interference between emissions of a soft gluon from different partons. We found that two other features of the shower algorithm are important. 

First, partons in a parton shower are treated as being on-shell, but it is not possible for an on-shell parton to split into two on-shell partons and still conserve momentum. There has to be a momentum mapping that takes a small amount of momentum from somewhere and supplies it to the daughter partons. In the case of an initial state splitting, the problem is more pronounced because the initial state partons are treated as having zero transverse momentum. This means that when a soft gluon is emitted from an initial state parton, some transverse momentum has to come from somewhere. We found that the momentum mapping has to be such that the $Z$-boson gets most of the recoil transverse momentum. In fact, the momentum mapping proposed in ref.~\cite{NSI} did not have this property and thus needed to be modified. Similarly, we noted that the momentum mapping used for perturbative subtractions in the Catani-Seymour dipole subtraction scheme does not give the desired exponentiation for $\sbra{1}{\cal Q}(\bm b, Y)\sket{\rho(t_\Lc)}$ in a dipole based shower.

Second, in a shower like that of ref.~\cite{NSI} that is based on color dipoles, there is a function that we call $A_{lk}'$ that specifies how much of the radiation from dipole $l$-$k$ is treated as being emitted from parton $l$ and how much is treated as being emitted from parton $k$. We found that, for initial state emissions, the choice of $A_{lk}'$ matters. A choice based on the angles between the momenta of partons $l$ and $k$ and an emitted soft gluon is satisfactory, while the choice used in the Catani-Seymour dipole subtraction scheme fails to give the proper exponentiation in a dipole based shower.

After having checked for the proper exponentiation of the large logarithms, the next question was which non-zero coefficients $D_{nm}$ are correctly produced by the shower algorithm. 

We found that the leading coefficient, $D_{12}$, is correctly reproduced. This result has a simple physical interpretation in the parton shower picture. The quantity $\sbra{1}{\cal Q}(\bm b, Y)\sket{\rho(t_\Lf)}$ is the cross section to produce the $Z$-boson with rapidity $Y$ but not radiate initial state gluons with transverse momentum bigger than approximately $1/\bm b^2$. The Sudakov exponent $S$ is the integral of the differential probability to radiate such a gluon, so $\exp(-S)$ is the probability not to radiate any such gluons. The leading term in the Sudakov exponent is $C_\LF\as/\pi$ times twice the area of the triangle in figure~\ref{fig:TrianglePlain}. This area is total phase space for gluon emission with $\bm k_\perp^2 > 1/\bm b^2$ from one of the two initial state partons; doubling the area gives the total phase space for emission from either line. Quantum coherence plays a role here. The initial emission is from a color dipole consisting of the two incoming quarks. Emissions with large positive rapidity count as emission from incoming quark ``a'' while emissions with large negative rapidity count as emission from incoming quark ``b''. 

We found also that the next-to-leading log coefficient, $D_{11}$, is correctly reproduced. This coefficient corresponds to the behavior of the parton splitting near the three edges of the triangle in figure~\ref{fig:TrianglePlain}. It thus represents physics that is more subtle than the physics behind the leading coefficient, $D_{12}$. 

We note that once the choices of momentum mapping and $A_{lk}'$ have been made, no adjustment of parameters is needed to get $D_{12}$ and $D_{11}$ to match the QCD result.

One could argue that matching the leading order coefficients $D_{12}$ and $D_{11}$ is all that one should expect from a leading order parton shower. Nevertheless, we found that it is possible to do better (as in ref.~\cite{CataniMCsummation}). With a suitable choice of the argument of $\as$ in the splitting function, it is possible to match all of the coefficients $D_{nm}$ with $m = n+1$ and with $m = n$. With this level of matching, the Sudakov exponent $S$ produced by the shower algorithm is a good approximation to the full QCD exponent $S_{\rm QCD}$ in the limit $\log(M^2 \bm b^2)\to \infty$ and $\alpha_s(M^2) \to 0$ with $\alpha_s(M^2)\log(M^2 \bm b^2)$ fixed.

We have investigated how well the virtuality ordered parton shower algorithm defined in refs.~\cite{NSI, NSII, NSIII} and this paper performs in summing large logarithms for the transverse momentum of a vector boson produced in the Drell-Yan process in hadron-hadron collisions. The same analysis would apply to the transverse momentum distribution of Higgs bosons produced in hadron-hadron collisions. There are other sorts of summations of large logarithms that are of importance for understanding experiments. We believe that the investigation of other cases, perhaps using methods developed in this paper, is an important goal, both for the virtuality ordered parton shower algorithm used in this paper and for other parton shower algorithms. 

\acknowledgments{ 
This work was supported in part by the United States Department of Energy and the Helmoltz Alliance ``Physics at the Terascale."
}
\appendix

\section{Structure of inclusive splitting}
\label{sec:inclusive}

In this appendix, we review formulas from refs.~\cite{NSI, NSII, NSIII} for inclusive splitting, that is the splitting operator ${\cal H}_\LI(t)$ times the inclusive measurement function $\sbra{1}$. We start with in eq.~(12.20) of ref.~\cite{NSI},
\begin{equation}
\label{eq:1Hencore}
\sbra{1}{\cal H}_\LI(t)\sket{\{p,f,s',c',s,c\}_{m}}
= 2\,\brax{\{s'\}_{m}}\ket{\{s\}_{m}}\,
\bra{\{c'\}_{m}}h(t,\{p,f\}_{m})\ket{\{c\}_{m}}
\;\;.
\end{equation}
The inclusive splitting function is diagonal in spin, but it has a non-trivial color structure. The structure of the splitting is contained in the operator $h$, which is given in eq.~(12.21) of ref.~\cite{NSI}. With a slightly modified notation and with one error corrected, this equation reads
\begin{equation}
\begin{split}
\label{eq:hdef}
\bra{\{c'\}_{m}}&h(t,\{p,f\}_{m})\ket{\{c\}_{m}}
=
\\&
\frac{1}{2}
\sum_l
\sum_{\zeta_{\rm f}\in \Phi_{l}(f_l)}
\int d\zeta_{\rm p}\
\theta(\zeta_{\rm p} \in \varGamma_{l}(\{p\}_{m},\zeta_{\rm f}))
\\&\times
\delta\!\left(
t - \log\left(\frac{Q_0^2}
{|(\hat p_l 
+(-1)^{\delta_{l,\La} + \delta_{l,\Lb}}
 \hat p_{m+1})^2 - m^2(f_l)|}\right)
\right)
\\&\times
\frac
{n_\Lc(a) n_\Lc(b)\,\eta_{\La}\eta_{\Lb}}
{n_\Lc(\hat a) n_\Lc(\hat b)\,
 \hat \eta_{\La}\hat \eta_{\Lb}}\,
\frac{
f_{\hat a/A}(\hat \eta_{\La},\mu^{2}_{F})
f_{\hat b/B}(\hat \eta_{\Lb},\mu^{2}_{F})}
{f_{a/A}(\eta_{\La},\mu^{2}_{F})
f_{b/B}(\eta_{\Lb},\mu^{2}_{F})}
\\ &\times
\biggl\{
\theta(\hat f_{m+1}\ne {\rm g})\
\brax{\{c'\}_{m}}\ket{\{c\}_{m}}\,
C(\hat f_l, \hat f_{m+1})\
\overline w_{ll}(\{\hat f,\hat p\}_{m+1})
\\ &\ \ -
\theta(\hat f_{m+1} = {\rm g})\sum_{k \ne l}
\bra{\{c'\}_{m}}
\bm T_k\cdot \bm T_l
\ket{\{c\}_{m}}
\\& \quad\quad\times
\big[
\overline w_{ll}(\{\hat f,\hat p\}_{m+1})
-
2\,A_{lk}(\{\hat p\}_{m+1})\,
\overline w_{lk}(\{\hat f,\hat p\}_{m+1})
\big]
\bigg\}
\;\;.
\end{split}
\end{equation}
Here possible quark masses are included. There is a sum over the index $l$ of the initial state or final state parton that splits. There is a sum over flavor choices $\zeta_\Lf$ for the splitting and an integration over the momentum choices, $\int\! d\zeta_\Lp$. Once we choose splitting variables $t$, $z$ and $\phi$, this becomes an integration over the splitting variables. There is a delta function that expresses the definition of the shower time $t$. Next, there is a ratio of parton luminosities, containing parton distribution functions, as in eq.~(\ref{eq:rhopertdef}). The main momentum and color dependence is contained inside the braces that follow. 

The second term inside the braces applies when the newly emitted parton with label $m+1$ is a gluon. There is a sum over partons $k$, where $k \ne l$. Partons $l$ and $k$ constitute a color dipole that can emit the gluon. There are functions $\overline w_{ll}$ and $\overline w_{lk}$ that describe the direct terms and the interference terms, respectively, for gluon emission. These functions are defined in ref.~\cite{NSI}. Convenient expressions for $\overline w_{ll}$ and $\overline w_{lk}$ as rational functions of dot products of the momenta involved are given in ref.~\cite{NSII}.\footnote{In ref.~\cite{NSII} and also in ref.~\cite{NSIII}, we use the notations $\overline W_{ll} = \overline w_{ll}$ and $\overline W_{lk} = 2 A_{lk}\overline w_{lk}$.} There is also a function $A_{lk}$ that expresses how the interference term is partitioned between a part considered to be a splitting of parton $l$ and a corresponding part considered to be a splitting of parton $k$. There is a color operator $\bm T_k\cdot \bm T_l$ that inserts a gluon color matrix $T^a$ on line $k$ and another on line $l$ and sums over the gluon color index $a$. This operator was denoted $g_{lk}(\{\hat f\}_{m+1})$ in the original equation.

The first term inside the braces applies when the newly emitted parton with label $m+1$ is not gluon. Then there is no interference term. There is a color factor
\begin{equation}
C(\hat f_l, \hat f_{m+1}) = 
\begin{cases}
T_{\LR} & \{\hat f_l, \hat f_{m+1}\} = \{q, \bar q\}\ {\rm or}\
\{\bar q, q\}
\\
C_\LF &   \{\hat f_l, \hat f_{m+1}\} = 
\{\Lg, q\}, \{\Lg, \bar q\}, 
\{q,\Lg\}\,{\rm or}\ \{\bar q,\Lg\}
\\
C_\LA &   \{\hat f_l, \hat f_{m+1}\} = 
\{\Lg, \Lg\}
\end{cases}
\;\;.
\end{equation}
The cases with $\hat f_{m+1} = \Lg$ do not appear in eq.~(\ref{eq:hdef}), but we will need these cases later. In eq.~(12.21) of ref.~\cite{NSI}, we failed to note the possibility of an initial state splitting with $\{\hat f_l, \hat f_{m+1}\} = \{\Lg, q\}\ {\rm or}\ \{\Lg, \bar q\}$, so we listed this factor simply as $T_{\LR}$.

We now manipulate eq.~(\ref{eq:hdef}) so as to obtain a more useful form. For the $\hat f_{m+1} = {\rm g}$ term, we divide the splitting function into two parts by adding and subtracting the eikonal approximation to $\overline w_{ll}$. (The eikonal approximation is the limiting form when the gluon momentum tends to zero. The function $\overline w_{lk}$ is already constructed in the eikonal approximation in ref.~\cite{NSI}.) Thus we write
\begin{equation}
\begin{split}
\overline w_{ll}(\{\hat f,\hat p\}_{m+1})
-&
2A_{l k}(\{\hat p\}_{m+1})\,
\overline w_{l k}(\{\hat f,\hat p\}_{m+1})
\\
={}& 
\big(
\overline w_{ll}(\{\hat f,\hat p\}_{m+1})
- \overline w_{ll}^{\rm eikonal}(\{\hat f,\hat p\}_{m+1})
\big)
\\&
+ 
\big(
\overline w_{ll}^{\rm eikonal}(\{\hat f,\hat p\}_{m+1})
- 2A_{l k}(\{\hat p\}_{m+1})\,
\overline w_{l k}(\{\hat f,\hat p\}_{m+1})
\big)
\;\;.
\end{split}
\end{equation}

The first term is important in only in the collinear limit. The momentum dependent factor $\overline w_{ll} - \overline w_{ll}^{\rm eikonal}$ is independent of $k$ and multiplies $\bra{\{c'\}_{m}} \bm T_k\cdot \bm T_l \ket{\{c\}_{m}}$. When we sum over $k$, we can use 
\begin{equation}
\begin{split}
\label{eq:colorsumagain}
\sum_{k\ne l} \bra{\{c'\}_{m}} \bm T_k\cdot \bm T_l \ket{\{c\}_{m}}
={}&
-\bra{\{c'\}_{m}} \bm T_l\cdot \bm T_l \ket{\{c\}_{m}}
\\
={}&
- C(\{\hat f_l, \hat f_{m+1}\})\ \brax{\{c'\}_{m}} \ket{\{c\}_{m}}
\;\;.
\end{split}
\end{equation}

For the second term, we take $A_{lk}$ to be of the form specified in eq.~(7.2) of ref.~\cite{NSIII}, in which $A_{lk}$ is expressed using another function $A'_{lk}$. This gives the result of in eq.~(7.10) of ref.~\cite{NSIII},
\begin{equation}
\begin{split}
\label{eq:AlktoAlkprime}
\overline w_{ll}^{\rm eikonal}(\{\hat f,\hat p\}_{m+1})
- & 2A_{l k}(\{\hat p\}_{m+1})\,
\overline w_{l k}(\{\hat f,\hat p\}_{m+1})
\\
={}& 4\pi\as\, A_{l k}'(\{\hat p\}_{m+1})\
\frac{-\hat P_{l k}^2}
{(\hat p_{m+1}\!\cdot\!\hat p_l\ \hat p_{m+1}\!\cdot\!\hat p_k)^2}
\;\;,
\end{split}
\end{equation}
where
\begin{equation}
\hat P_{lk} = \hat p_{m+1}\!\cdot\!\hat p_l\ \hat p_k
-\hat p_{m+1}\!\cdot\!\hat p_k\ \hat p_l
\;\;.
\end{equation}
This is for general masses. For the analysis of the $Z$-boson $\bm p_\perp$ distribution, we want all of the partons to be massless. Then
\begin{equation}
\begin{split}
\big(
\overline w_{ll}^{\rm eikonal}(\{\hat f,\hat p\}_{m+1})
- & 2A_{l k}(\{\hat p\}_{m+1})\,
\overline w_{l k}(\{\hat f,\hat p\}_{m+1})
\big)
\\
={}& 4\pi\as\, A_{l k}'(\{\hat p\}_{m+1})\
\frac{2 \hat p_l\!\cdot\!\hat p_k}
{\hat p_{m+1}\!\cdot\!\hat p_l\ \hat p_{m+1}\!\cdot\!\hat p_k}
\;\;.
\end{split}
\end{equation}
Our choice in this paper for the function $A_{lk}'$ is given in eq.~(\ref{eq:Alkangle}).

With these rearrangements, we have
\begin{equation}
\begin{split}
\label{eq:inclusiveH}
\sbra{1}{\cal H}_\LI(t)&\sket{\{p,f,s',c',s,c\}_{m}}
\\
={}& \brax{\{s'\}_{m}}\ket{\{s\}_{m}}\,
\sum_l
\sum_{\zeta_{\rm f}\in \Phi_{l}(f_l)}
\int d\zeta_{\rm p}\
\theta(\zeta_{\rm p} \in \varGamma_{l}(\{p\}_{m},\zeta_{\rm f}))
\\&\times
\delta\!\left(
t - \log\left(\frac{Q_0^2}
{|(\hat p_l 
+(-1)^{\delta_{l,\La} + \delta_{l,\Lb}}
 \hat p_{m+1})^2 - m^2(f_l)|}\right)
\right)
\\&\times
\frac
{n_\Lc(a) n_\Lc(b)\,\eta_{\La}\eta_{\Lb}}
{n_\Lc(\hat a) n_\Lc(\hat b)\,
 \hat \eta_{\La}\hat \eta_{\Lb}}\,
\frac{
f_{\hat a/A}(\hat \eta_{\La},\mu^{2}_{F})
f_{\hat b/B}(\hat \eta_{\Lb},\mu^{2}_{F})}
{f_{a/A}(\eta_{\La},\mu^{2}_{F})
f_{b/B}(\eta_{\Lb},\mu^{2}_{F})}
\\ &\times
\biggl\{
\brax{\{c'\}_{m}}\ket{\{c\}_{m}}\,
C(\hat f_l, \hat f_{m+1})
\\&\qquad\qquad \times
\Big[\overline w_{ll}(\{\hat f,\hat p\}_{m+1})
- \theta(\hat f_{m+1} = {\rm g})\,
\overline w_{ll}^{\rm eikonal}(\{\hat f,\hat p\}_{m+1})
\Big]
\\ &\ \ -
\theta(\hat f_{m+1} = {\rm g})\sum_{k \ne l}
\bra{\{c'\}_{m}}
\bm T_k\cdot \bm T_l
\ket{\{c\}_{m}}
\\& \quad\quad\times
4\pi\as\, A_{l k}'(\{\hat p\}_{m+1})\
\frac{-\hat P_{l k}^2}
{(\hat p_{m+1}\!\cdot\!\hat p_l\ \hat p_{m+1}\!\cdot\!\hat p_k)^2}
\bigg\}
\;\;.
\end{split}
\end{equation}

We can now specialize to initial state splittings with massless partons. In the sum over the label $l$ of the parton that splits, we take $l = \La$. The integration measure $d\zeta_\Lp$, with the choice of variables $\{t,z,\phi\}$ used in this paper, is
\begin{equation}
\begin{split}
\label{eq:dzetap2}
d\zeta_\Lp ={}& 
2p_\La\cdot p_\Lb\,
(2\pi)^{-2} dt\,dz\,
\frac{d\phi}{2\pi}\,
\frac{y}{4 z }
= \frac{\hat p_{m+1}\!\cdot\!\hat p_\La}{8\pi^2\,z}\,
dt\,dz\,\frac{d\phi}{2\pi}\,
\;\;.
\end{split}
\end{equation}
The integration over $t$ is performed using the delta function that defines $t$. We have defined the matrix element of ${\cal H}^{\rm pert}_\La(t,z,\phi,f')$ in eqs.~(\ref{eq:HtoHpert}) and (\ref{eq:Hexpansion}) to be the coefficient of $dz\,d\phi/(2\pi)$ in the integrand and sum over flavors, leaving out the ratio of parton luminosities. {Cf.} eqs.~(\ref{eq:Hexpansionnonpert}) and (\ref{eq:HtoHpert2}). Thus 
\begin{equation}
\begin{split}
\label{eq:inclusiveHapert}
\sbra{1}z{\cal H}^{\rm pert}_\La(t,z,\phi,f')&\sket{\{p,f,s',c',s,c\}_{m}}
\\
={}& 
\frac{\hat p_{m+1}\!\cdot\!\hat p_\La}{8\pi^2}\,
\brax{\{s'\}_{m}}\ket{\{s\}_{m}}
\\ &\times
\biggl\{
\brax{\{c'\}_{m}}\ket{\{c\}_{m}}\,
C(\hat f_\La, \hat f_{m+1})
\\&\qquad \times
\Big[\overline w_{\La \La}(\{\hat f,\hat p\}_{m+1})
- \theta(\hat f_{m+1} = {\rm g})\,
\overline w_{\La \La}^{\rm eikonal}(\{\hat f,\hat p\}_{m+1})
\Big]
\\ &\ \ -
\theta(f' = {\rm g})\sum_{k \ne \La}
\bra{\{c'\}_{m}}
\bm T_k\cdot \bm T_\La
\ket{\{c\}_{m}}
\\& \quad\quad\times
4\pi\as\, A_{\La k}'(\{\hat p\}_{m+1})\
\frac{2 \hat p_\La\!\cdot\!\hat p_k}
{\hat p_{m+1}\!\cdot\!\hat p_\La\ \hat p_{m+1}\!\cdot\!\hat p_k}
\bigg\}
\;\;.
\end{split}
\end{equation}

We can use the results of ref.~\cite{NSII} to state the functions $\overline w_{\La \La}$ in terms of the variables $y,z,\phi$ used in this paper. For $\{\hat f_\La, \hat f_{m+1}\} = \{q,\Lg\} \ {\rm or}\ \{\bar q,\Lg\}$, we find, using eq.~(2.29) of ref.~\cite{NSII},
\begin{equation}
\label{eq:waa1}
\frac{\hat p_{m+1}\!\cdot\!\hat p_\La}{8\pi^2}
\Big[\overline w_{\La \La}
- 
\overline w_{\La \La}^{\rm eikonal}
\Big]
=
\frac{\as}{2\pi}\,\frac{1}{z}\,
(1-z)(1+y)
\;\;.
\end{equation}
For $\{\hat f_\La, \hat f_{m+1}\} = \{\Lg,\Lg\}$, we find, using eq.~(2.57) of ref.~\cite{NSII},
\begin{equation}
\label{eq:waa2}
\frac{\hat p_{m+1}\!\cdot\!\hat p_\La}{8\pi^2}
\Big[\overline w_{\La \La}
- 
\overline w_{\La \La}^{\rm eikonal}
\Big]
=
\frac{\as}{2\pi}\,\frac{2}{z}
\left[
\frac{1-z}{z}
+ z(1-z)
-\frac{z(1-z)y}
{(1-z) + y}
\right]
\;\;.
\end{equation}
For $\{\hat f_\La, \hat f_{m+1}\} = \{q,\bar q\} \ {\rm or}\ \{\bar q,q\}$, we find, using eq.~(A.2) of ref.~\cite{NSII},
\begin{equation}
\label{eq:waa3}
\frac{\hat p_{m+1}\!\cdot\!\hat p_\La}{8\pi^2}\
\overline w_{\La \La}
=
\frac{\as}{2\pi}\,\frac{1}{z}\,
\frac{1 + (1-z)^2}{z}
\;\;.
\end{equation}
For $\{\hat f_\La, \hat f_{m+1}\} = \{\Lg,\bar q\} \ {\rm or}\ \{\Lg,q\}$, we find, using eq.~(A.3) of ref.~\cite{NSII},
\begin{equation}
\label{eq:waa4}
\frac{\hat p_{m+1}\!\cdot\!\hat p_\La}{8\pi^2}\
\overline w_{\La \La}
=
\frac{\as}{2\pi}\,\frac{1+y}{z}
\left[
z^2 + (1-z)^2
\right]
\;\;.
\end{equation}

The contribution to $\sbra{1}z{\cal H}^{\rm pert}_\La(t,z,\phi,f')\sket{\{p,f,s',c',s,c\}_{m}}$ from the term proportional to $A'_{\La k}$ in the last line of eq.~(\ref{eq:inclusiveHapert}) vanishes if the emitted parton is not a gluon and if $f' = {\rm g}$ it is
\begin{equation}
\begin{split}
\sbra{1}z&{\cal H}^{\rm pert}_\La(t,z,\phi,\Lg)
\sket{\{p,f,s',c',s,c\}_{m}}_{\rm eikonal}
\\
={}&\brax{\{s'\}_{m}}\ket{\{s\}_{m}}
(-1)\sum_{k \ne \La}
\bra{\{c'\}_{m}}
\bm T_k\cdot \bm T_\La
\ket{\{c\}_{m}}\,
\frac{\as}{2\pi}\, A_{\La k}'(\{\hat p\}_{m+1})\
\frac{2 \hat p_\La\!\cdot\!\hat p_k}
{\hat p_{m+1}\!\cdot\!\hat p_k}
\;\;.
\end{split}
\end{equation}
Let $r_k$ be the rapidity of parton $k$ in the $p_\La + p_\Lb$ rest frame and let $\phi_k$ its azimuthal angle, while $\phi$ denotes the azimuthal angle of  parton $m+1$. Then, using the definition (\ref{eq:Alkangle}) of $A_{lk}'$ from eq.~(7.12) of ref.~\cite{NSIII}, we find
\begin{equation}
\begin{split}
\sbra{1}z&{\cal H}^{\rm pert}_\La(t,z,\phi,\Lg)
\sket{\{p,f,s',c',s,c\}_{m}}_{\rm eikonal}
\\
={}&\brax{\{s'\}_{m}}\ket{\{s\}_{m}}
(-1)\sum_{k \ne \La}
\bra{\{c'\}_{m}}
\bm T_k\cdot \bm T_\La
\ket{\{c\}_{m}}\,
\frac{\as}{2\pi}\, 
\frac{2}{1-z+y}\, f(z,y,\phi,r_k)
\;\;,
\end{split}
\end{equation}
where
\begin{equation}
\begin{split}
f(z,y,\phi,r_k)
={}&
\bigg[
\frac{1}{1+y}
- e^{r_k} \frac{2\sqrt{(1-z)y}}{1-z+y}\,\frac{z}{1+y}\cos(\phi - \phi_k)
\\&
+ e^{2r_k}\frac{2 y}{(1-z+y)}\,\frac{z^2}{(1+y)^2}
\bigg]^{-1} 
\;\;.
\end{split}
\end{equation}

In the limit that the emitted gluon is soft, we can neglect $(1-z)$ compared to 1 and $y$ compared to 1. We do not, however, neglect $y$ compared to $(1-z)$. Then
\begin{equation}
\begin{split}
\label{eq:fapprox}
f(z,y,\phi,r_k)
\approx{}&
\bigg[
1
- e^{r_k} \frac{2\sqrt{(1-z)y}}{1-z+y}\,\cos(\phi - \phi_k)
+ e^{2r_k}\frac{2 y}{(1-z+y)}
\bigg]^{-1} 
\;\;.
\end{split}
\end{equation}
We use this result in section \ref{sec:1mzll1}.

When the emitted gluon becomes collinear, $y \to 0$ with fixed $z$, we get
\begin{equation}
f(z,0,\phi,r_k) = 1
\;\;.
\end{equation}
Then we can use eq.~(\ref{eq:colorsumagain}), so that
\begin{equation}
\begin{split}
\sbra{1}z&{\cal H}^{\rm pert}_\La(t,z,\phi,\Lg)
\sket{\{p,f,s',c',s,c\}_{m}}_{\rm eikonal}
\\
\approx{}&\brax{\{s'\}_{m}}\ket{\{s\}_{m}}
\brax{\{c'\}_{m}} \ket{\{c\}_{m}}\,
C(\hat f_l, \hat f_{m+1})\,
\frac{\as}{2\pi}\, \frac{2}{z}\,
\frac{z}{1-z+y}
\;\;.
\end{split}
\end{equation}
The behavior of the splitting probability in the collinear limit is obtained by combining this result with the results in eqs.~(\ref{eq:waa1}), (\ref{eq:waa2}), (\ref{eq:waa3}), and (\ref{eq:waa4}).

\section{An integral of Bessel functions}
\label{sec:Bessel}

In this appendix, we prove a theorem about the $J_0$ Bessel function that is needed in section~\ref{sec:small1mzsimplified}. The theorem concerns an integral of the form
\begin{equation}
F(x_2,x_1) = \int_{x_1}^{x_2} \frac{dx}{x}\
f(\log x)\,
[J_0(x) - 1]
\;\;.
\end{equation}
where $x$ here stands for $|\bm k_\perp| |\bm b|$. We imagine that $x_1 \ll 1$ and $x_2 \gg 1$. We suppose that the function $f(\log x)$ is slowly varying in the sense that its second derivative is small: $f''(\log x) \ll f(\log x)$. In our application, $f$ is $\as(\bm k_\perp^2) = \as(x^2/\bm b^2)$. This is slowly varying because its second derivative with respect to $\log(x)$ is proportional to $\as^3$. If the integration extends down to small values of $\bm k_\perp^2$, then $\as(\bm k_\perp^2)$ is not slowly varying if we use the perturbative evolution equation, so the theorem is useful only if we use an approximation in which $\as(\bm k_\perp^2)$ stops varying for small values of $\bm k_\perp^2$. 

The theorem concerns the approximation of $F(x_2,x_1)$ by
\begin{equation}
F(x_2,x_1)_{\rm approx} = -\int_{x_1}^{x_2} \frac{dx}{x}\
f(\log x)\,
\theta(x > x_0)
\;\;,
\end{equation}
where
\begin{equation}
x_0 = 2 e^{-\gamma_{\rm E}}
\;\;.
\end{equation}
We thus analyze the difference,
\begin{equation}
\Delta F(x_2,x_1) = F(x_2,x_1) - F(x_2,x_1)_{\rm approx}
\;\;,
\end{equation}
which is given by
\begin{equation}
\Delta F(x_2,x_1) = \int_{x_1}^{x_2} \frac{dx}{x}\
f(\log x)\,
[J_0(x) - \theta(x < x_0)]
\;\;.
\end{equation}
We expect that $\Delta F(x_2,x_1)$ is small because for the integration range $x \gg 1$, $J_0(x)$ oscillates, so that its integral is small, while for the integration range $x \ll 1$, $J_0(x) \approx 1$, so that it cancels the term $\theta(x < x_0)$.

To prove that $\Delta F(x_2,x_1)$ is small, we define
\begin{equation}
I_1(x) = \int_0^x \!\frac{d\bar x}{\bar x}\ 
[J_0(\bar x) - \theta(\bar x < x_0)]
\;\;.
\end{equation}
Now, $I_1(x)$ vanishes as $x \to 0$. The approach to the limit is given by
\begin{equation}
I_1(x) \sim -\frac{x^2}{8} \;\;,\;\; x \to 0
\;\;.
\end{equation}
The integral converges for $x \to \infty$ and, with our special choice of $x_0$, its value is $I_1(\infty) = 0$.  The approach to the limit is given by
\begin{equation}
I_1(x) \sim
-\sqrt{\frac{2}{\pi}}\, \frac{1}{x^{3/2}}\sin \left(\frac{\pi}{4} -  x \right)
\;\;,\;\; x \to \infty
\;\;.
\end{equation}
We rewrite $\Delta F(x_2,x_1)$ using an integration by parts:
\begin{equation}
\begin{split}
\Delta F(x_2,x_1) ={}& \int_{x_1}^{x_2} \frac{dx}{x}\
f(\log x)\,
x\frac{d}{dx}I_1(x)
\\
={}& 
\left[f(\log x_2)\,I_1(x_2) - f(\log x_1)\,I_1(x_1)\right]
\\&
-\int_{x_1}^{x_2} \frac{dx}{x}\
f'(\log x)\,
I_1(x)
\;\;.
\end{split}
\end{equation}

Now define
\begin{equation}
I_2(x) = \int_0^x \!\frac{d\bar x}{\bar x}\ 
\log(x/\bar x)\
[J_0(\bar x) - \theta(\bar x < x_0)]
\;\;.
\end{equation}
Note that the derivative of $I_2(x)$ with respect to $\log x$ is $I_1(x)$. Again, $I_2(0) = 0$.  The approach to the limit is given by
\begin{equation}
I_2(x) \sim -\frac{x^2}{16}
\;\;,\;\; x \to 0
\;\;.
\end{equation}
Again, the integral converges for $x \to \infty$ and its value is, perhaps surprisingly, $I_2(\infty) = 0$. This can be proved using the generating function in eq.~(4.12) of ref.~\cite{Ellis1981}. The approach to the limit is given by
\begin{equation}
I_2(x) \sim
-\sqrt{\frac{2}{\pi}}\, \frac{1}{x^{5/2}}\cos \left(\frac{\pi}{4}- x \right)
\;\;,\;\; x \to \infty
\;\;.
\end{equation}
We can write $\Delta F(x_2,x_1)$ using
\begin{equation}
\begin{split}
\Delta F(x_2,x_1) ={}& \left[f(\log x_2)\,I_1(x_2) - f(\log x_1)\,I_1(x_1)\right]
\\&
-\int_{x_1}^{x_2} \frac{dx}{x}\
f'(\log x)\,
x\frac{d}{dx}I_2(x)
\;\;.
\end{split}
\end{equation}
This gives
\begin{equation}
\begin{split}
\label{eq:Fremainder}
\Delta F(x_2,x_1) ={}& \left[f(\log x_2)\,I_1(x_2) - f(\log x_1)\,I_1(x_1)\right]
\\&
-\left[f'(\log x_2)\,I_2(x_2) - f'(\log x_1)\,I_2(x_1)\right]
\\&
+\int_{x_1}^{x_2} \frac{dx}{x}\
f''(\log x)\,
I_2(x)
\;\;.
\end{split}
\end{equation}

Now we can see the conditions under which $\Delta F(x_2,x_1)$ is small. There are terms proportional to $f(\log x_2)$ and $f'(\log x_2)$. Using the large $x$ behavior of $I_1(x_2)$ and $I_2(x_2)$, we see that these contributions are power suppressed for large $x_2$ as long as $f(\log x_2)$ and $f'(\log x_2)$ do not grow for large $x_2$. There are terms proportional to $f(\log x_1)$ and $f'(\log x_1)$. Using the small $x$ behavior of $I_1(x_1)$ and $I_2(x_1)$, we see that these contributions are power suppressed for small $x_1$ as long as $f(\log x_1)$ and $f'(\log x_1)$ do not grow for small $x_1$. Finally, there is an integral of $f''(\log x)\,I_2(x)$. Since $I_2(x)$ falls off for large $x$ and for small $x$, it is the behavior of $f''(\log x)$ around $x = 1$ that is important in the integration (as long as $f''(\log x)$ is bounded everywhere). If $f''(\log x)$ is small, $\Delta F(x_2,x_1)$ will be small.

Additionally, $\Delta F(x_2,x_1)$ is small when both $x_1$ and $x_2$ are large compared to 1. In this case, all of the terms in eq.~(\ref{eq:Fremainder}) are suppressed by powers of $1/x_1$ and $1/x_2$.


\end{document}